\documentstyle[epsfig]{mn2e}

\def\Msun{\hbox{$\rm\thinspace M_{\odot}$}}

\def\ltsima{$\; \buildrel < \over \sim \;$}
\def\simlt{\lower.5ex\hbox{\ltsima}}
\def\gtsima{$\; \buildrel > \over \sim \;$}
\def\simgt{\lower.5ex\hbox{\gtsima}}

\title[SMBHs and Their Environments]{Supermassive Black Holes and Their Environments}
\author[J.M.\ Colberg \& T.\ Di Matteo]
       {J\"org M.\ Colberg$^{1,2}$, Tiziana Di Matteo$^{1}$\\
        $^1$ Carnegie Mellon University, Department of Physics,
             5000 Forbes Avenue, Pittsburgh PA 15213, USA\\
        $^2$ Department of Astronomy, University of Massachusetts at 
             Amherst, 710 North Pleasant Street, Amherst MA 01002, 
             USA}
\date{Accepted 200? ???? ??.
      Received 2007 ???? ??;
      in original form 2007  xx}

\pagerange{\pageref{firstpage}--\pageref{lastpage}}
\pubyear{200?}

\begin{document}

\maketitle

\label{firstpage}

\begin{abstract}
We make use of the first high--resolution hydrodynamic simulations of
structure formation which self-consistently follows the build up of
supermassive black holes introduced in Di Matteo et al. (2007) to 
investigate the relation between black holes (BH),
host halo and large--scale environment. There are well--defined relations
between halo and black hole masses and between the activities of galactic
nuclei and halo masses at low redshifts. A large fraction of black holes forms
anti--hierarchically, with a higher ratio of black hole to halo mass at high
than at low redshifts. At $z=1$, we predict group environments (regions of enhanced
local density) to contain the highest mass and most active (albeit with a
large scatter) BHs while the rest of the BH population to be spread over all
densities from groups to filaments and voids. Density dependencies are more
pronounced at high rather than low redshift. These results are consistent with
the idea that gas rich mergers are likely the main regulator of quasar
activity. We find star formation to be a somewhat stronger and tighter
function of local density than BH activity, indicating some difference in the
triggering of the latter versus the former. There exists a large number of
low--mass black holes, growing slowly predominantly through accretion, which
extends all the way into the most underdense regions, i.e. in voids.
\end{abstract}


\section{Introduction}

Remnant supermassive black holes (BHs) from an early quasar phase are now
being found ubiquitously in the centers of local galaxies. Even more
remarkably, a number of tight correlations have been discovered between the
black hole mass $m_{\rm BH}$ and properties of the host, such as
the bulge mass, $m_{\rm bulge}$, or K-band luminosity or its velocity dispersion,
$\sigma$ (e.g.;Ferrarese and Merritt 2000, Gebhardt et al. 2000, Marconi \&
Hunt 2003, Haring \& Rix 2004). These relations demonstrate a strong link
between BHs and galaxy formation and have thus motivated a large theoretical
effort to address their origin and evolution (e.g., Kauffmann \& Haehnelt
2000, Adams et al. 2001, Di Matteo et al. 2003, Hatziminaoglou et al. 2003, 
Hopkins et al. 2005, Bower et al. 2006, Croton et al. 2006, Marulli et al. 
2007 ; for a review of semi--analytical models especially Baugh 2006). 

Recent ultra-deep {\em Chandra} X-ray observations of submillimeter galaxies
(SMGs; Alexander et al. 2005) imply that the time of rapid black hole growth
is related to activity at sites of intensive star formation and hence to
massive flows of gas at the center of galaxies. SMGs are indeed roughly coeval
with the peak of the quasar phase. UV images from the {\em Hubble Space
Telescope} show that a considerable fraction of SMGs are undergoing a major
merger (Conselice et al. 2003, Pope et al. 2005). In the local universe, 
ultraluminous infrared (ULIRG) systems associated with merger driven starburst 
activity have now been directly shown to be associated with black hole growth 
(Komossa et al. 2003), supporting the same picture for a link between quasar 
activity and merger induced starbursts in galaxies and formation of bulges. 

Quasars with inferred black hole masses of around $10^9$\,M$_\odot$ at
redshifts of $z \sim 6$ (Fan et al. 2003, Jiang et al. 2007, Kurk et al. 2007) 
have been discovered. In view of our standard picture of structure formation, 
the implications from the local $m_{\rm BH}$--$\sigma$ and the association of 
black hole fuelling with a major merger and a starburst, it is challenging 
to understand the presence of such massive objects at $z \sim 6$. Outstanding 
questions remain as to where and how these first BH form and what their 
descendants in our local universe are. Furthermore, it is now known that the 
space density of low luminosity X--ray selected AGN peaks at lower redshifts 
than that of high luminosity ones (e.g. Steffen et al. 2003, Ueda et al. 2003, 
Hasinger et al. 2005), an effect called ``cosmic downsizing''. If the build--up 
of galaxies proceeds from smaller ones to larger ones, it is hard to explain 
why the bulk of the mass in the largest black holes and brightest AGN/quasars 
was acquired first, whereas today the lower mass black holes are being 
assembled.
 
These and other lines of evidence firmly demonstrate that even though the
growth of black holes is intimately linked to the formation and evolution 
of galaxies, it is not directly linked to the growth of the halo mass function.
Naturally, the growth of black holes and galaxies has to be addressed in the 
framework of the standard $\Lambda$ cold dark matter ($\Lambda$CDM) cosmology.
In that context, semi--analytical models have been developped (e.g. Cattaneo 
et al. 1999, Kauffmann \& Haehnelt 2000, Granato et al. 2001, Wyithe \& Loeb 
2003), which at first linked black hole directly to the growth of dark matter 
halos and quasar activity to major mergers. The latest generation of these
models has started to incorporate AGN feedback, which truncates ongoing star 
formation and suppresses cooling in more massive halos (see, e.g; Granato et 
al. 2004, Monaco \& Fontanot 2005, Bower et al. 2006, Croton et al. 2006, 
De Lucia et al. 2006, Malbon et al. 2007). Such models often suggest that 
the fundamental relations between galaxy structure and black holes may arise 
if feedback energy expels (in the form of strong outflows) the nearby gas and 
shuts down the accretion phase (Ciotti \& Ostriker 1997, Silk \& Rees 1998, 
Fabian 1999, Wyithe \& Loeb 2003). This picture has now been confirmed in 
studies employing detailed galaxy simulations (Di Matteo et al. 2005; 
Robertson et al. 2006; Di Matteo et al. 2007).

The coupling of mergers, star formation with black hole growth/quasar activity
in the context of galaxy evolution is difficult to treat on the basis of
analytical estimates alone. Such methods unavoidably neglect the dynamics of
quasar evolution in galaxies, and they cannot predict time--dependent effects
such as the characteristic lifetime of the accretion phase prior to its
self-termination, the available fueling rates driven by gas inflows that
regulate the black hole accretion, and their dependence on the environment.

In this paper we present a detailed study of the cosmological evolution of the
demographics of supermassive black holes as a function of their local
environment. We use the first full hydrodynamical high--resolution simulation
of a cosmological volume that incorporates black hole growth and associated
feedback self-consistently (Di Matteo et al. 2007, Sijacki et al. 2007), based
on the methodology developed and explored in hydrodynamical simulation of
galaxy mergers (Di Matteo, Springel \& Hernquist 2005; Springel, Di Matteo \&
Hernquist 2005). Using a direct cosmological SPH simulation, besides treating
the hierarchical assembly of black holes and their haloes, allows us to
consider the physical conditions of the gas inflows that drive star formation
and lead to the growth of the central black holes in galaxies as well as their
interaction with associated feedback processes.

In Di Matteo et al. (2007), we discussed some promising results from this
simulation method, which include its ability to reproduce the local value of
the black hole mass density, $\rho_{BH}$, and its extrapolation to higher
redshift ($z< 2.5$), as well as a peak in the global black hole accretion rate
history at $z \sim 2-3$, the expected peak of the quasar phase, and a sharp
drop at higher redshifts (see also Sijacki et al. 2007). These trends are
consistent with the general picture, in which gas is available for star
formation and eventually gets to the central regions to ignite the quasar
activity in major mergers. In addition, the locations of our cosmological
galaxies and black holes agree very well with the local $m_{\rm BH} -
m_{\sigma}$ and $m_{\rm BH}-m_{*}$ relations over a very large dynamic range
and predict an evolution consistent with recent observational studies (Woo et
al. 2006, Shields et al. 2006, Peng et al. 2006). 
Here, we carry out a detailed study of the local environments of black holes
as a function of redshift. We will look at the relations between black hole
hole mass and accretion rate for BHs in the simulation with halo mass,
relevant also for quasar clustering measurements. If indeed gas rich mergers
are expected to be primarily responsible for triggering quasar activity we
also expect a dependence of BH accretion on local environment. We will look at
the relationship between black hole mass and accretion and local density.
Along the lines of the topics just outlined, we will address how and where BH
form.

This paper is organized as follows. In the following Section
(\ref{sec:simulation}), we first briefly introduce the numerical code and
simulation data. In Section~\ref{sec:connection}, we study the connection
between BH and their host haloes, including the co--evolution of black holes
and their host haloes (\ref{sec:coevolution}), the relations between black
hole mass and host halo mass (\ref{sec:hosts}), and formation epochs
(\ref{sec:formation}). Section~\ref{sec:environments} deals with the
large--scale environment, in particular the dependence of black hole mass,
accretion rate, and of host galaxy star formation rate on large--scale
environment (\ref{sec:largescale}), a short check whether our definition of
environment introduces a systematic effect (\ref{sec:density}), a study of the
lowest mass black holes in the simulation volume (\ref{sec:starvation}), and
the dependence of the assembly mode (accretion versus mergers) on environment
(\ref{sec:assembly}). With the summary in Section~\ref{sec:summary}, we
conclude our work.

%
%
\begin{figure*}
  \begin{center}
    \begin{tabular}{cc}
      \begin{minipage}{85mm}
        \begin{center}
          \includegraphics[width=70mm]{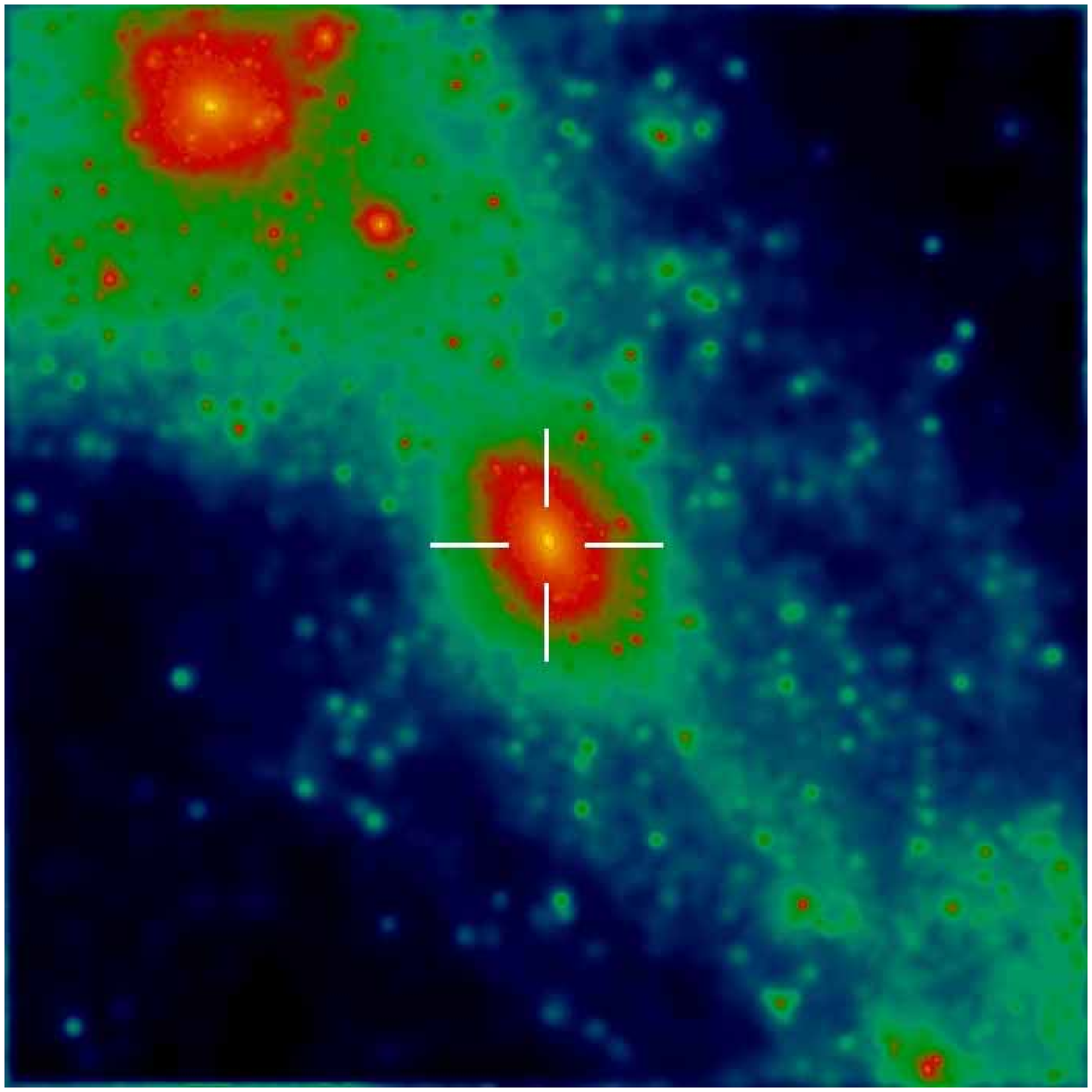}
        \end{center}
      \end{minipage}
      \hspace{0.2cm}
      \begin{minipage}{85mm}
        \begin{center}
          \includegraphics[width=70mm]{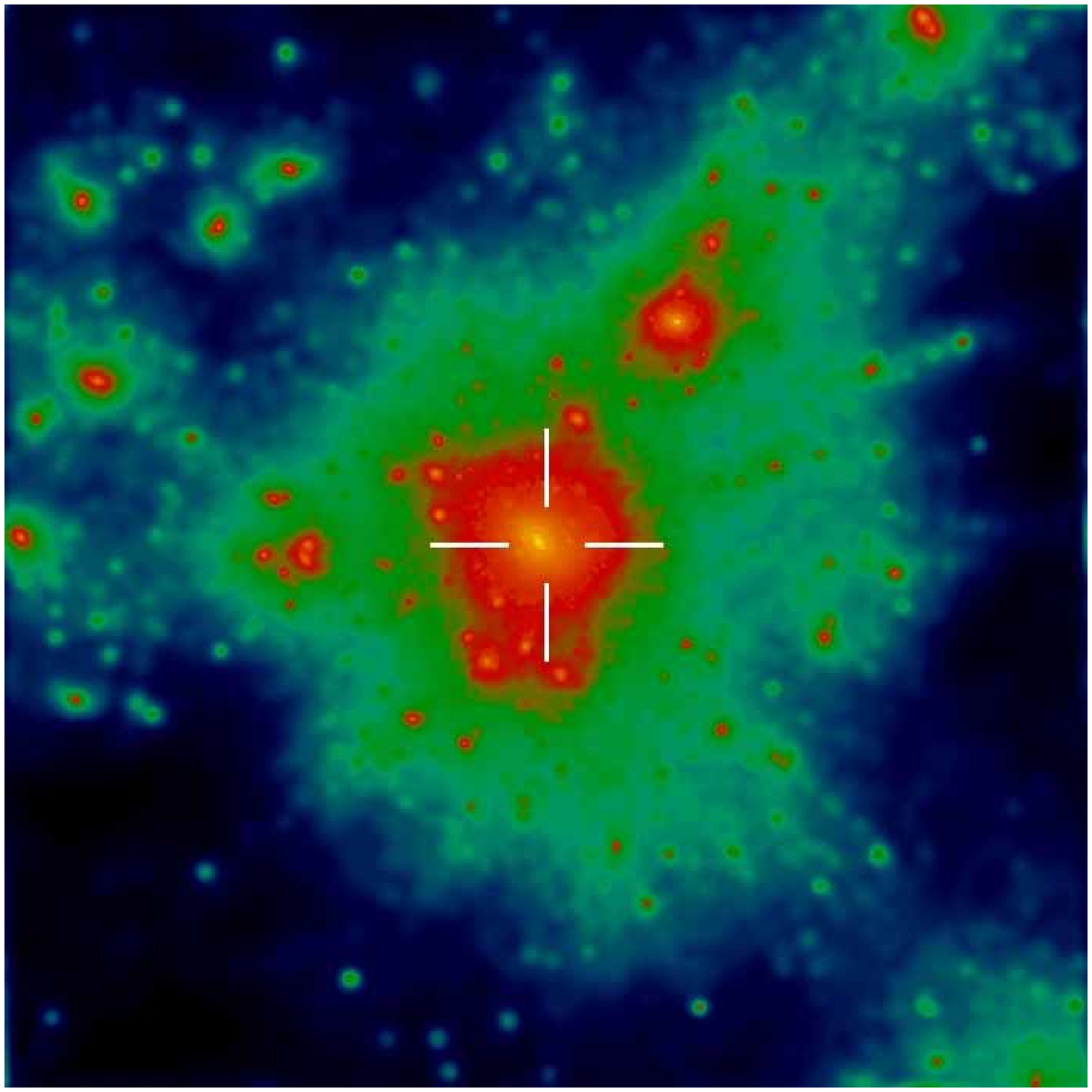}
        \end{center}
      \end{minipage}
    \end{tabular}
    \vspace{2mm}
    \begin{tabular}{cc}
      \hspace{-1.3cm}
      \begin{minipage}{85mm}
        \begin{center}
          \includegraphics[width=80mm]{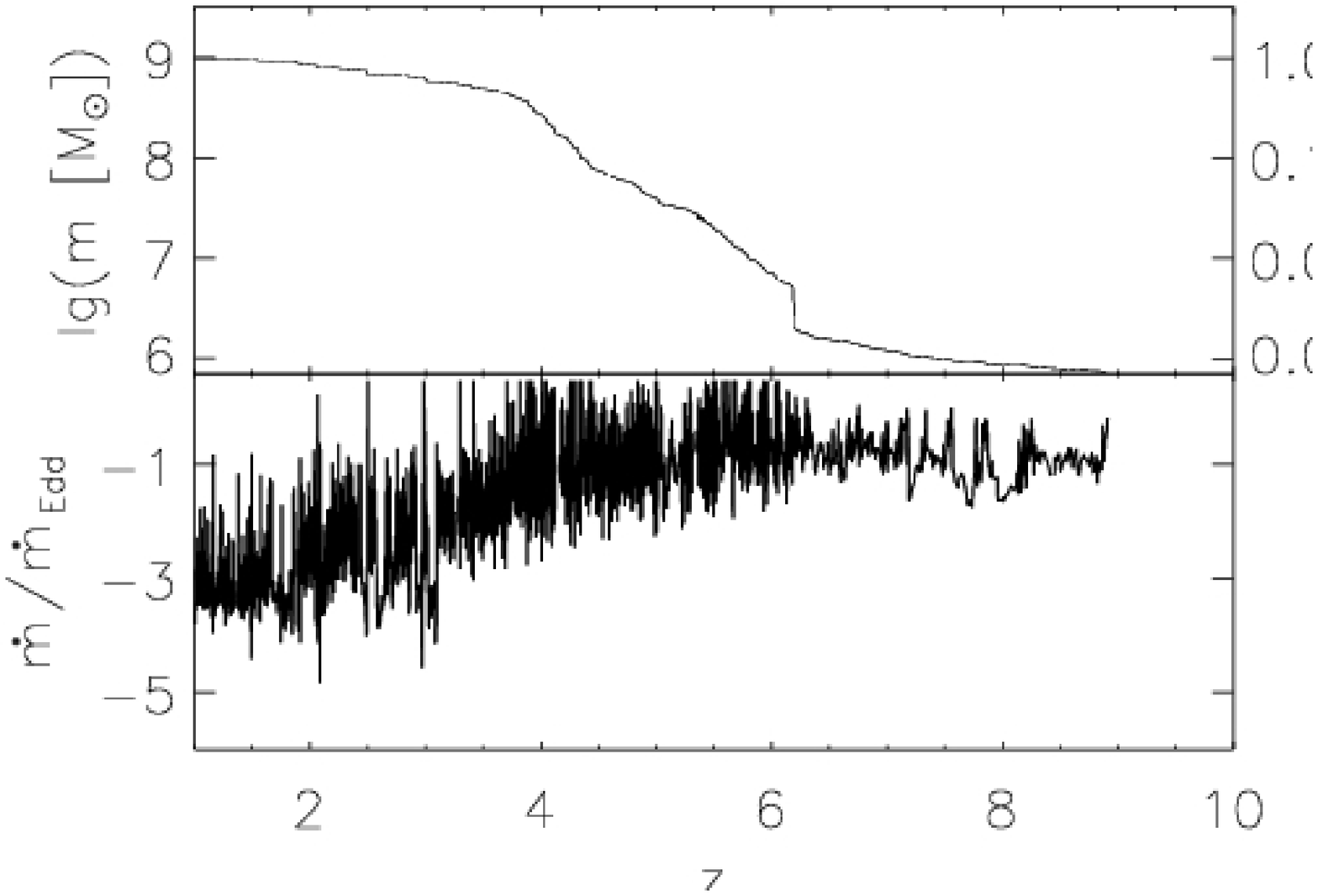}
        \end{center}
      \end{minipage}
      \hspace{0.3cm}
      \begin{minipage}{85mm}
        \begin{center}
          \includegraphics[width=80mm]{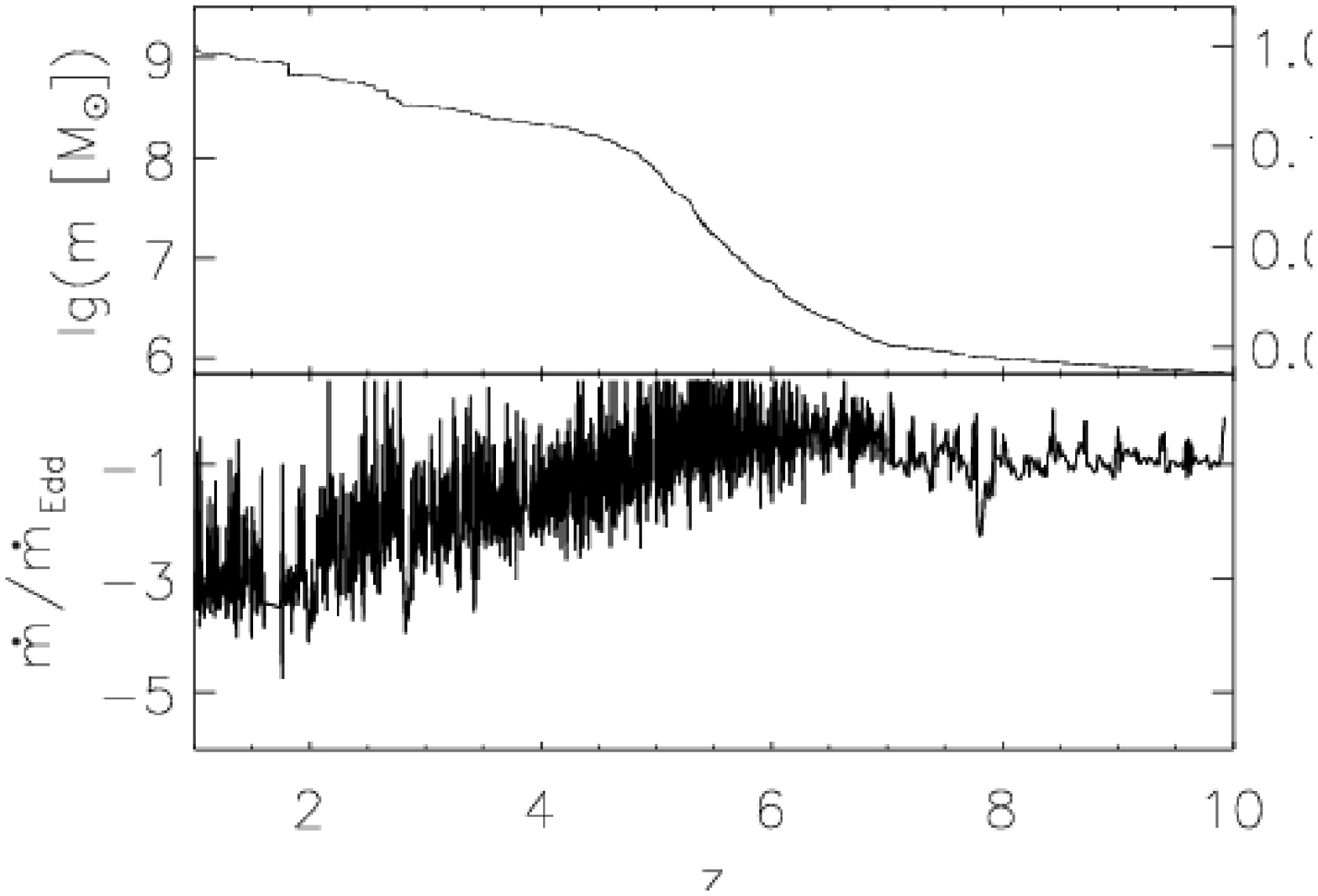}
        \end{center}
      \end{minipage}
    \end{tabular}
    \vspace{1mm}
    \begin{tabular}{cc}
      \hspace{-0.4cm}
      \begin{minipage}{88mm}
        \begin{center}
          \includegraphics[width=88mm]{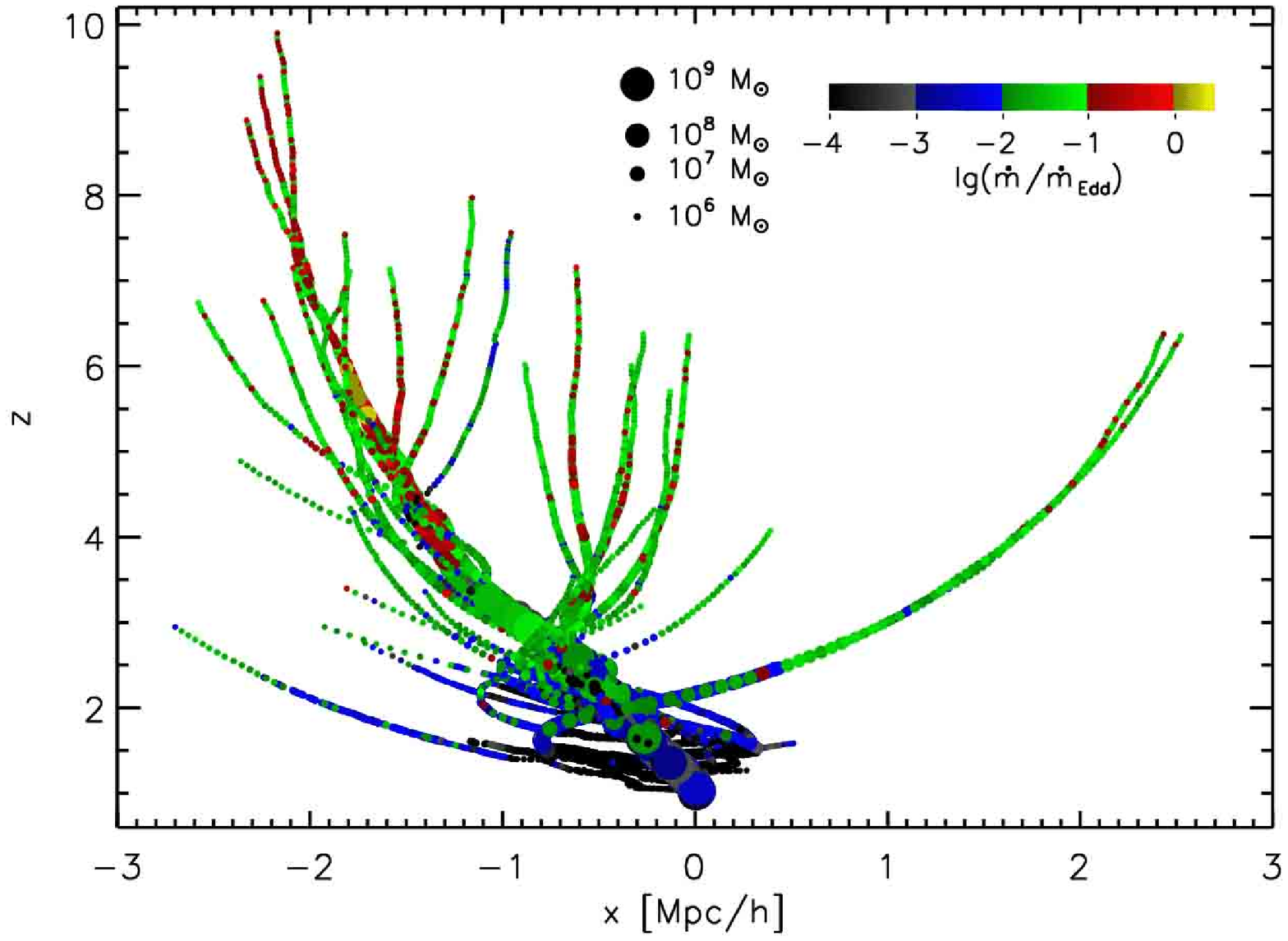}
        \end{center}
      \end{minipage}
      \hspace{-0.2cm}
      \begin{minipage}{88mm}
        \begin{center}
          \includegraphics[width=88mm]{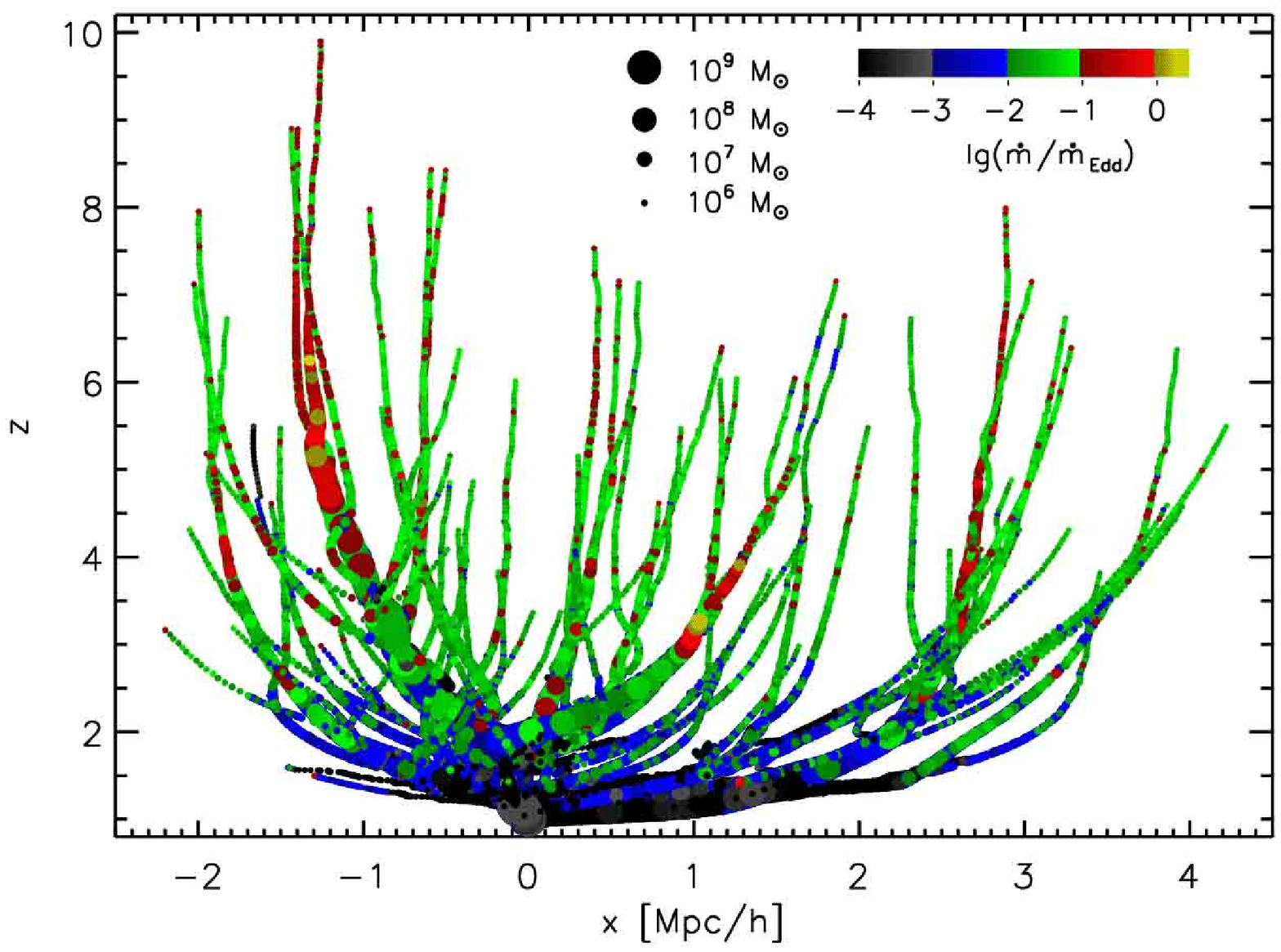}
        \end{center}
      \end{minipage}
    \end{tabular}
   \end{center}
  \vspace{-0.2cm}
  \caption{The $z = 1$ environments of two high--mass BHs and their evolution. 
           The images (top row) show the adaptively smoothed dark--matter 
           density in volumes of size ($3.0\,h^{-1}$\,Mpc)$^3$. Each image 
           is centered on the location of the black hole, which is also 
           marked with cross hairs. In the middle row, the evolution of the
           black holes' most massive progenitors' masses with time, and the 
           corresponding accretion rates, are shown. For the mass panels, we
           also show the fraction $f$ of the final mass. The two bottom panels 
           show the merger trees, with redshift $z$ on the $y$--axis, and 
           symbol sizes and colours depicting masses and accretion rates 
           (see legend). The halo masses and the local densities are
           $1.44\cdot 10^{13}$\,M$_\odot$ and $r_{13}=0.44\,h^{-1}$\,Mpc 
           (left column), and $4.07\cdot 10^{13}$\,M$_\odot$ and 
           $r_{13}=0.26\,h^{-1}$\,Mpc (right column).}
\label{fig:massive}
\end{figure*}

\section{The Simulation} \label{sec:simulation}

In this Section, we present a brief description of the simulation code and the
simulation used in this work. We refer to Di Matteo et al. (2007) for detailed
description of our method.

\subsection{Methodology}

\subsubsection{Numerical Code}
We use the massively parallel cosmological TreePM--SPH code {\small GADGET2}
(Springel 2005), with the addition of a multi--phase modelling of the ISM,
which allows treatment of star formation (Springel \& Hernquist 2003), and
black hole accretion and associated feedback processes (Springel et al. 2005).
The {\small GADGET2} implementation of smoothed particle hydrodynamics (SPH;
Monaghan 1992) uses a formulation which conserves energy and entropy despite
the use of fully adaptive SPH smoothing lengths (Springel \& Hernquist
2002). Radiative cooling and heating process (Katz et al. 1996) and
photoheating due to an imposed ionizing UV background are included. To model
star formation the multiphase model for star-forming gas developed by Springel
\& Hernquist (2003) is used. It has two principal ingredients, a star
formation prescription and an effective equation of state. For the former, we
a adopt a rate motivated by observations and given by the Schmidt--Kennicutt
Law (Kennicutt 1989), where the star formation rate is proportional to the
density of cold clouds divided by the local dynamical time, and it
is normalized to reproduce the star formation rates observed in isolated spiral
galaxies (Kennicutt 1989, Kennicutt 1998). The latter includes the
self--regulated nature of star formation due to supernovae feedback in a
simple model for a multiphase ISM.

\subsubsection{Accretion and Feedback from Supermassive Black Holes}
Analogous to the treatment of star formation and supernova feedback, a
sub--resolution model for the accretion of matter onto a black hole and for
the associated feedback is adopted. Full details of the model can be found in
Di Matteo et al.~(2005), Springel et al.~(2005), and Di Matteo et al.~(2007).

Black holes are represented by collisionless particles that grow in mass
through accretion of gas from their environments. A fraction $\epsilon_{\rm
f}$ of the energy released through radiation is put back into the system as it
couples thermally to nearby gas to influence its motion and thermodynamic
state. The underlying idea is that while the physics of the accretion disks
around the black holes is not resolved, the large--scale feeding of galactic
nuclei with gas, which is resolved, is the {\it critical} process that
determines the growth of supermassive black holes.

BH collisionless 'sink' particles are placed at the centers of galaxies. In
the cosmological simulation discussed here, a friends--of--friends group
finder (e.g.; Davis et al. 1985), called at regular intervals (the time intervals
are equally spaced in log $a$, with $\Delta \log{a} = \log{1.25}$), is
employed to find groups of particles. Each group that does not already contain
a black hole is provided with one by turning its densest particles into a sink
particle with a seed black hole mass of $10^5 h^{-1}$\,M$_\odot$. The black
hole particle then grows in mass through either the accretion of gas or
through mergers with other black hole particles. Note that the black hole seed
mass is negligible compared to the accreted mass. For black holes with $z=1$
masses of $M_{\rm BH} = 10^7$\,M$_{\odot}$, on which the bulk of the following
analysis is based, the seed mass contributes around 1\% of the final mass or
less. For a more detailed discussion of the motivation behind the choice of 
seed mass see Di Matteo et al. (2007).

The accretion onto the black hole from the large--scale gas distribution is
modeled using a Bondi--Hoyle--Lyttleton parameterization (Hoyle \& Lyttleton
1939, Bondi \& Hoyle 1944, Bondi 1952), which relates the accretion rate onto
the black hole, $\dot{M}_{\rm B}$, to the density $\rho$ and sound speed $c_s$ 
of the surrounding gas.
For the simulation used in this work, it was assumed that accretion is limited
to three times the Eddington rate, $\dot{M}_{\rm Edd}$. The radiated
luminosity, $L_{\rm r}$, is related to the accretion rate, $\dot {M}_{\rm
BH}$, through the radiative efficiency $\eta = L_{\rm r}/(\dot{M}_{\rm BH} \,
c^2)\,,$ which gives the amount of energy that can be extracted from the
innermost stable orbit of the accretion disk. Here, a value of $\epsilon_{\rm
r} = 0.1$ was used, the mean value for a radiatively efficient accretion disk
(Shakura \& Sunyaev 1973) onto a non-rapidly rotating black hole. The model
assumes that a fraction $\epsilon_{\rm f} = 0.5$ of the released radiation couples
to the surrounding gas in the form of feedback energy.
For reasons of simplicity, this process is modeled as thermal energy 
deposited {\it isotropically} in the region around the black hole. 
In Di Matteo et al. (2005), $\epsilon_{\rm f}$ was fixed by fitting the $z=0$
$m_{\rm BH}-\sigma$ relation, and the value of $\epsilon_{\rm f} = 0.5$ found
there was also adopted in this work. This is the only parameter for the BH
model, and our cosmological simulations have all parameters fixed from our
previous work. Note that even though we do not run the cosmological
simulation to $z=0$, we do use a value for $\epsilon_{\rm f}$ determined to
match the $z=0$ $m_{\rm BH}-\sigma$ relation. A more detailed discussion of
the motivation behind these choices plus a discussion of the predictions for
the $m_{\rm BH}-\sigma$ relation at various redshifts (incl.  comparisons with
the available observational constraints) can be found in Di Matteo et
al. (2008).

\subsubsection{Mergers of Supermassive Black Holes}

When galaxies merge their central black holes are also expected to merge 
at some stage. Thus, mergers generated through the hierarchical buildup of
haloes and galaxies contribute to the growth of the central black holes. 
Some of the details of this process are still a matter of debate. Given
the resolution of the simulations it is impossible to treat this problem 
in detail. Instead, it is assumed that two black hole particles merge if 
they come within the spatial resolution of the simulation and if their 
relative speed lies below the gas sound speed. Note that we also cannot
resolve (nor accurately model) the ejection of black holes by gravitational 
recoil (see, for example, Hoffman \& Loeb 2006), or by three--body 
sling--shot ejection of black holes in triple systems (see, for example, 
Baker et al. 2006).

\subsection{Details of the Simulation}
We use the standard $\Lambda$CDM cosmological model, with parameters chosen to
match the first year Wilkinson Microwave Anisotropy Probe measurements
(Spergel et al. 2003). The parameters are $\Omega_0 = 0.3$, $\Omega_{\Lambda}
= 0.7$, Hubble constant $h =0.7$, expressed in units of $H_0 =
100\,h$\,km\,sec$^{-1}$\,Mpc$^{-1}$, a scale--invariant primordial power
spectrum with index $n=1$, and a normalization of the amplitude of
fluctuations of $\sigma_{8} = 0.9$. The more recent (third year) WMAP results,
with their considerably lower value of $\sigma_8$, were published after the
simulation had progressed quite far already (see however Evrard et al. 2007).
For a discussion of the repercussions of the larger value of $\sigma_8$ on our
cosmological simulations with black holes see also Sijacki et al. (2007).

The simulation is of a periodic box of size 33.75\,$h^{-1}$\,Mpc, using
$486^3$ gas and dark--matter particles each. The size of the volume was
motivated by the goal to achieve the highest possible resolution, while, at
the same time, matching the suite of simulations in Springel \& Hernquist
(2003), to allow detailed comparisons between the physical properties of
simulations with and without black holes.  In addition, the box size and
particle number were chosen such that the physical resolution at $z\sim 6$ is
comparable to that of some of the previous work of galaxy mergers runs with
black holes (see Di Matteo et al.  2005, Springel et al. 2005, Robertson et
al. 2006). Following Di Matteo et al. (2007) we hereafter refer to this
simulation as the {\it BHCosmo} run. Di Matteo et al. (2007) contains a 
very detailed discussion of, for example, the $m_{\rm BH}$--$\sigma$
and $m_{\rm BH}$--$m_{\rm Bulge}$ relations from the simulation, and we thus
refer the reader interested in the details -- which provide ample justification
for the choices in the simulation methodology -- to that work.

From a starting redshift of $z=157$, the simulation was evolved to a final
redshift of $z=1$. A total set of 36 outputs was saved. In addition, whenever
a black hole particle was accreting matter, its mass, accretion rate, and
position was stored, as was information about which black hole merged with
what other black hole at what redshift. For the 3,547 black holes that
are contained in the simulation volume at $z=1$, a total of more than 
four and a half million progenitor black holes exists. From this raw data,
the merger trees of all $z=1$ black holes were generated, with the most
massive black hole counting more than one hundred twenty thousand progenitor
black holes. 

In addition to the {\it BHCosmo} box, there is a second simulation with the
same cosmological parameters but a larger volume (its box size is
50.0\,$h^{-1}$\,Mpc), called E6, which, however, was only evolved until
$z=3.82$ due to constraints on the available computing time. We will make use 
of E6 only for comparisons with {\it BHCosmo} at $z=5$, one of the times that will be
studied in the following.

%
%
\begin{figure}
\begin{tabular}{cc}
  \hspace{-0.6cm}
  \begin{minipage}{90mm}
    \begin{center}
      \includegraphics[width=90mm]{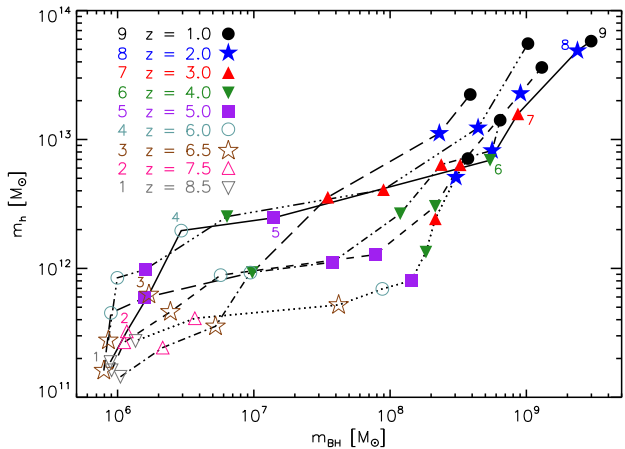}
    \end{center}
  \end{minipage}
\end{tabular}
\vspace{-0.4cm}
\caption{Black hole mass $m_{\mbox{\scriptsize{BH}}}$ versus host halo 
         mass $m_{\mbox{\scriptsize{h}}}$ for six high--mass black holes, at
         nine different redshifts. We use different colours and symbols for the different
         redshifts, as shown in the legend. In addition, for one example, 
         the numbers near the circles show how the graph has to be read. 
         The most massive high--redshift black hole (dotted line) clearly 
         stands out: It is more than one order of magnitude more 
         massive at $z = 6.5$ than the other black holes. And unlike all the other
         black holes, it only gains one order of magnitude in mass until 
         the final redshift in the plot.}
\label{fig:BH_vs_halo_2}
\end{figure}

%
%
\begin{figure*}
  \begin{center}
    \begin{tabular}{cc}
      \hspace{-0.9cm}
      \begin{minipage}{95mm}
        \begin{center}
          \includegraphics[width=95mm]{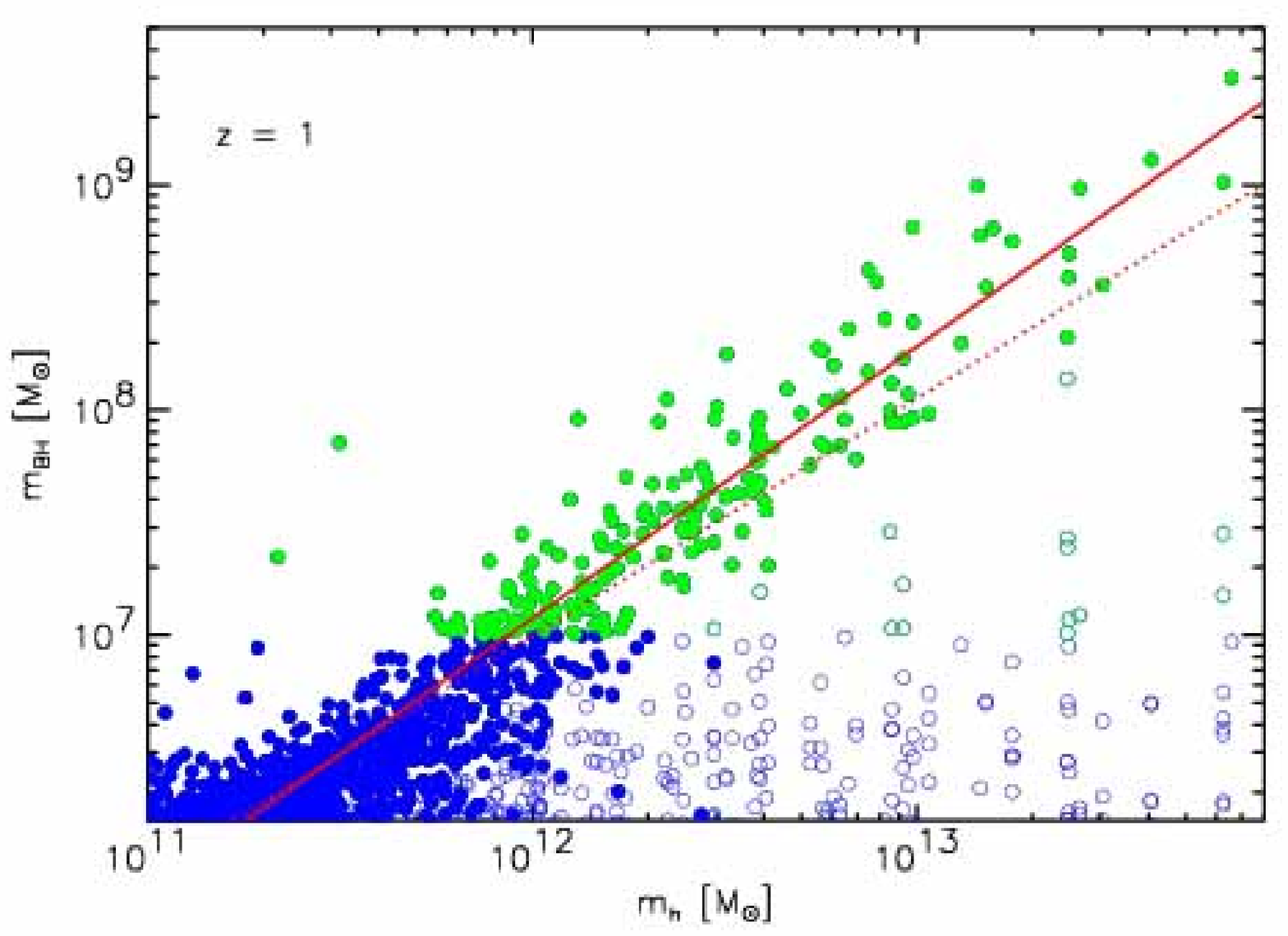}
        \end{center}
      \end{minipage}
      \hspace{-0.8cm}
      \begin{minipage}{95mm}
        \begin{center}
          \includegraphics[width=95mm]{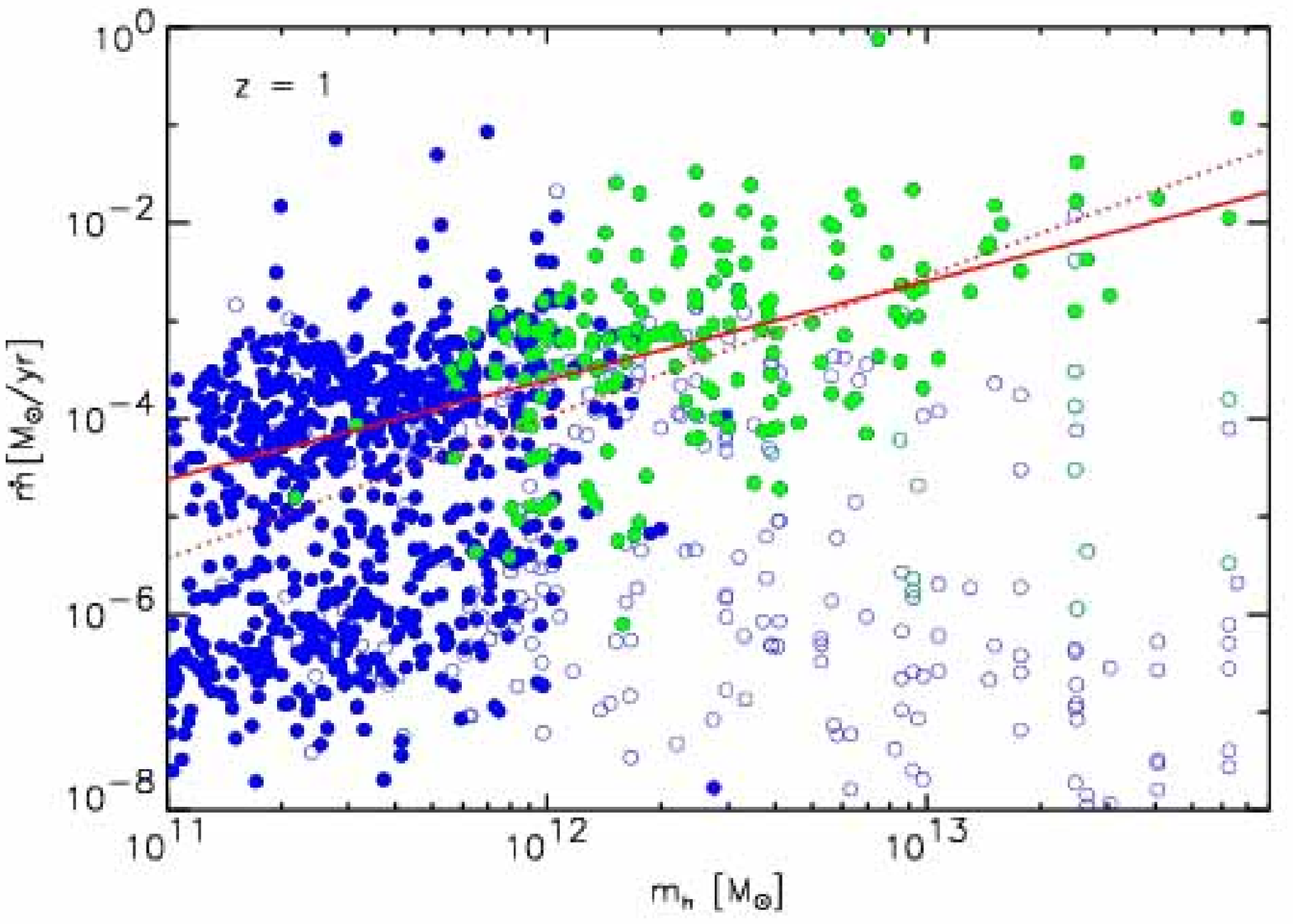}
        \end{center}
      \end{minipage}
    \end{tabular}
    \vspace{-5mm}
    \begin{tabular}{cc}
      \hspace{-0.9cm}
      \begin{minipage}{95mm}
        \begin{center}
          \includegraphics[width=95mm]{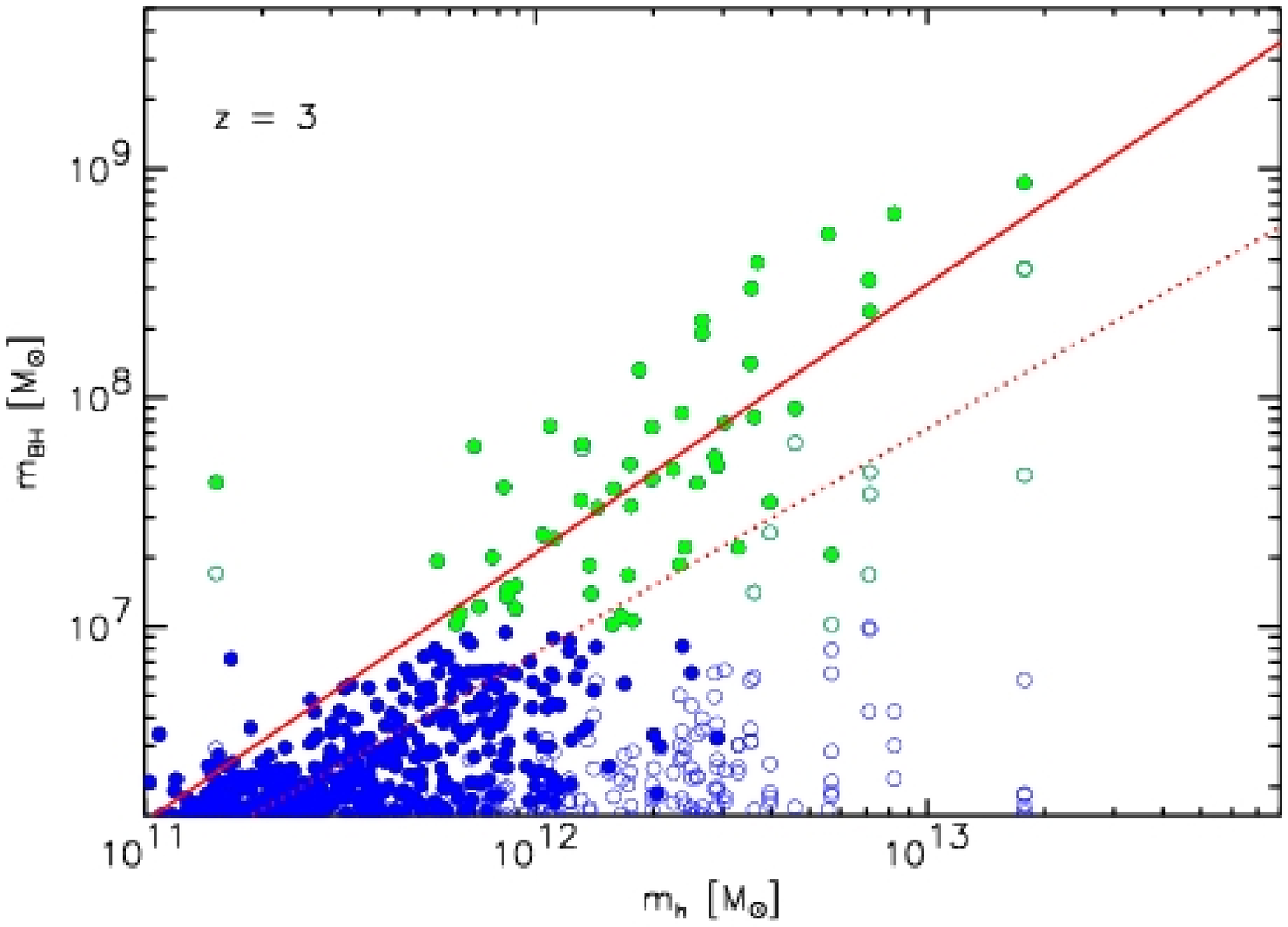}
        \end{center}
      \end{minipage}
      \hspace{-0.8cm}
      \begin{minipage}{95mm}
        \begin{center}
          \includegraphics[width=95mm]{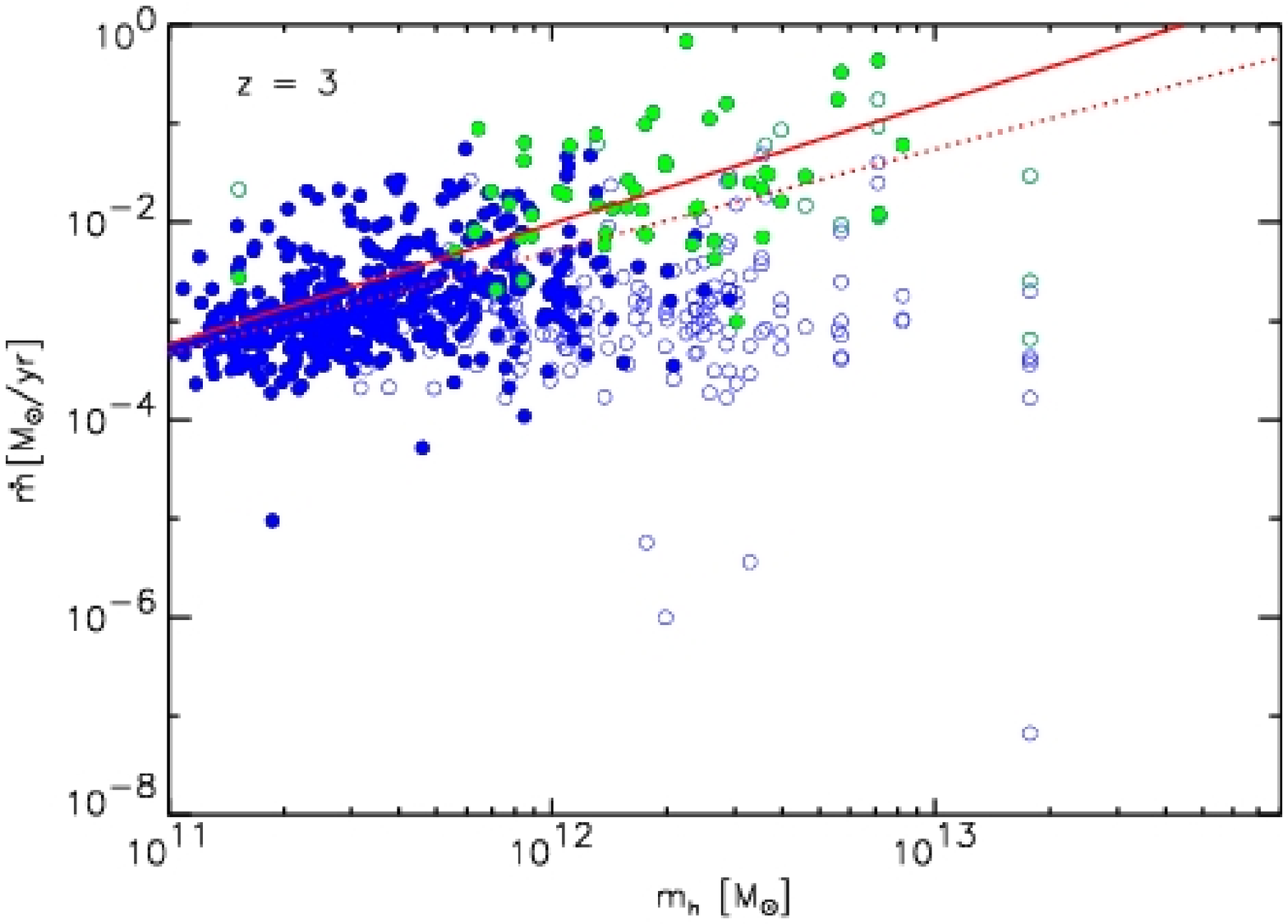}
        \end{center}
      \end{minipage}
    \end{tabular}
    \vspace{-5mm}
    \begin{tabular}{cc}
      \hspace{-0.9cm}
      \begin{minipage}{95mm}
        \begin{center}
          \includegraphics[width=95mm]{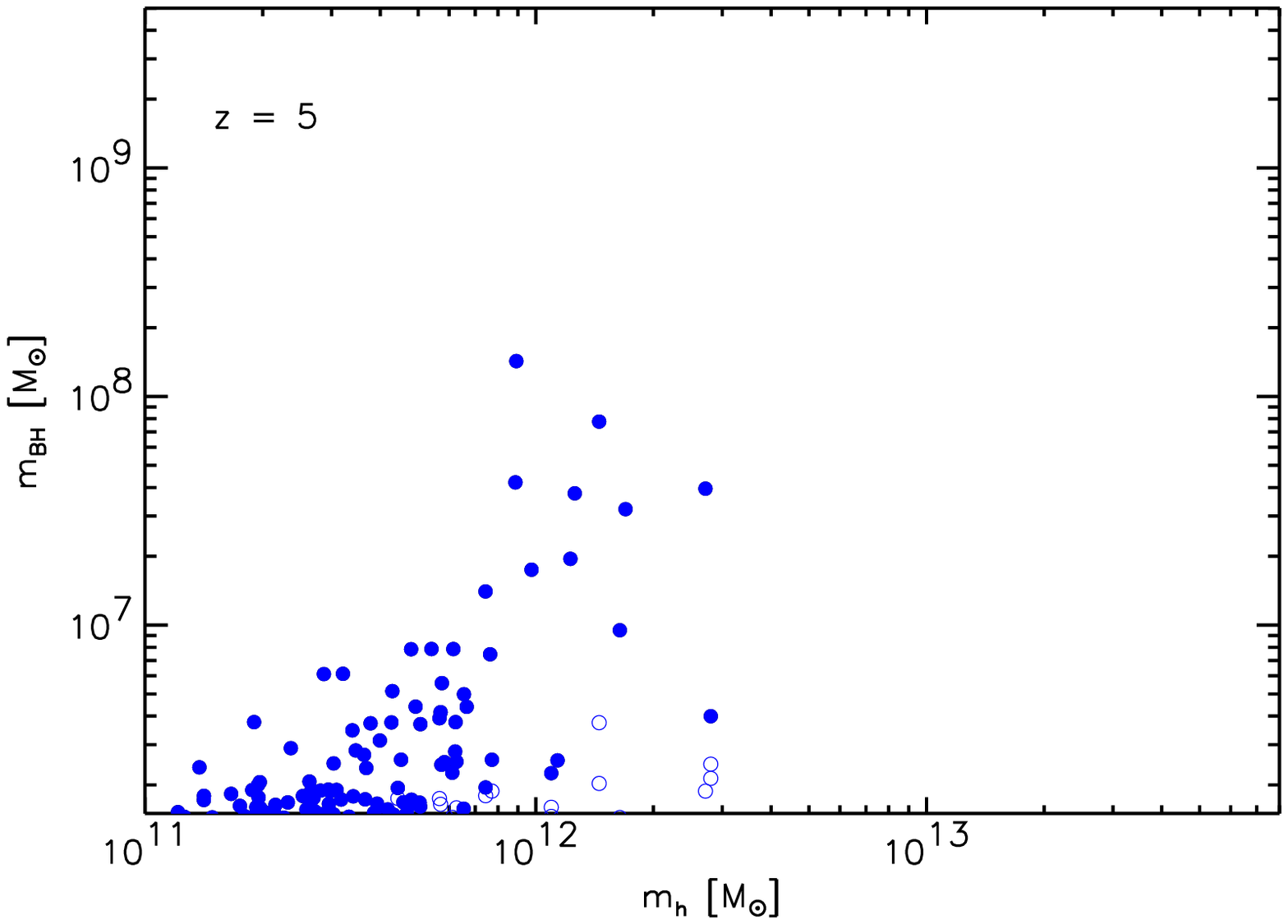}
        \end{center}
      \end{minipage}
      \hspace{-0.8cm}
      \begin{minipage}{95mm}
        \begin{center}
          \includegraphics[width=95mm]{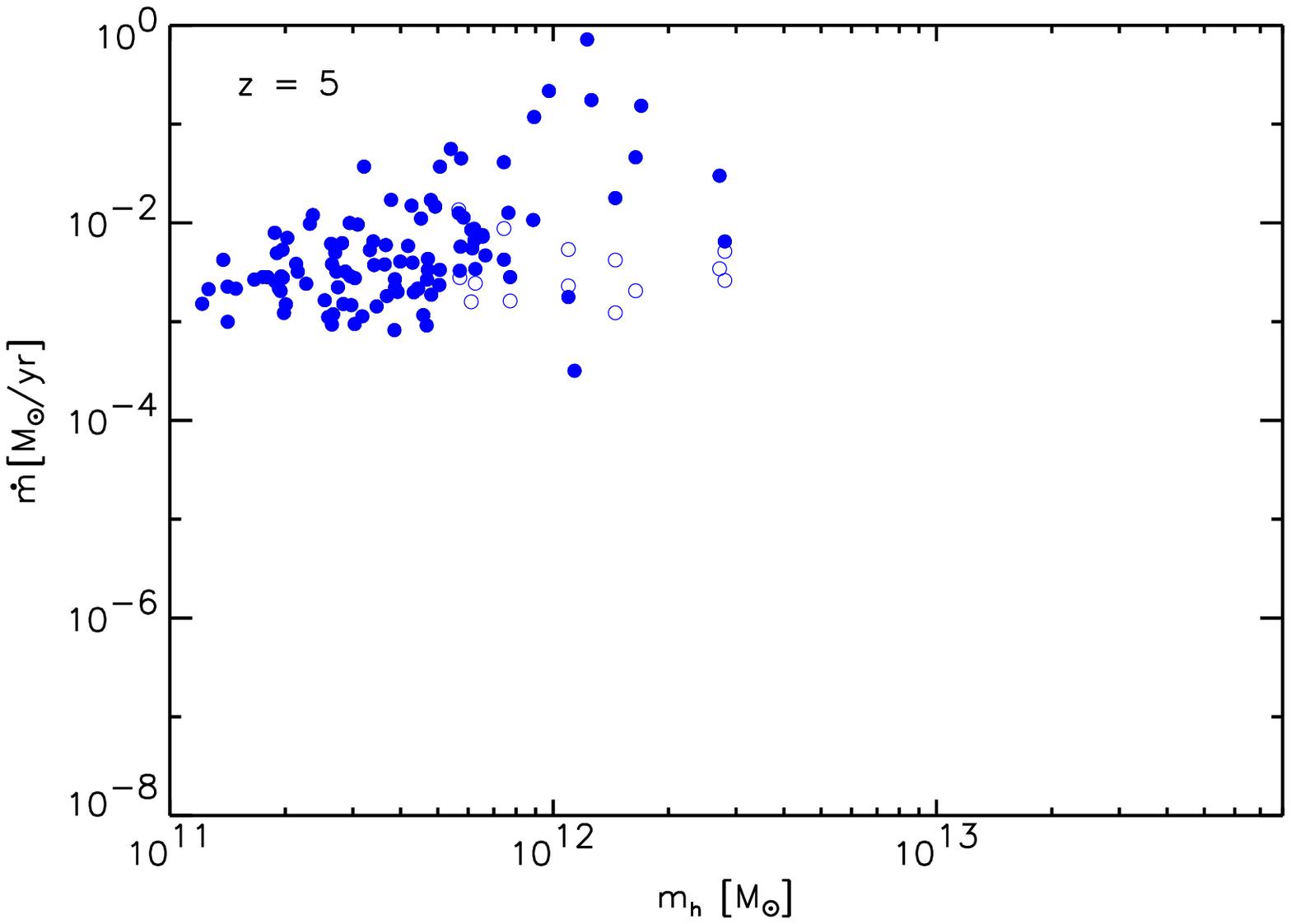}
        \end{center}
      \end{minipage}
    \end{tabular}
    \end{center}
    \vspace{-4mm}
    \caption{Left panels: Black hole mass $m_{\mbox{\scriptsize{BH}}}$ versus host halo mass
         $m_{\mbox{\scriptsize{h}}}$, with green (blue) symbols for BHs above
         (below) $1.0\cdot\,10^7$\,M$_\odot$. Solid and open circles show main and
         satellite black holes, respectively. The solid (dotted) red line
         shows the best (linear) fit for the relation between
         $m_{\mbox{\scriptsize{BH}}}$ and $m_{\mbox{\scriptsize{h}}}$ for BHs
         more massive than $10^7\,\mbox{M}_\odot$ (for the full sample).
         Right panels: Black hole accretion rate $\dot{m}$ versus host halo
         mass $m_{\mbox{\scriptsize{h}}}$, using the same convention as in the
         left panels. The solid (dotted) red line shows the best (linear) fit 
         for the relation between $\dot{m}$ and $m_{\mbox{\scriptsize{h}}}$
         for BHs more massive than $10^7\,\mbox{M}_\odot$ (for the full sample).
         From top to bottom, data at redshifts $z=1$, $z=3$, and $z=5$ are
         shown. Because of the paucity of data points at $z=5$, we do not
         attempt to fit those data.}
\label{fig:BH_vs_halo_1}
\end{figure*}

\section{The BH--Halo Connection} \label{sec:connection}

\subsection{The Co--Evolution of BHs and Their Host Haloes and the
            Anti--hierarchical formation of the most massive BHs} \label{sec:coevolution}

The most immediate environment of a black hole is its host galaxy and, by
extension, its host halo. The formation of a dark matter halo, unlike that of
a galaxy, is solely governed by gravity. Haloes grow through the accretion of
matter and through mergers. Both processes also contribute to the growth of
the baryonic component of galaxies, as gas is added, some of which then forms
new stars. But unlike in the case of the dark matter, the growth of the
baryonic component and of BHs depends on complex hydrodynamical processes,
including gas cooling, mergers, feedback due to supernovae and AGNs which may
expel some of the gas from the galaxies' potentials. The interplay between
these processes and their relative importance are subject to numerous studies
such as, for example, those involving semi--analytical models of galaxy
formation (for a detailed review of such models, and many references, see
Baugh 2006). Di Matteo et al.~(2007) already contains detailed descriptions of
the relationship between the properties of BHs and of the subhaloes or
galaxies they reside in, through a study of the evolution of the $m_{\rm
BH}-\sigma$ and $m_{\rm BH}-m_{*}$ relations. For a discussion of those
relationships we refer to that work. Unless noted otherwise, in the following
we will present results from the {\it BHCosmo} simulation.

In the top panels of Figure~\ref{fig:massive}, we show two examples of the
dark matter distribution in the immediate surrounding of two of the most
massive black holes at $z = 1$. The BH and halo mass on the left (right) are
$9.89\cdot 10^{8}$ M$_\odot$ and $1.44 \cdot 10^{13}$M$_\odot$ ($1.30\cdot
10^{9}$\,M$_\odot$ and $4.1\cdot 10^{13}$ M$_\odot$), respectively.  The
locations of the black holes within their host halos are marked with cross
hairs. Each picture shows the adaptively smoothed dark--matter density in
volumes of size ($3.0\,h^{-1}$\,Mpc)$^3$.

In the middle row of Figure~\ref{fig:massive}, we illustrate the mass assembly
histories of the black holes shown in the top panels. We plot the
instantaneous BH mass and the fraction of the final BH mass, and accretion rate 
(in units of Eddington) as a function of redshift. It is evident that these 
objects experienced major critical accretion phase (and hence BH mass growth) 
epochs at high redshifts, in the range of $4 < z < 7$. For $z < 4$, the BH 
accretion rates decrease, showing far more sporadic episodes of critical 
Eddington growth and a drop of the  average accretion rates of about two 
orders of magnitude for $z \simeq 2$ and later. A BH mass of 
$\sim 4-5 \cdot 10^{8}$ is assembled by $z\sim 4$, which roughly doubles 
by $z\sim 1$.

Lastly, in the bottom row of Figure~\ref{fig:massive} we show the merger 
trees of the two black holes. Redshift is given on the $y$--axis; the 
$x$--axis is one of the actual coordinates (both black holes are set to 
be at the coordinate origin at $z=1$). Symbol sizes and colours are scaled 
to reflect masses and accretion rates, respectively, with each decade in 
accretion rate using a different colour. Just like in the cases discussed 
in Di Matteo et al. (2007), the histories of these massive black holes 
are quite complex. Even though the two black holes have comparable masses 
as a result of a roughly similar accretion history, the BH shown in the 
right column experiences many more merger events than the one in the left 
column. Since the two black holes appear to be residing in somewhat different 
large--scale environments, this indicates that the properties of the BHs 
can be affected by the large--scale matter configuration. We will investigate 
the influence of large--scale environment on black hole properties in 
Section~\ref{sec:environments}.

It is instructive to compare Figure~\ref{fig:massive}
with Figures~4 and 5 in Malbon et al. (2007), who use a semi--analytical
galaxy formation model to study the growth of supermassive black holes (note
the difference in the BH masses). While here we only trace the evolution of
the BHs, our direct simulation provides us with an amount of detail unmatched
by the semi--analytical {\it ansatz}, since our simulation method stores
information about the black holes whenever the are active or merging
(resulting in many thousands of different periods of activity for the most
massive of our BHs). In contrast, semi--analytical galaxy formation models are
restricted to constructing the histories of BHs on top of merger trees of dark
matter haloes which are typically a lot sparser (fifty or one hundred output
times).

Figure~\ref{fig:massive} indicates noticeable differences in the growth
histories of these massive black holes. To further characterize the evolution
of haloes and their black holes, Figure~\ref{fig:BH_vs_halo_2} shows the growth
history of BH mass, $m_{\mbox{\scriptsize{BH}}}$, and host halo mass,
$m_{\mbox{\scriptsize{h}}}$, for a sample of six high--mass
($m_{\mbox{\scriptsize{BH}}} \ge 10^8$\,M$_\odot$) black holes.  Shown are BH
and respective host halo mass at nine different redshifts, starting from $z =
8.5$ and ending at $z = 1.0$. The growth curves in Figure~\ref{fig:BH_vs_halo_2} 
show the histories of the same sample of massive black holes discussed in 
Di Matteo et al. (2007; see their Figure~13). The figure shows that at first 
the black holes' masses grow relatively slowly compared with their host halo 
masses (for $z > 7$), followed by rather rapid growth between redshifts of 
$z \simeq 6$ and $z \simeq 3$. During this time most black holes in this 
sample have increased their masses by two orders of magnitude while their 
host haloes' masses increase by most an order of magnitude. This is depicted 
by the rather flat region in the curves in Figure~\ref{fig:BH_vs_halo_2} at 
redshifts $6 < z < 3$. The curves steepen again at $z < 3$ as the growth of 
the parent halo offsets that of its BH. The dashed line shows the black hole 
and halo in the right column of Figure~\ref{fig:massive}. As is clear from this
Figure, the growth of haloes and black holes is quite different from each
other. As discussed earlier, BH gas accretion in these high mass systems is 
the highest at $z > 4 $, leading to the build--up of the bulk of the BH mass, 
which is later supplemented by a large number of merger events at lower 
redshifts.  Note that, as discussed in Di Matteo et al.~(2007), the most 
massive black hole in the simulation at $z = 6.5$ is more than an order 
of magnitude more massive at than any of the other black holes. Its 
black--hole--to--halo mass ratio is also the largest all the way to 
$z\sim 4$ (as shown by the dotted line in Fig.~\ref{fig:BH_vs_halo_2}). 
However, below this redshift the BH only gains one order of magnitude in 
mass and is overtaken by most of the other objects.

In a $\Lambda$CDM cosmology, dark matter haloes -- such as those in
Figure~\ref{fig:BH_vs_halo_2} -- form hierarchically. Small haloes assemble
first, and through mergers and accretion, larger and larger haloes are
built. Maulbetsch et al. (2007) provide an instructive study of the mass
assembly histories of dark matter haloes in a very large simulation volume,
which here, and in the following sections, can be used to compare black hole
assembly histories with dark matter assembly histories\footnote{While their
definition of environment differs from the one we will adopt, the comparison
nevertheless is instructive.}. The complex relationship between halo and BH
masses in Figure~\ref{fig:BH_vs_halo_2} indicates that the massive black holes
in those haloes do {\it not} follow this pattern. Instead, high--mass black
holes are assembled early compared to their host halos. We will further
discuss this 'anti--hierarchical' behaviour in Section~\ref{sec:formation}.

\subsection{BH Host Halo Masses} \label{sec:hosts}

After studying the co-evolution of the most massive black holes and their host
haloes, we now broaden the investigation and look at the demographics of black
holes of all masses in their dark matter haloes. In the left panels of
Figure~\ref{fig:BH_vs_halo_1} we show the relation between host halo mass,
$m_{\mbox{\scriptsize{h}}}$, and BH mass, $m_{\mbox{\scriptsize{BH}}}$, at
redshifts $z=1$, $z=3$, and $z=5$ (from top to bottom) with green (blue) 
symbols for BHs above (below) $1.0\cdot\,10^7$\,M$_\odot$. We show main (or
central) BHs with solid circles and satellite BHs with open circles. For each
halo, the main black hole is considered to be the one that resides at the 
center of the most massive subhalo. Satellite BHs are those that reside at 
the center of galaxies that orbit around the central, typically most massive 
galaxy (see, e.g., Fig.~7 in Di Matteo et al. 2007). Note that the smallest 
haloes that contain a BH in the simulation always consist of two thousand 
or more particles in the {\it BHCosmo} run and are therefore very well resolved.

Following Ferrarese (2002), we determine the relation between
$m_{\mbox{\scriptsize{BH}}}$ and $m_{\mbox{\scriptsize{h}}}$ in
\begin{equation}
\frac{m_{\mbox{\scriptsize{BH}}}}{10^8\,\mbox{M}_\odot} = c(z) 
  \left( \frac{m_{\mbox{\scriptsize{h}}}}{10^{12}\,\mbox{M}_\odot}
  \right)^\alpha
  \label{eq:mBH-mh}
\end{equation}
We let the fit cover black holes with masses larger than
$10^7\,\mbox{M}_\odot$ (the range for which large numbers of observational
data are available) and all black holes. We allow for $\alpha$ and $c$ to 
be redshift dependent. For the $m_{\rm BH} \ge 10^7\,\mbox{M}_\odot$ sample
and redshifts $z=1$, $z=3$  we find $\alpha=1.21 \pm 0.05$,
$\alpha=1.17 \pm 0.14$, and $c=0.12 \pm 0.01$, $c=0.21 \pm 0.03$,
respectively. The full sample yields $\alpha=1.05 \pm 0.02$, 
$\alpha=0.98 \pm 0.04$, and $c=0.10 \pm 0.01$, $c=0.08 \pm 0.02$
for the redshifts $z=1$, $z=3$, respectively. In the left panels of 
Figure~\ref{fig:BH_vs_halo_1} we show the fits superimposed as a solid 
(dotted) red line for the $m_{\rm BH} \ge 10^7\,\mbox{M}_\odot$ (full)
sample. Because of the small number of objects at $z=5$ and their large 
scatter, we do not attempt a fit.

Observationally, the $m_{\mbox{\scriptsize{BH}}}-m_{\mbox{\scriptsize{h}}}$
relation (eq.~\ref{eq:mBH-mh}) can only be determined indirectly, since the
masses of the dark matter haloes cannot be measured directly. For example,
Ferrarese (2002) uses an $m_{\mbox{\scriptsize{h}}}-v_{\scriptsize{vir}}$
relation (where $v_{\scriptsize{vir}}$ is the halo velocity at its virial
radius) from numerical simulations and then connects the haloes'
$v_{\scriptsize{vir}}$ with the circular velocities, $v_c$, of the BH host
galaxies, using models for halo profiles. That way, the black hole masses can
be related to the masses of their host haloes. For a local set of BHs,
Ferrarese (2002) finds $\alpha=1.82$ and $c=0.03$, $\alpha=1.65$ and $c=0.1$,
and $\alpha=1.82$ and $c=0.67$ for an isothermal dark matter profile, an NFW  
profile (Navarro et al. 1997), and a profile based on weak lensing results by
Seljak (2002), respectively. An example of a more recent result is provided by
Shankar \& Mathur (2007) who find $\alpha = 1.39$ and $c=(58/700)(1+z)$, or
$c=0.17$, $c=0.33$, and $c=0.50$ for $z=1$, $z=3$, and $z=5$, respectively.

Finally, in the right panels of Figure~\ref{fig:BH_vs_halo_1} we show the relation
between host halo mass, $m_{\mbox{\scriptsize{h}}}$, and BH accretion rate, 
$\dot{m}$ (in M$_\odot$/year), at redshifts $z=1$, $z=3$, and $z=5$ (from top
to bottom). Just like in the panels on the left--hand side, satellite
BHs constitute a distinguishable population, predominantly at lower accretion 
rates than the main BHs in the same halo. In the same fashion as for the
black hole masses, we determine the relation between $\dot{m}$ and
$m_{\mbox{\scriptsize{h}}}$ in
\begin{equation}
\dot{m} = d\,m_{\mbox{\scriptsize{h}}}^\beta
\label{eq:mdot-mh}
\end{equation}
For the $m_{\rm BH} \ge 10^7\,\mbox{M}_\odot$ sample and redshifts $z=1$ and
$z=3$ we find $\beta=1.01 \pm 0.18$ and $\beta=1.22 \pm 0.37$,
respectively. The full sample yields $\beta=1.45 \pm 0.08$ and $\beta=1.04 \pm
0.06$ for the redshifts $z=1$ and $z=3$, respectively.  The slopes are quite
similar to those found for the
$m_{\mbox{\scriptsize{BH}}}-m_{\mbox{\scriptsize{h}}}$ relation above, albeit
with larger scatter. However, when we divide the sample using accretion rate
instead of black hole mass a different picture emerges. Using only the most
active black holes (with $\dot{m} \ge 10^{-2}$\,M$_\odot$/year), at $z=1$,
$\beta = -0.02 \pm 0.14$, whereas at $z=3$, $\beta = 0.66 \pm 0.12$. That is,
at $z=1$, there is virtually no relation between BH accretion rates and their
halo masses, and at $z=3$, there is a very slight dependence. This result is
fully consistent with the lack of a luminosity dependence in the clustering of
AGNs in the DEEP2 sample as measured by Coil et al. (2007).

%
%
\begin{figure}
\begin{tabular}{cc}
  \hspace{-0.6cm}
  \begin{minipage}{90mm}
    \begin{center}
      \includegraphics[width=90mm]{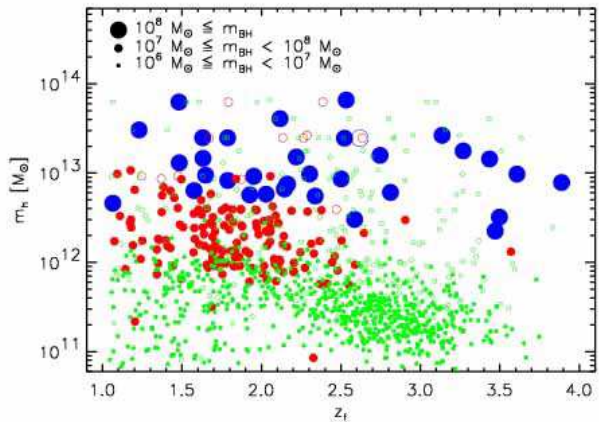}
    \end{center}
  \end{minipage}
\end{tabular}
\vspace{-0.4cm}
\caption{Formation redshift $z_f$ versus host halo mass $m_h$ for all black
         holes. The symbol sizes (and colours) depict different masses 
         (see legend). In addition, filled and empty circles show central and 
         satellite BH, respectively.}
\label{fig:zf_vs_halo_mass}
\end{figure}

%
%
\begin{figure}
\begin{tabular}{cc}
  \hspace{-0.6cm}
  \begin{minipage}{90mm}
    \begin{center}
      \includegraphics[width=90mm]{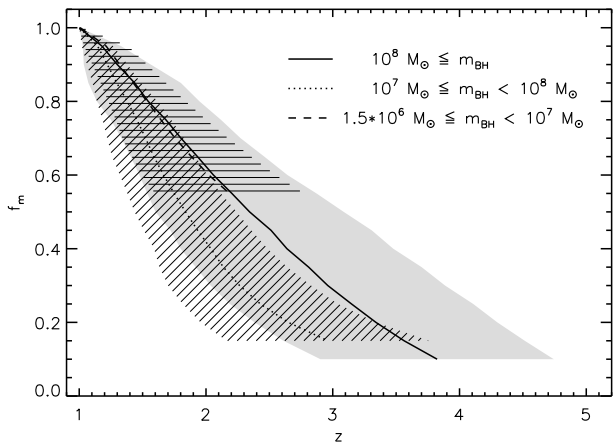}
    \end{center}
  \end{minipage}
\end{tabular}
\vspace{-0.4cm}
\caption{Mean assembly redshifts $z$ at which fractions $f_m$ of the largest
         progenitor masses were assembled for the mass ranges $1.5\cdot 
         10^6$\,M$_\odot \le m_{\mbox{\scriptsize{BH}}} < 10^7$\,M$_\odot$ 
         (dashed/green), $10^7$\,M$_\odot \le m_{\mbox{\scriptsize{BH}}} 
         < 10^8$\,M$_\odot$ (dotted/red), and $m_{\mbox{\scriptsize{BH}}} \le
         10^8$\,M$_\odot$ (solid/blue). See main text for more details.}
\label{fig:BH_mean_formation_z}
\end{figure}

\subsection{Formation Epochs} \label{sec:formation}

As already outlined in Section~\ref{sec:coevolution}, the evolution of
the most massive black holes does not mirror that of their host haloes.
Is the same true for all black holes? For this and the following Sections, 
we divide the black--hole sample into three mass ranges, which we will refer to 
as the low, intermediate, and high--mass sample. The masses covered in 
these samples are $10^6$\,M$_\odot \le m_{\rm BH} < 10^7$\,M$_\odot$, 
$10^7$\,M$_\odot \le m_{\rm BH} < 10^8$\,M$_\odot$, and $10^8$\,M$_\odot 
\le m_{\rm BH}$, respectively.

Figure~\ref{fig:zf_vs_halo_mass} shows the formation redshift $z_f$ versus host 
halo mass $m_h$ for all black holes. Here, $z_f$ is defined as the redshift 
at which a black hole's most massive progenitor for the first time contained half 
of its $z=1$ mass\footnote{In order for $z_f$ to be determined a BH must have 
acquired at least twice the seed mass by $z=1$.}. Different symbol sizes (and 
colours) give the three different mass bins as indicated in the legend. As in 
previous Figures, filled and open circles indicate main and satellite BHs, 
respectively. 

As expected in the standard hierarchical scenario, a large number of small 
black hole systems forms first, and these BHs are followed by intermediate and 
high mass systems at lower redshifts. However, Figure~\ref{fig:zf_vs_halo_mass} 
also confirms our earlier finding (Section~\ref{sec:coevolution}), namely
that a large fraction of the high mass black holes have formation redshifts 
$z>2.5$. The formation redshifts of the largest BHs cover the full range of 
$z_{f}$. Thus, while very massive haloes form last in a hierarchical structure 
formation scenario, massive black holes are decoupled from this process.

Figure~\ref{fig:BH_mean_formation_z} relaxes the condition that the formation
redshift is the redshift at which half the $z=1$ was first assembled in the
largest progenitor. Instead, here the fraction $f_{\rm m}$ of final mass is
now given on the y--axis, and the redshift at which that mass fraction was
first assembled can be read of from the x--axis (the $f_{\rm m} = 0.5$ case 
corresponds to our earlier definition of formation redshift). The shaded and 
dashed areas around the mean are the {\it rms} around that mean. Regardless 
of the mass fraction used to define $z_f$, the most massive black holes always 
have the highest average formation redshifts (note that because of the choice 
of seed black hole mass in the simulation, for the lowest mass black holes in
Figure~\ref{fig:BH_mean_formation_z}, a formation redshift can only be
determined down to $f_{\rm m} = 0.5$). This clear antihierachical assembly of
BH mass in our simulation is consistent with the presence of high--$z$ quasars
and thus provides a promising result of the self--consistent modelling of
black hole growth in our cosmological simulations.

A comparison with Malbon et al.'s (2007)
semi--analytical results is instructive, even though it is extremely important
to remember the differences in the final redshifts (unlike in our case, Malbon
et al.  evolved their model until $z=0$). Their Figure~11 (right--hand
column) can be compared with Figures~\ref{fig:zf_vs_halo_mass} and
\ref{fig:BH_mean_formation_z}. At $z=0$, these authors find hierarchical
assembly of BH mass (using 50 or 95 percent of the final BH mass). Our results
albeit at $z=1$, see hierarchical assembly for all but the most massive black
holes. Given the formation histories of dark matter haloes, where the most
massive haloes merge at redshifts below $z=1$, our results do not appear to
contradict Malbon et al.'s: One can expect the massive BHs in very massive
haloes to merge at $z<1$, thus shifting the formation redshifts of $z=0$ to
the values seen in Malbon et al. It is interesting to note also that the BH
formation redshifts are also very similar to the formation redshifts of the
brightest cluster galaxies (BCGs) in De Lucia \& Blaizot (2007).

%
%
\begin{figure}
\begin{tabular}{cc}
  \hspace{-1.0cm}
  \begin{minipage}{85mm}
    \begin{center}
      \includegraphics[width=85mm]{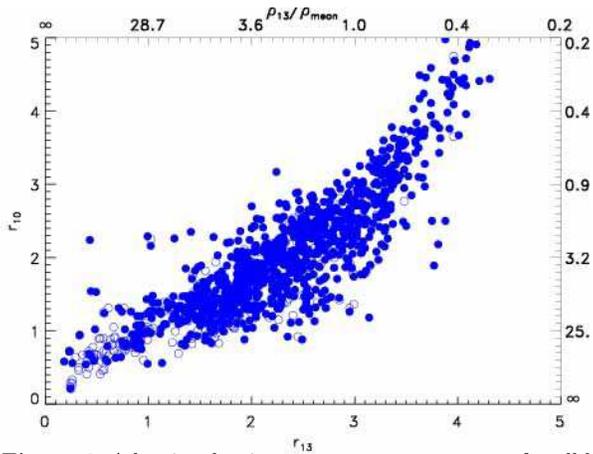}
    \end{center}
  \end{minipage}
\end{tabular}
\vspace{-0.4cm}
\caption{Adaptive density measures $r_{13}$ versus $r_{10}$ for all
         black holes in the sample.}
\label{fig:density_measures}
\end{figure}

%
%
\begin{figure}
\begin{tabular}{cc}
  \hspace{-0.6cm}
  \begin{minipage}{90mm}
    \begin{center}
      \includegraphics[width=95mm]{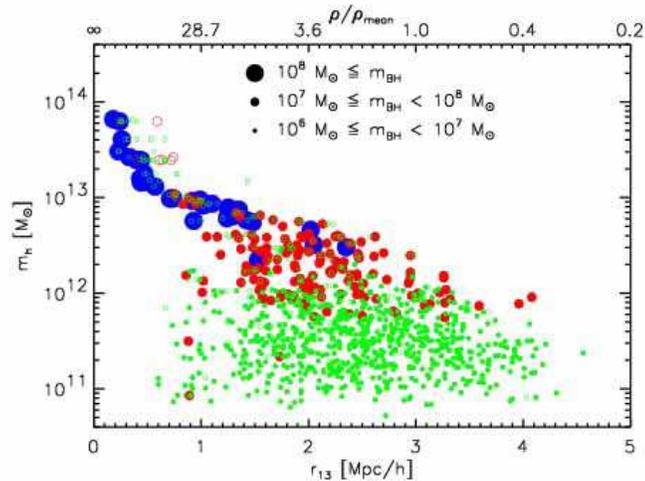}
    \end{center}
  \end{minipage}
\end{tabular}
\vspace{-0.4cm}
\caption{$z=1$ BH local density $r_{13}$ versus host halo mass $m_h$, using 
         different size symbols (and colours) for three different mass
         ranges. In addition, we use closed and open circles for main and 
         satellite BHs.}
\label{fig:density_halo_mass}
\end{figure}

%
%
\begin{figure*}
  \begin{center}
    \begin{tabular}{cc}
      \hspace{-0.9cm}
      \begin{minipage}{95mm}
        \begin{center}
          \includegraphics[width=95mm]{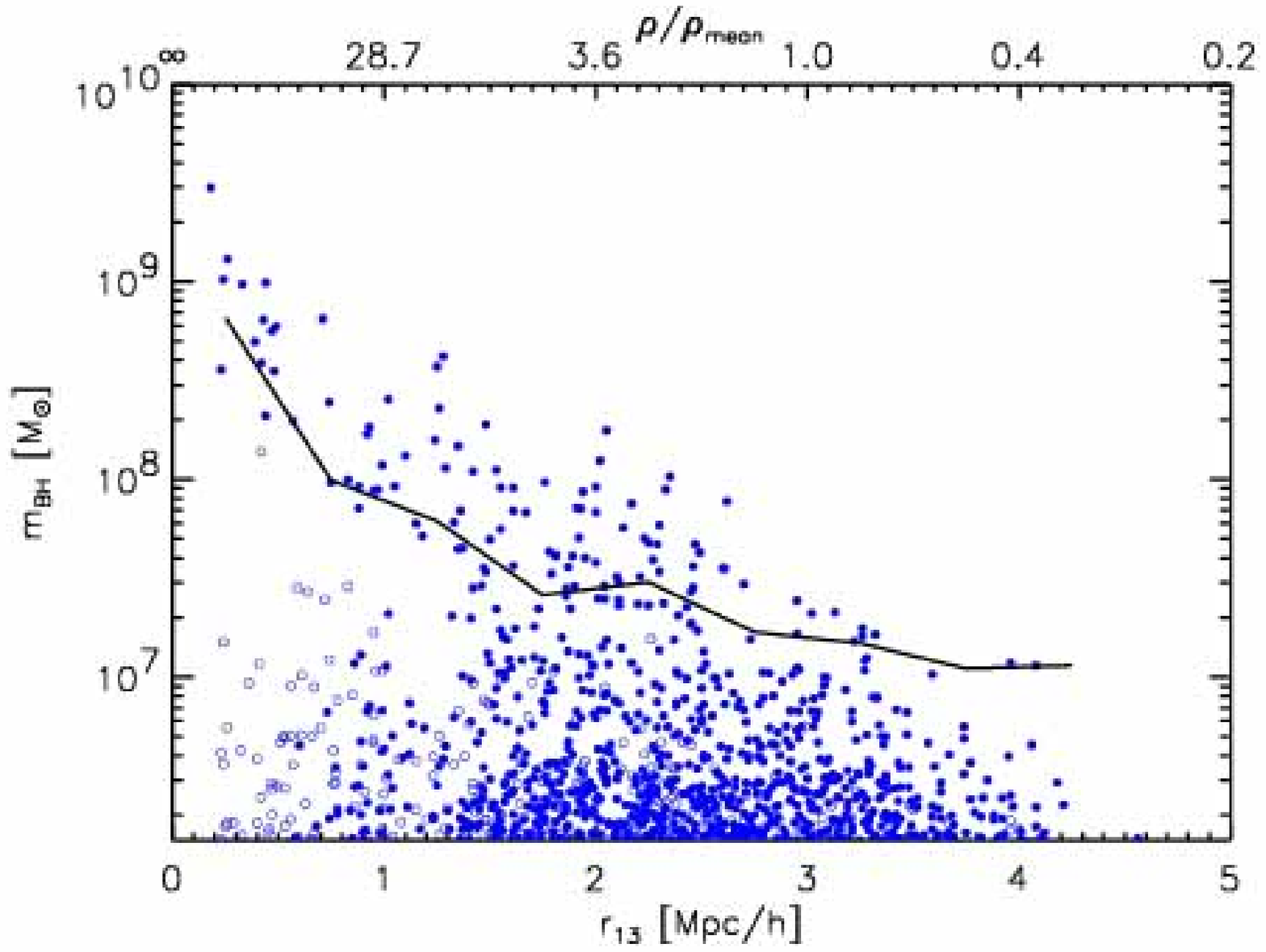}
        \end{center}
      \end{minipage}
      \hspace{-0.9cm}
      \begin{minipage}{95mm}
        \begin{center}
          \includegraphics[width=95mm]{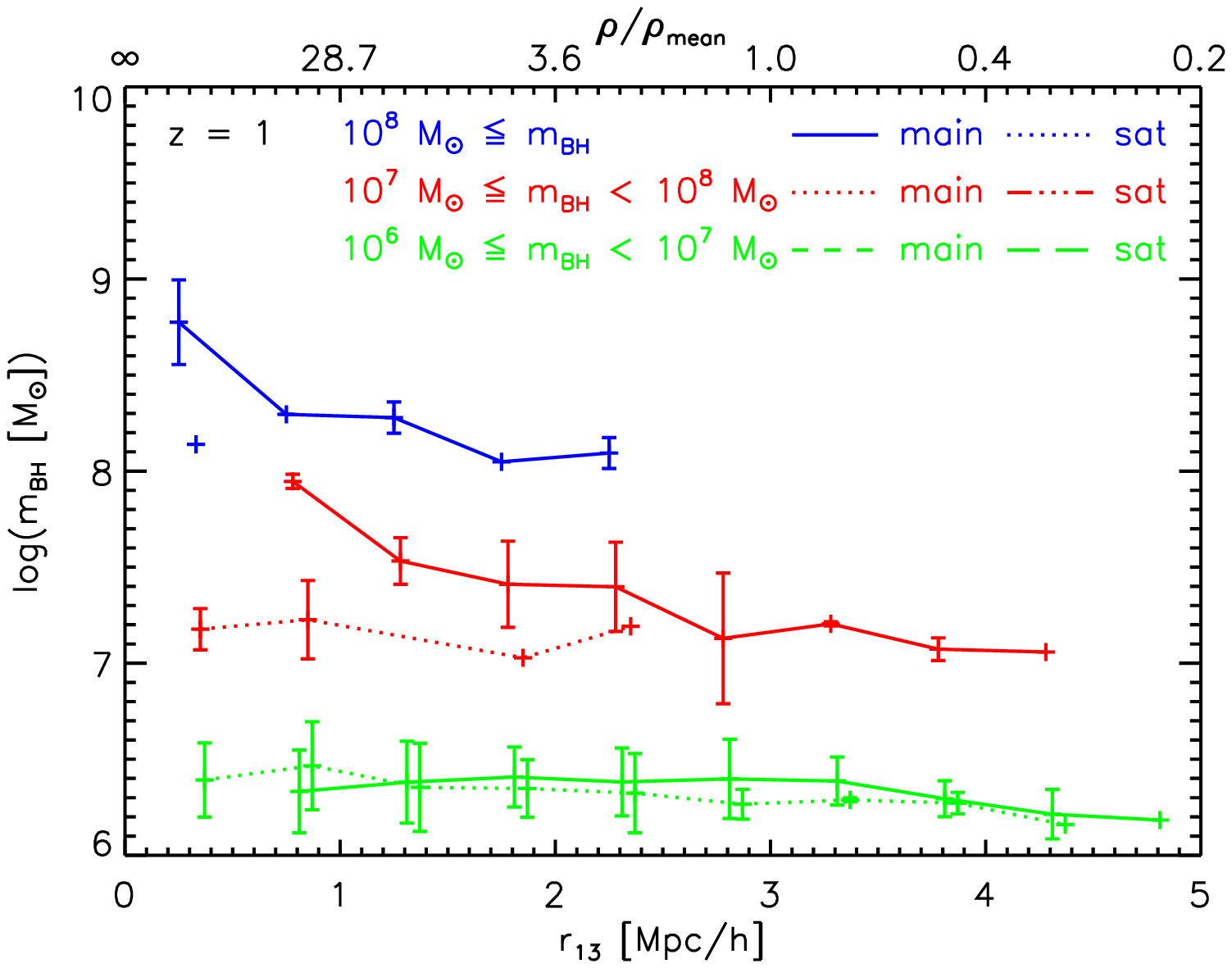}
        \end{center}
      \end{minipage}
    \end{tabular}
    \vspace{-4mm}
    \begin{tabular}{cc}
      \hspace{-0.9cm}
      \begin{minipage}{95mm}
        \begin{center}
          \includegraphics[width=95mm]{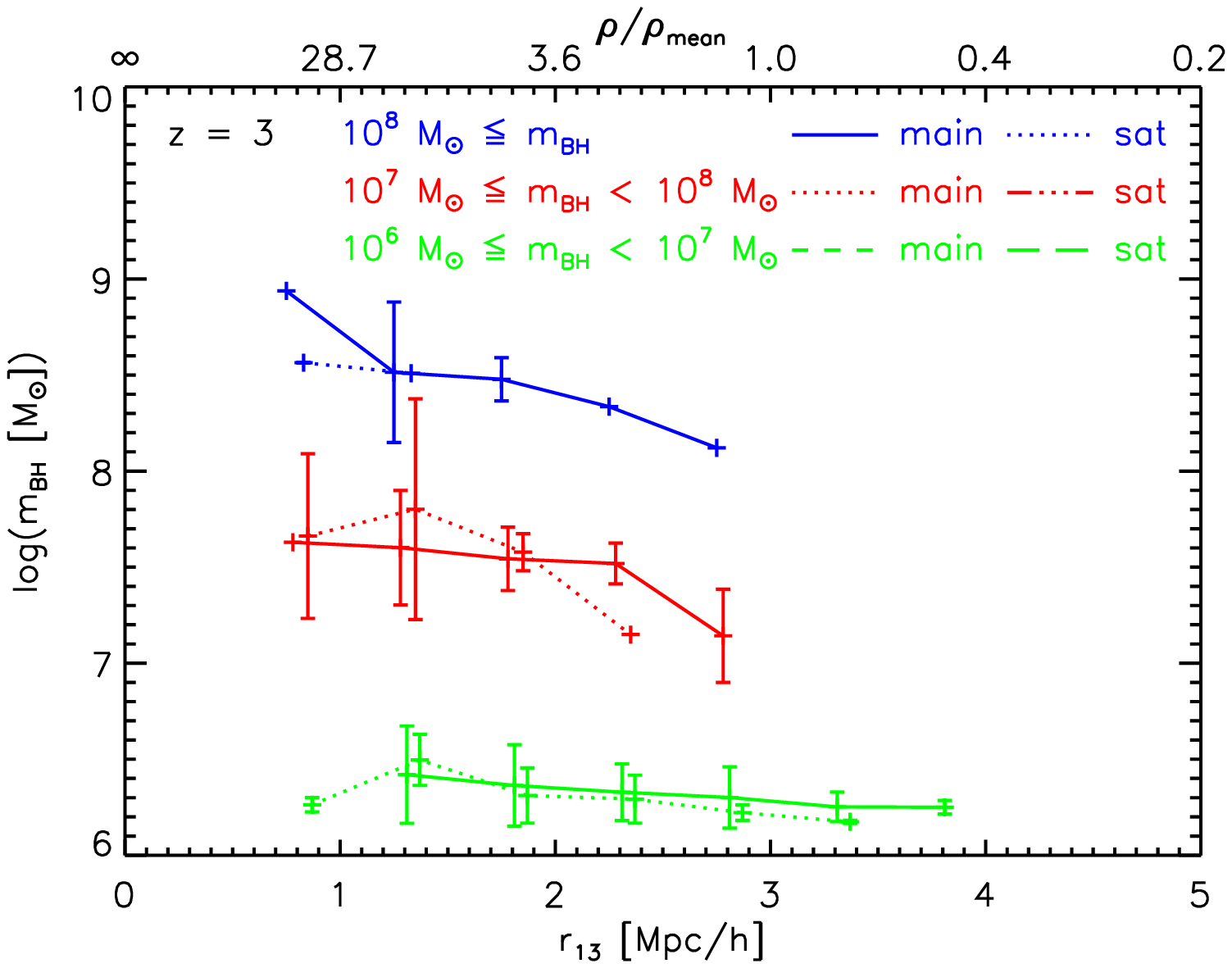}
        \end{center}
      \end{minipage}
      \hspace{-0.9cm}
      \begin{minipage}{95mm}
        \begin{center}
          \includegraphics[width=95mm]{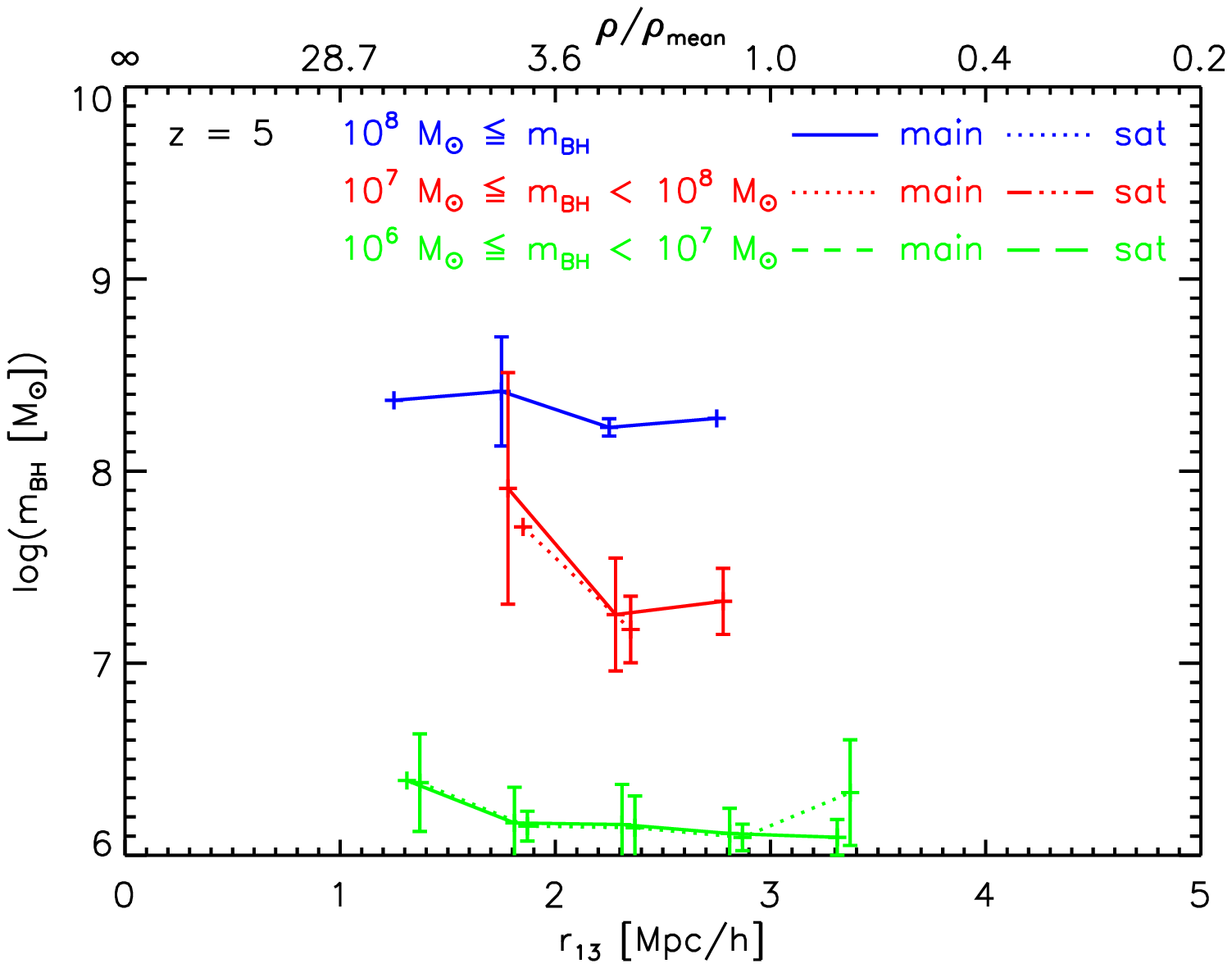}
        \end{center}
      \end{minipage}
    \end{tabular}
    \end{center}
    \vspace{-4mm}
  \caption{Upper left panel: $z=1$ BH local density $r_{13}$ versus black hole mass 
           $m_{\rm BH}$, using closed and open circles for main and satellite
           BHs, respectively. Superimposed as a solid line is the mean for all BHs
           with $M_{\rm BH} \ge $10$^{7}\,$M$_\odot$. Upper right and bottom panels:
           Distributions of the median values for three different redshifts,
           using different size symbols (and colours) for three
           different mass ranges and with error bars showing a 25\% spread.}
\label{fig:density_halomass}
\end{figure*}

%
%
\begin{figure*}
  \begin{center}
    \begin{tabular}{cc}
      \hspace{-0.9cm}
      \begin{minipage}{95mm}
        \begin{center}
          \includegraphics[width=95mm]{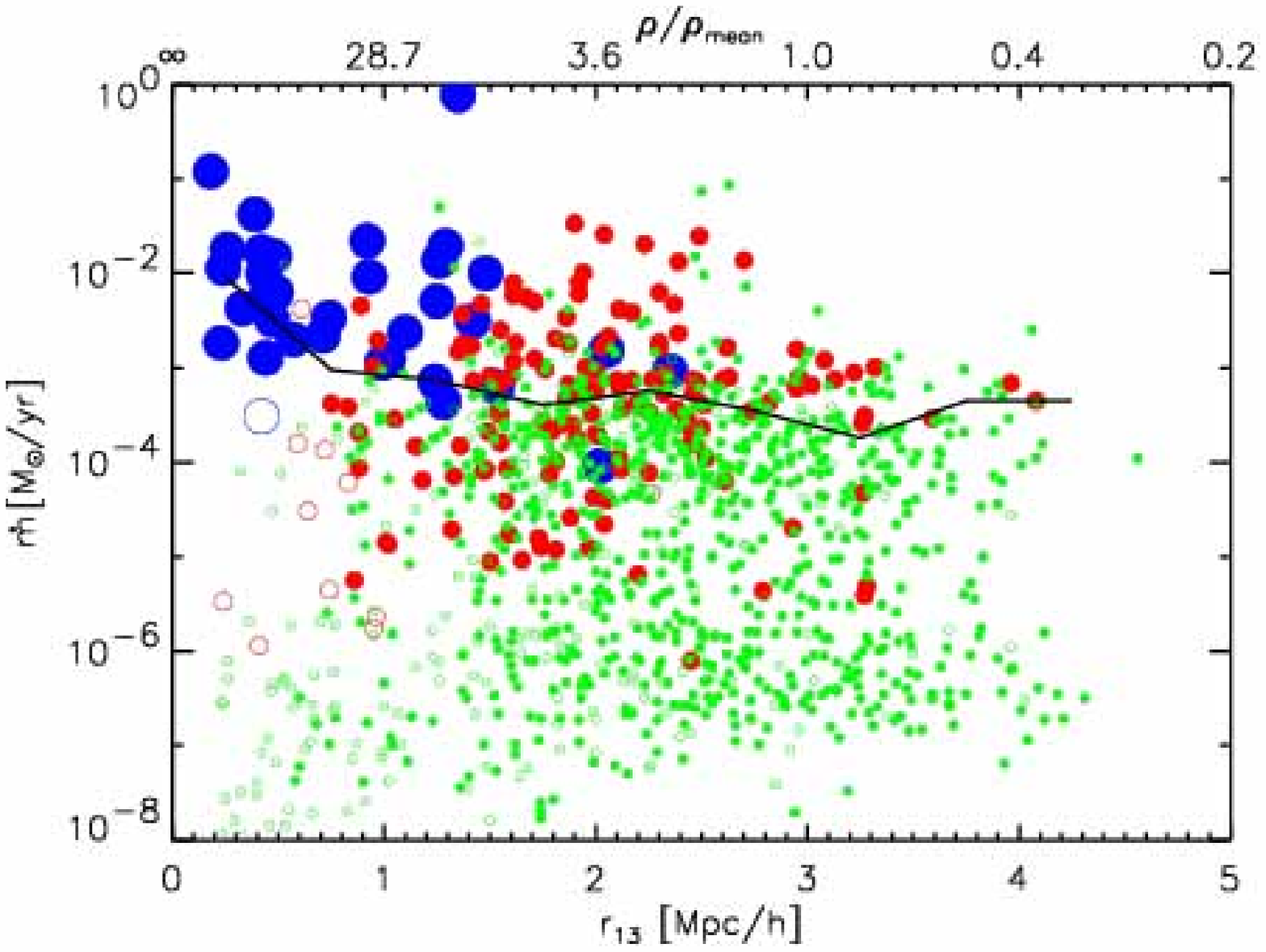}
        \end{center}
      \end{minipage}
      \hspace{-0.9cm}
      \begin{minipage}{95mm}
        \begin{center}
          \includegraphics[width=95mm]{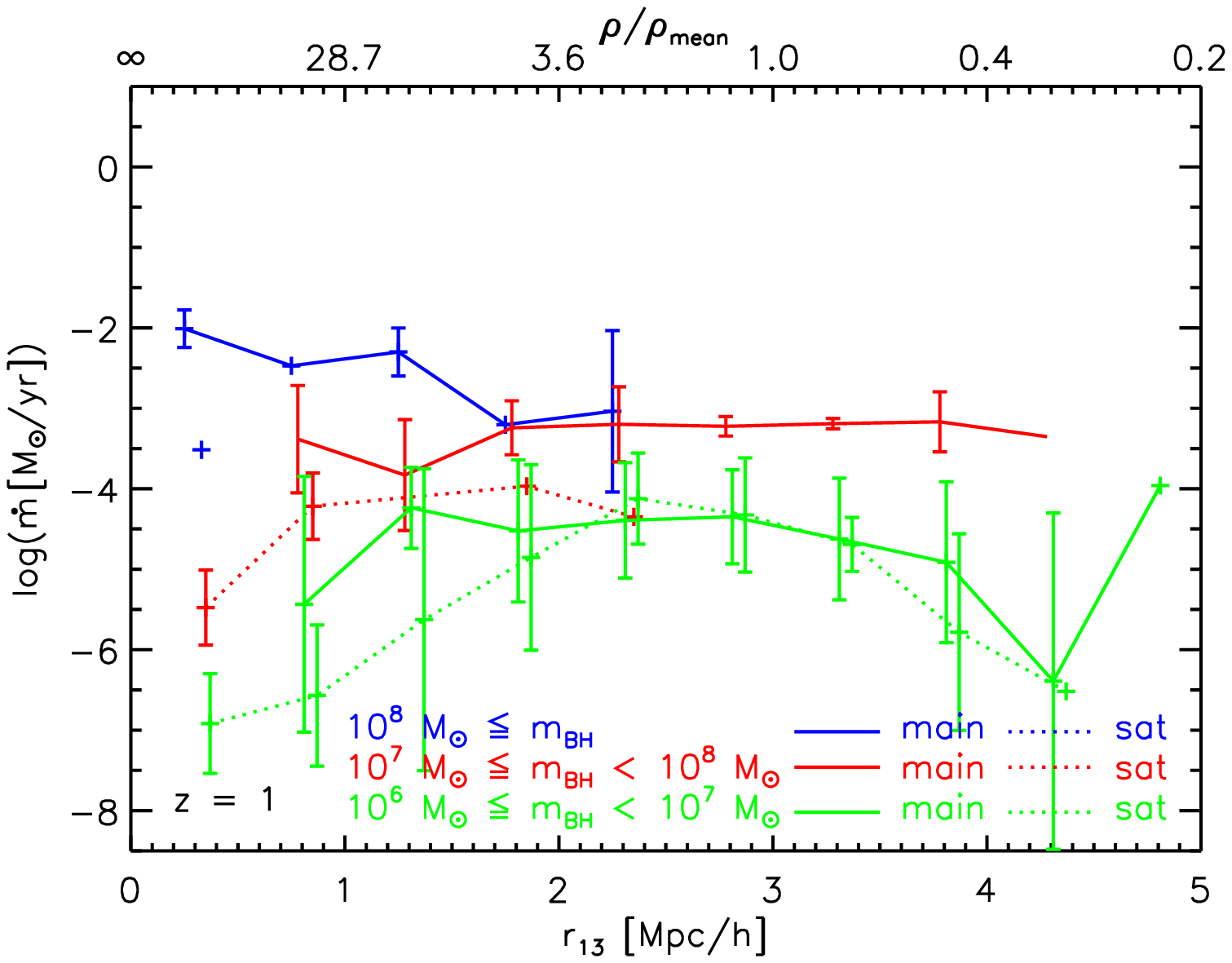}
        \end{center}
      \end{minipage}
    \end{tabular}
    \vspace{-4mm}
    \begin{tabular}{cc}
      \hspace{-0.9cm}
      \begin{minipage}{95mm}
        \begin{center}
          \includegraphics[width=95mm]{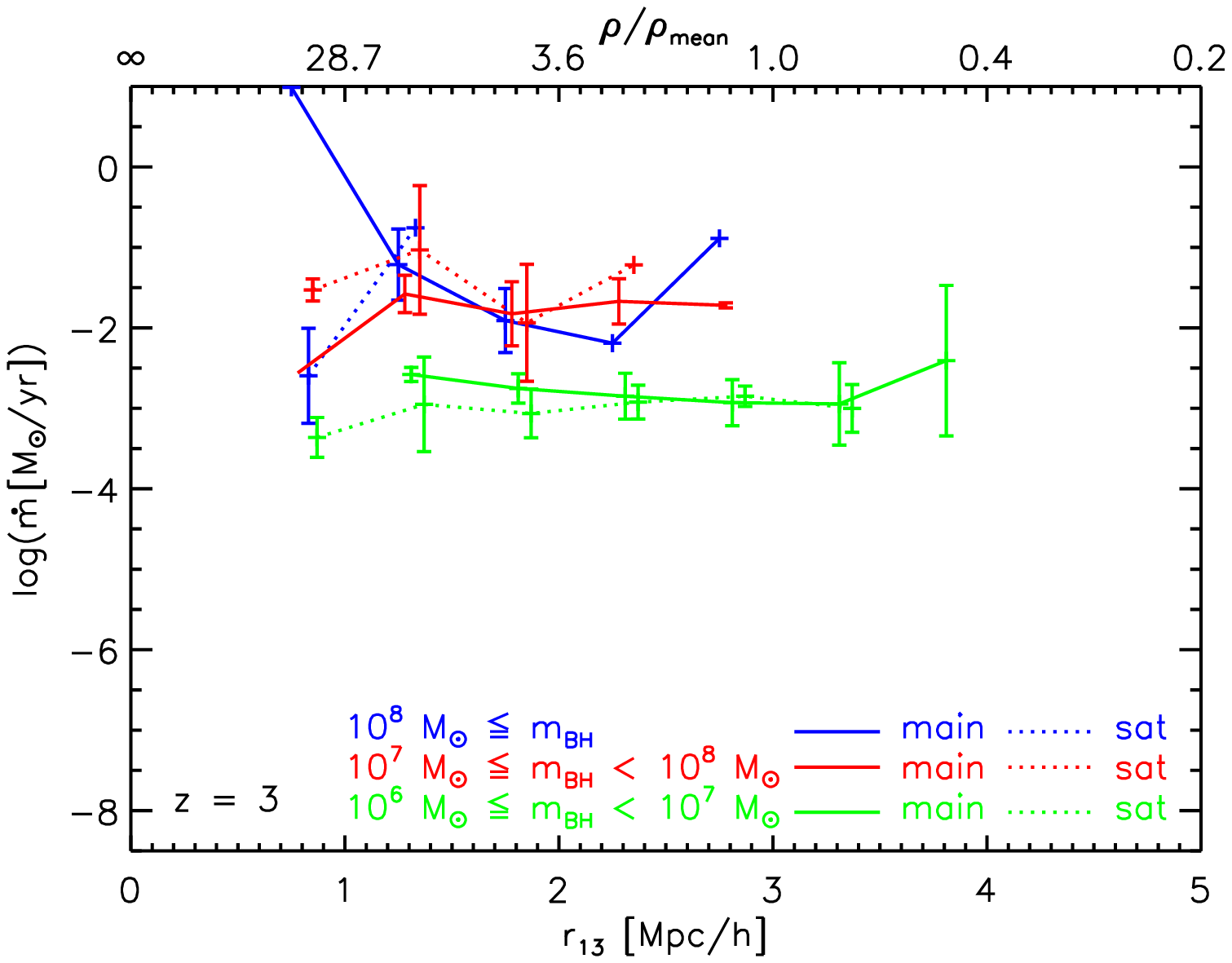}
        \end{center}
      \end{minipage}
      \hspace{-0.9cm}
      \begin{minipage}{95mm}
        \begin{center}
          \includegraphics[width=95mm]{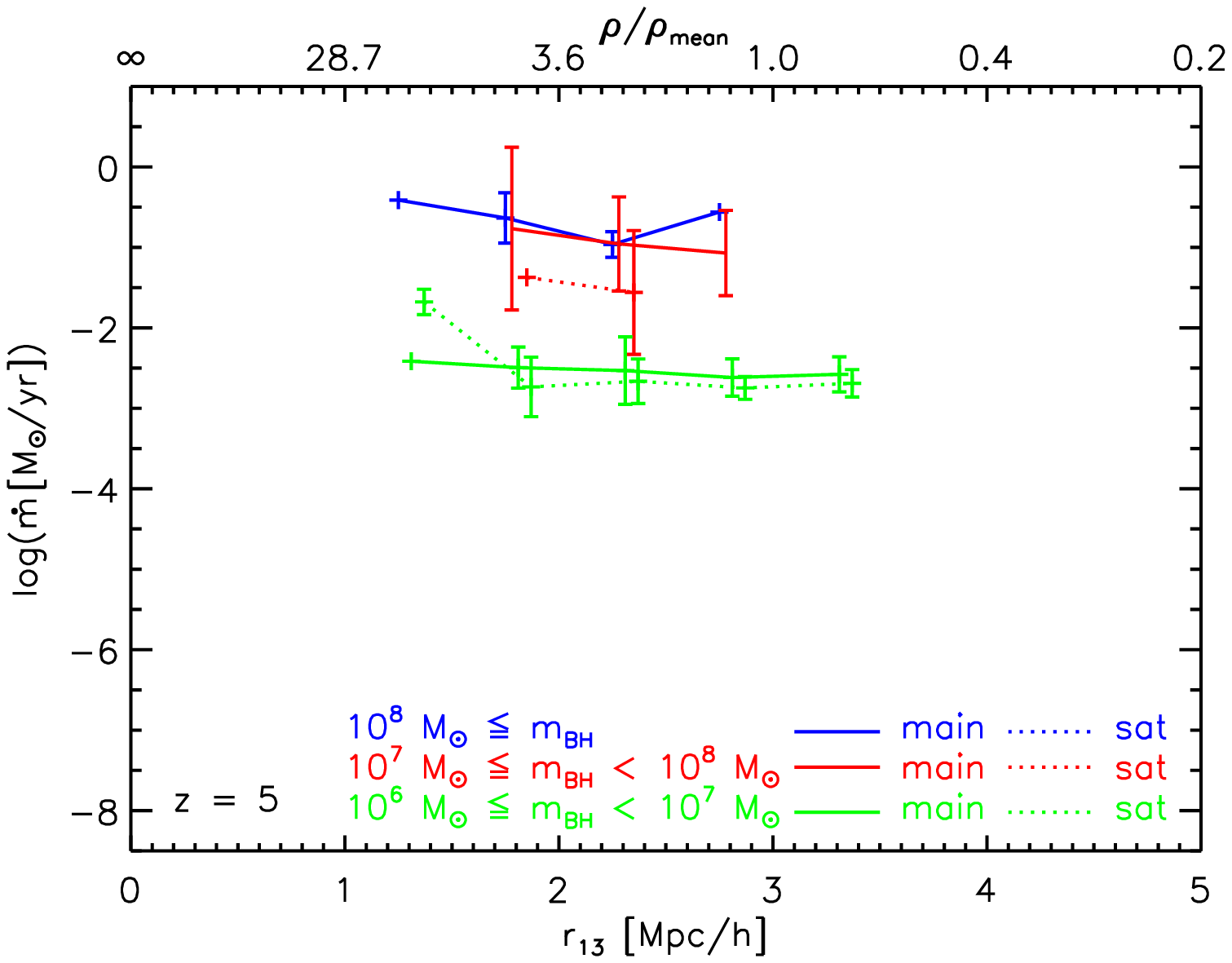}
        \end{center}
      \end{minipage}
    \end{tabular}
    \end{center}
    \vspace{-4mm}
  \caption{Upper left panel: $z=1$ BH accretion rate as a function of the local 
           density $r_{13}$, using different size symbols (and colours) 
           for three different mass ranges. In addition, we use closed 
           and open circles for main and satellite BHs. Superimposed as a 
           solid line is the mean for all BHs with $M_{\rm BH} \ge 
           $10$^{7}\,$M$_\odot$. Upper right 
           and bottom panels: Distributions of the median values for three 
           different redshifts, using the same colours for the BH mass 
           ranges as in the upper left panel and with error bars showing 
           a 25\% spread.}
\label{fig:density_accretion}
\end{figure*}

%
%
\begin{figure*}
  \begin{center}
    \begin{tabular}{cc}
      \hspace{-0.9cm}
      \begin{minipage}{95mm}
        \begin{center}
          \includegraphics[width=95mm]{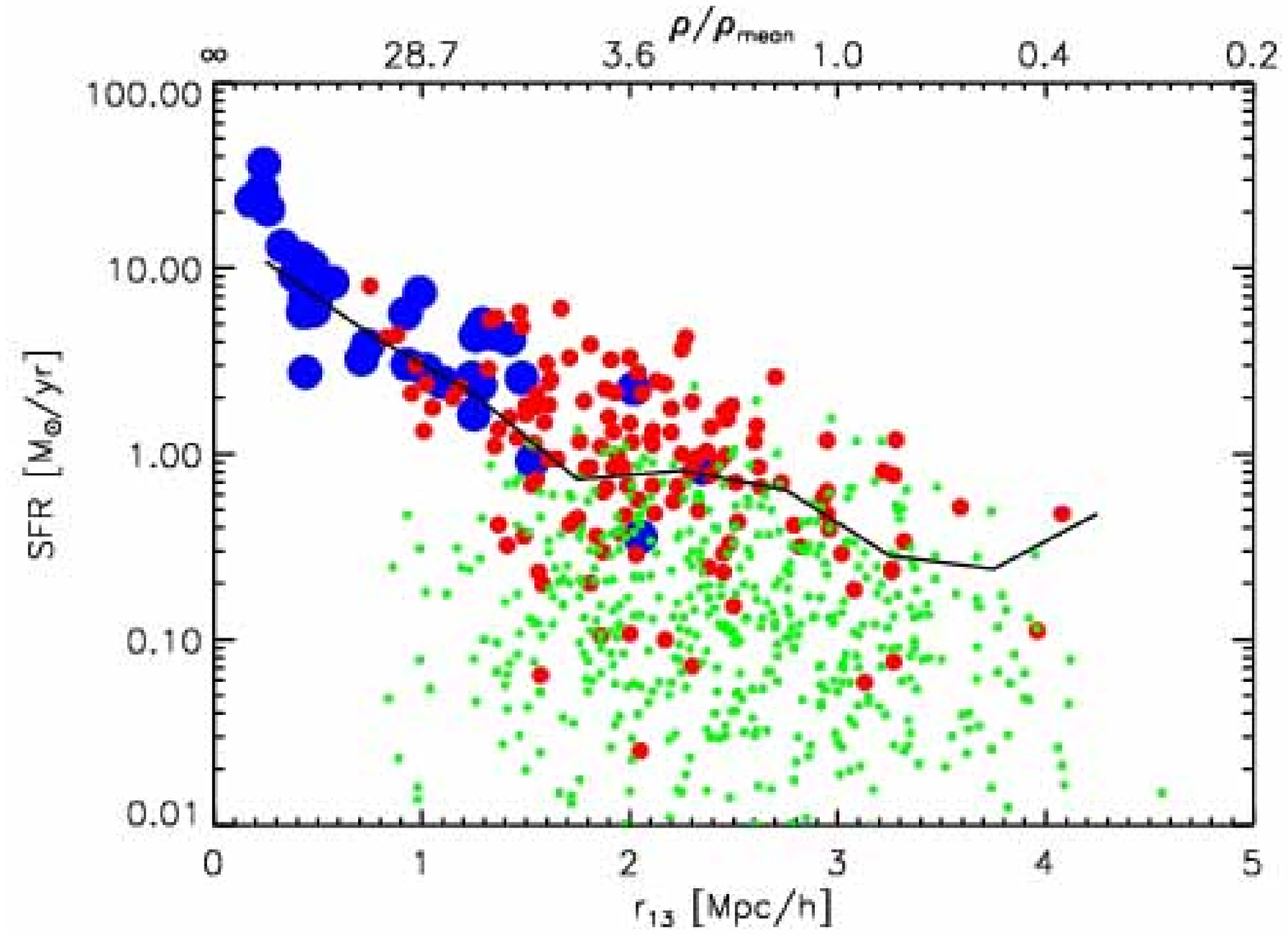}
        \end{center}
      \end{minipage}
      \hspace{-0.9cm}
      \begin{minipage}{95mm}
        \begin{center}
          \includegraphics[width=95mm]{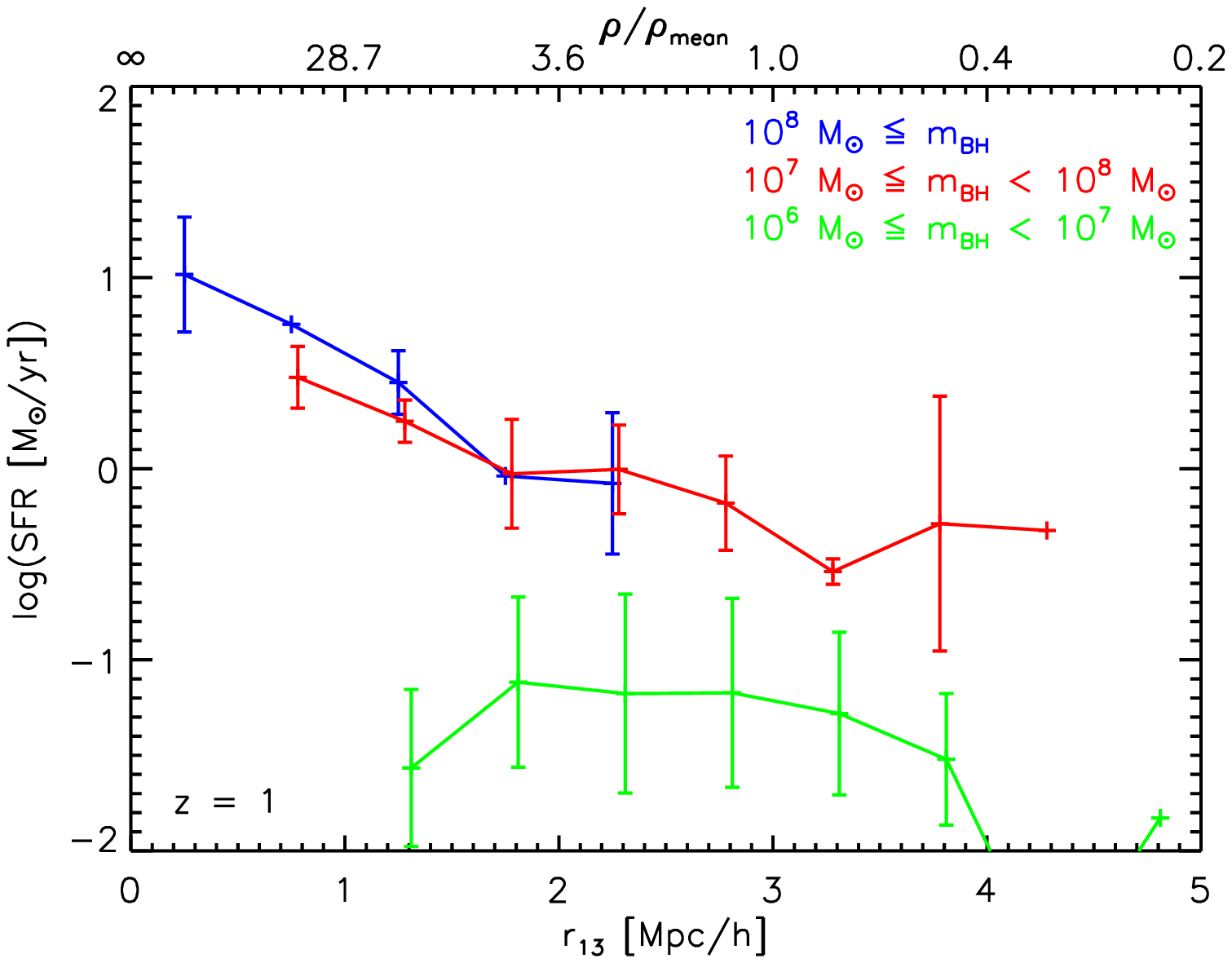}
        \end{center}
      \end{minipage}
    \end{tabular}
    \vspace{-4mm}
    \begin{tabular}{cc}
      \hspace{-0.9cm}
      \begin{minipage}{95mm}
        \begin{center}
          \includegraphics[width=95mm]{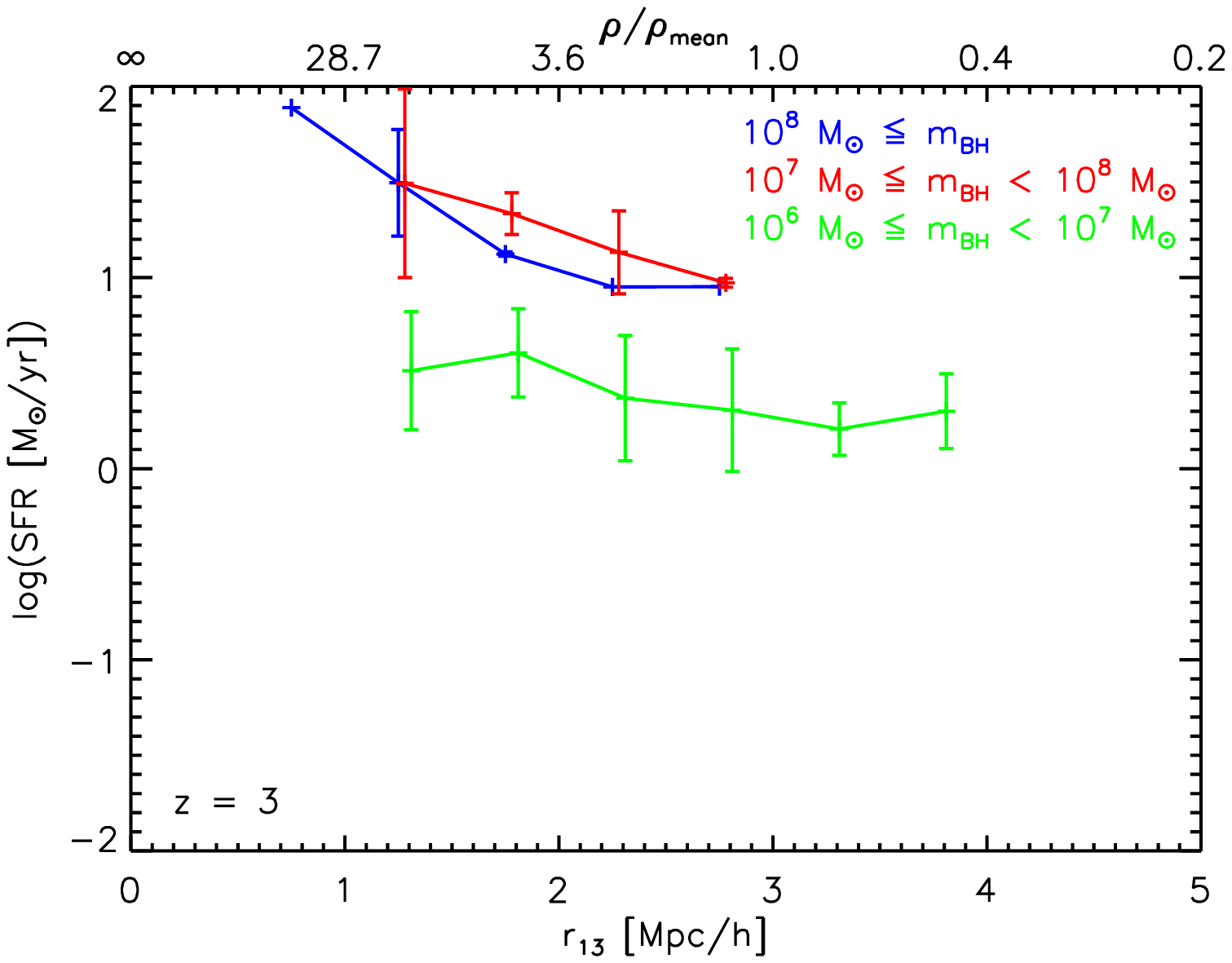}
        \end{center}
      \end{minipage}
      \hspace{-0.9cm}
      \begin{minipage}{95mm}
        \begin{center}
          \includegraphics[width=95mm]{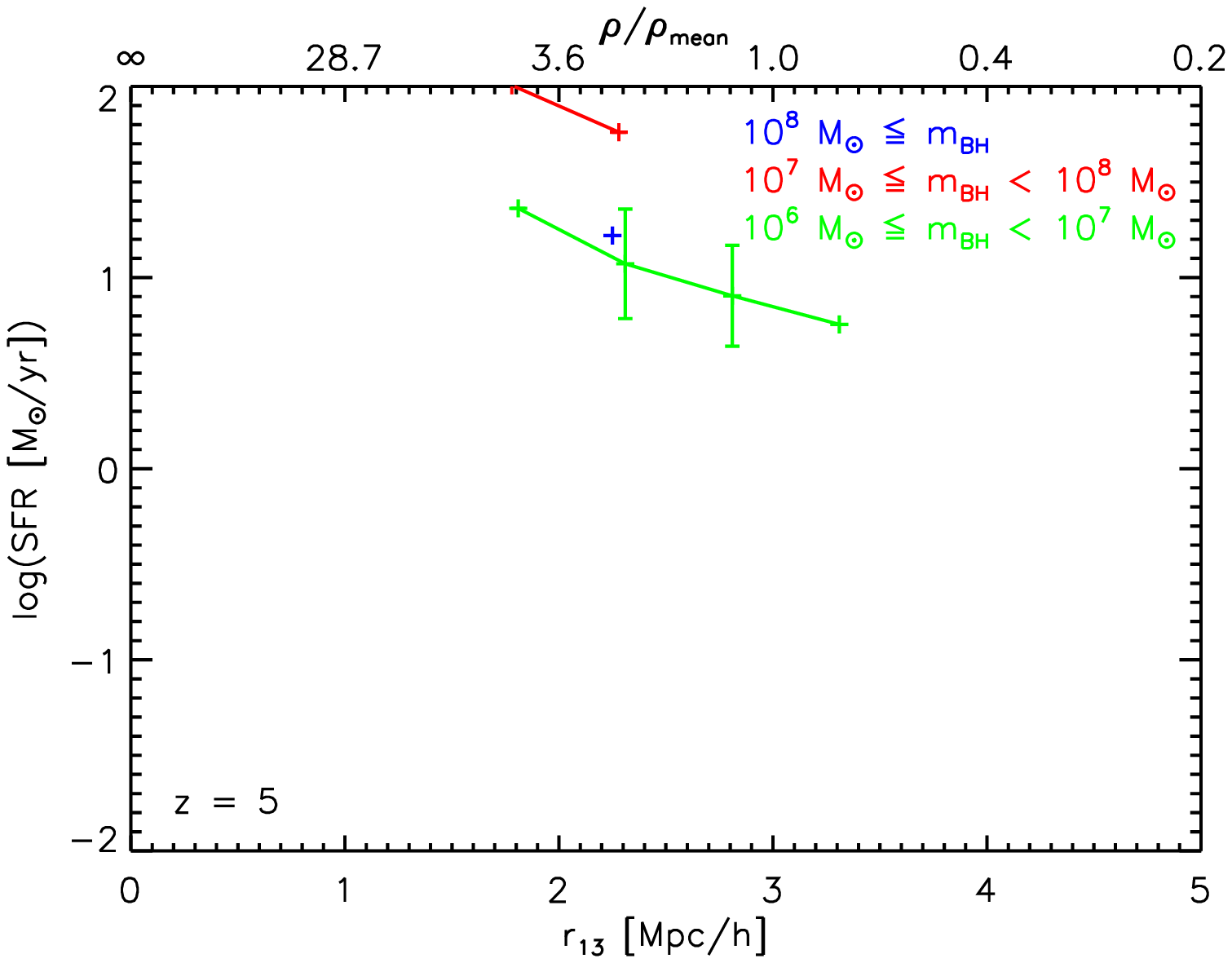}
        \end{center}
      \end{minipage}
    \end{tabular}
    \end{center}
    \vspace{-4mm}
  \caption{Upper left panel: $z=1$ star formation rates in galaxies that host
           BHs as a function of the local density $r_{13}$, using different
           size symbols (and colours) for three different mass ranges. 
           Superimposed as a solid line is the mean for all BHs
           with $M_{\rm BH} \ge $10$^{7}\,$M$_\odot$. Upper right 
           and bottom panels: Distributions of the median values for three 
           different redshifts, using the same colours for the BH mass 
           ranges as in the upper left panel and with error bars showing a 
           25\% spread.}
\label{fig:density_SFR}
\end{figure*}

\section{BHs and Their Environments} \label{sec:environments}

\subsection{The Large--Scale Environments of BHs} \label{sec:largescale}

Having explored the connection between black holes and their host haloes, we
now look at the influence of the large--scale environment on the growth of
supermassive black holes. Images of slices through our simulation volume (see
Di Matteo et al. 2007) show a plethora of environments, ranging from small
clusters/groups to voids and filaments. Our goal is to define a measure of
density that preserves these different environments and that makes use of
the fact that because of the large number of particles in our simulation
volume we sample the underlying density field extremely well. Inspired by the
SPH treatment, where the local density around a particle is
determined by the distance to its $n$th neighbour (with $n$ typically of the
order of 32 or 64), we quantify the large--scale environment through the
quantity $r_{13}$, which for each black hole gives the radius of the sphere
centered on that black hole that contains a mass of $10^{13}$\,M$_\odot$. Mean
density corresponds to about $r_{13} \approx 3.0\,h^{-1}$\,Mpc. This is an
adaptive density measure, which has the advantage of not blurring clusters or
filaments into the voids (both theoretical and observational studies of voids
have shown that their edges are rather steep; see Benson et al. 2003 or
Colberg et al. 2005 for examples).  A drawback of using $r_{13}$ is that it
cannot be easily interpreted in a linear--theory framework.

Adaptive measurements of local environment, such as the one adopted here, are
recently becoming increasingly common in observational studies, too --
essentially, for the same reasons that we just outlined. For example, for
volume--limited samples drawn from the SDSS, Park et al. (2007) define
environment via a Spline kernel containing 20 $L_*$ galaxies. And Cooper et
al. (2005) finds that the projected $n$--th nearest neighbour distance is the
most accurate estimate of galaxy density for the DEEP2 galaxy samples. To show
that $r_{13}$ is in fact quite similar to measuring the distance to an $n$--th
neighbour, in Figure~\ref{fig:density_measures}, we show the relationship
between $r_{13}$ and $r_{10}$, where for each black hole $r_{10}$ is the
distance to the 10th--nearest neighbour (solid and open circles show main and
satellite BHS, respectively). As can be seen $r_{13}$ and $r_{10}$ are pretty
well correlated and hence provide consistent measure of local enviroment. Here
we adopt $r_{13}$ as that makes use of the full density field in the
simulations.

In Figure~\ref{fig:density_halo_mass}, we show how black holes populate 
their host haloes and environments at $z=1$. Each black hole is represented 
by a circle (as before open and closed for satellite and main BHs, 
respectively), and plotted in the $m_{\rm h}$--$r_{13}$ ($\rho/\rho_{\rm 
mean}$) plane. The most massive haloes, which reside in the highest density 
regions, contain the most massive black holes. These populate the regions 
of $\rho /\rho_{mean} >$ a few, i.e. what we would identify as cluster/group 
like environments and filaments.  Regions of low density (with $r_{13} > 
3\,h^{-1}$\,Mpc or, equivalently, $\rho/\rho_{mean} < 1.0$) contain only 
haloes with masses lower than $10^{12}$\,M$_\odot$ and BHs with masses lower 
than $10^{7}$\,M$_\odot$.  It is interesting to note that black holes with 
masses in the range $10^7\Msun \ge M_{\rm BH} \le 10^{8} \Msun$ can be 
found across the whole range of densities (see also
Section~\ref{sec:starvation}). Also note that satellite black holes populate 
the full range of densities.

While Figure~\ref{fig:density_halo_mass} contains information about the BHs as
a function of host halo mass and large--scale environment, in the following we
will restrict our focus to large--scale environment only, as we have already
discussed the explicit one--to--one relation of $m_{\rm BH}$ versus $m_{\rm h}$
in Figure~\ref{fig:BH_vs_halo_2}. The upper left panel in 
Figure~\ref{fig:density_halomass} shows black hole masses $m_{\rm BH}$ as a
function of local density, $r_{13} \; (\rho /\rho_{\rm mean})$ and the mean
for all black holes with $M_{\rm BH} \ge $10$^{7}\,$M$_\odot$. Overall, there 
is a rough mean dependency with $m_{\rm BH} \sim
(\rho/\rho_{\rm mean})^{-3/2}$ for black holes with $M_{\rm BH} \ge
$10$^{7}\,$M$_\odot$. As noted above, larger mass BHs live in denser
environments, while intermediate to low mass BHs reside across a large range of
densities.

In the three additional panels in Figure~\ref{fig:density_halomass} we show
the median $m_{\rm BH}$ as a function of $r_{13}$ for the three mass bins
introduced before at three different redshifts $z=1$ (upper right panel),
$z=3$ (lower left panel), and $z=5$. For $z=5$, we take black holes from the
E6 simulation, which -- due to its larger volume -- contains more objects.
As before, we also divide the BH samples into main (solid lines) and satellite 
(dotted lines) BHs, and the error bars show the 25\% percentiles around the median.  
At both $z=1$ and $z=3$ the density dependence decreases across BH mass bins, 
with the highest BH mass bin (blue symbols) showing the strongest dependence 
and the lowest BH mass bin (green symbols) having none. Also note 
that the three panels reflect the overall growth of structure between 
$z=5$ and $z=1$: as cosmic time progresses, an increasingly larger range of 
densities gets covered.

Using the same panel structure as in Figure~\ref{fig:density_halomass},
Figure~\ref{fig:density_accretion} shows the density dependency of black hole
accretion rates, again with E6 data used for $z=5$. $\dot{m}$ spans a much
larger dynamic range than $m_{\rm BH}$. However, its overall mean density
dependence is similar to that of $M_{\rm BH}$. We find that approximately
$\dot{m} \propto (\rho/\rho_{\rm mean})^{-3/2}$ for $M_{BH} \ge 10^{7} \Msun$,
albeit with much larger scatter across all densities. In addition, at $z=1$,
the most active black holes (with $\dot{m} \ge
1.0\cdot10^{-2}$\,M$_\odot$/year) avoid underdense regions, that is they can
be found in environments with $r_{13} < 3.0\,h^{-1}$\,Mpc (or $\rho/\rho_{\rm
mean} > 1.0$). Enhanced accretion in higher density regions is consistent with
the overall picture that gas rich mergers (which are occuring in such regions)
are the main trigger of quasar phase as also supported from the analysis in Di
Matteo et al. (2007). Note that if we expressed the black hole accretion rates
in Eddington units we would be washing out some of our result because of the
additional BH mass dependence in Eddington rates.

At $z=1$, satellite black holes have lower accretion rates than central black
holes. This is because their host haloes had their gas stripped when they fell
into the large halo (Gunn \& Gott 1972, see Moore et al. 2000 for a
review). We find that the hosts of satellite black holes -- themselves
satellites inside their host halo -- either have sharply reduced gas fractions
or no gas left at all.  The upper right and the two bottom panels (redshifts
$z=1$, $z=3$, and $z=5$) show that black holes in each of the three mass
samples have larger accretion rates at earlier times. What is more, the
density dependence of the accretion rates of the most massive black holes is
more pronounced at $z=3$ than at $z=1$. This result is also expected as gas
fraction in galaxies and the mergers rates, responsible for the strong quasar
evolution decline with decreasing redshift (see also Di Matteo et al. 2007).

Most of the discussion above concentrated on the sample of black holes with
masses $M_{\rm BH} \ge 10^7 \Msun$. Resolution effects are a concern at the
low--mass end of the black hole sample. While haloes in the mass range
considered throughout this work are very well resolved -- a halo of mass
$10^{11}$\,M$_\odot$ consists of almost three thousand particles -- accretion
onto the black holes with masses close to the seed mass are likely to be not
so well resolved. For the black holes considered here, we require each BH to
have at least twice the seed hole mass -- in part so that we can reliably
determine a formation redshift $z_f$. But this requirement does not
necessarily prevent problems due to resolution effects.  Some investigations
we performed by splitting up the low BH mass bins in subsample did not show
indications for resolution issues biasing our density dependence result. To
test for this directly however, a higher resolution simulation would be
needed.

We now compare the density dependence of black hole masses and
accretion rates with that of the star--formation rates (SFRs) of their
host  galaxies (see also Croft et al. 2008). In
Figure~\ref{fig:density_SFR}, we show SFRs in the BH host galaxies,
using the same notations as in Figures~\ref{fig:density_halomass} and
\ref{fig:density_accretion}. We first note that in
Figure~\ref{fig:density_SFR}, satellite galaxies (galaxies that
contain what we earlier called satellite BHs) are absent.  With the
exception of a handful of galaxies in the lowest BH mass range, which
have very  small SFRs, satellite galaxies in the simulation form no
stars, a fact that -- again -- supports the picture of most of the gas
being stripped from haloes/galaxies after they have fallen into a
larger halo. There are, of course, two gas components, namely cold gas
and hot gas. Simulations by McCarty et al. (2008) indicate that haloes
might retain around 30\% of their hot gas after falling into a larger
halo. The hot gas would then have to cool, however, to be able to form
stars -- something which does not appear to be happening in our
simulation (recall though that the simulation was only run until $z=1$, which
severely limits the actual time available to cool gas). This picture
is also supported by the simulations run by and analyzed by Kere\v{s}
et al. (in preparation),  where star formation in satellite
haloes/galaxies is severely quenched after redshifts of $z\approx3$
(Katz, private communication).

Figure~\ref{fig:density_SFR} shows that at all redshifts, SFR is perhaps the
strongest function of density, with SFR\,$\propto (\rho /\rho_{\rm
mean})^{-2.0}$.  SFRs are also highest in galaxies with the most massive
central BHs and increase with increasing redshift. Although a detailed
discussion of SFR in the simulation is beyond the scope of this work and will
be done elsewhere we are interested in its relation to BH accretion. It is
important, however, to point out that the dependence of SFR on density in the
simulations is {\it is completely consistent with the $z = 1$ results}
reported by Elbaz et al. (2007), who find an increase in SFR with local
density in the GOODS data (see their Figure~8), and by Cooper et al. (2007)
using DEEP2 data. Note this trend is reversed at $z=0$, (see, for example,
Kauffmann et al. 2004).  Note that semi--analytical galaxy model in the
Millennium Run (Croton et al. 2006) fail to reproduce this increase in SFR
with density (Fig.'s 8 and 9 in Elbaz et al. 2007). This agreement of the
dependence of SFR on local density at $z=1$ provides further support for the
validity of the model used in our simulations and the reliability of the
trends discussed here.  

\begin{figure}
\begin{tabular}{cc}
  \hspace{-0.6cm}
  \begin{minipage}{90mm}
    \begin{center}
      \includegraphics[width=90mm]{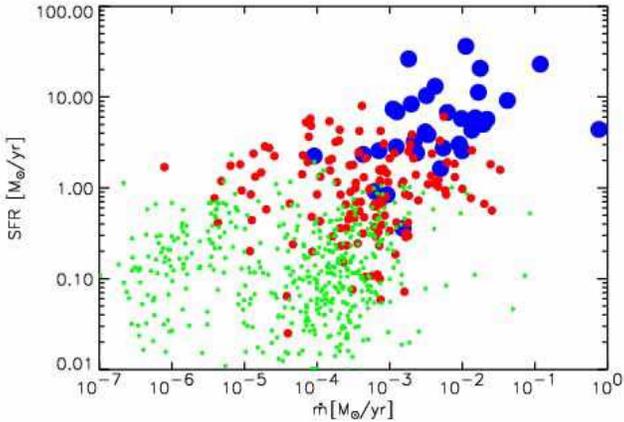}
    \end{center}
  \end{minipage}
\end{tabular}
\vspace{-0.4cm}
\caption{Accretion rate $\dot{m}$ versus SFR, both in M$_\odot$/yr for main
         BHs. As in the previous Figures, the symbol sizes and colours reflect
         the three BH samples discussed throughout this work.}
\label{fig:accretion_SFR}
\end{figure}

In particular, as it is already apparent from
Figures~\ref{fig:density_accretion} and \ref{fig:density_SFR}, there is no one
to one relationship between black hole accretion and star--formation rates of
the host galaxies. While BH accretion does depend somewhat on environment,
star formation rates show a stronger dependence. Such results are consistent
with quasar enviroments corrresponding to group/small group scales, where major
gas rich mergers will be most efficient, whereas the majority of the star
formation simply occurs in regions of high density and high concentration, and
most of it in a quiescent mode (see also Di Matteo et al. 2007). In
Figure~\ref{fig:accretion_SFR}, we compare directly the BH accretion rates and
the SFRs of their hosts directly. For fixed accretion (SFR), there is a wide
range in SFR (accretion), for all black hole mass ranges considered here.

%
%

%
%
\begin{figure*}
  \begin{center}
    \begin{tabular}{cc}
      \hspace{-0.9cm}
      \begin{minipage}{95mm}
        \begin{center}
          \includegraphics[width=95mm]{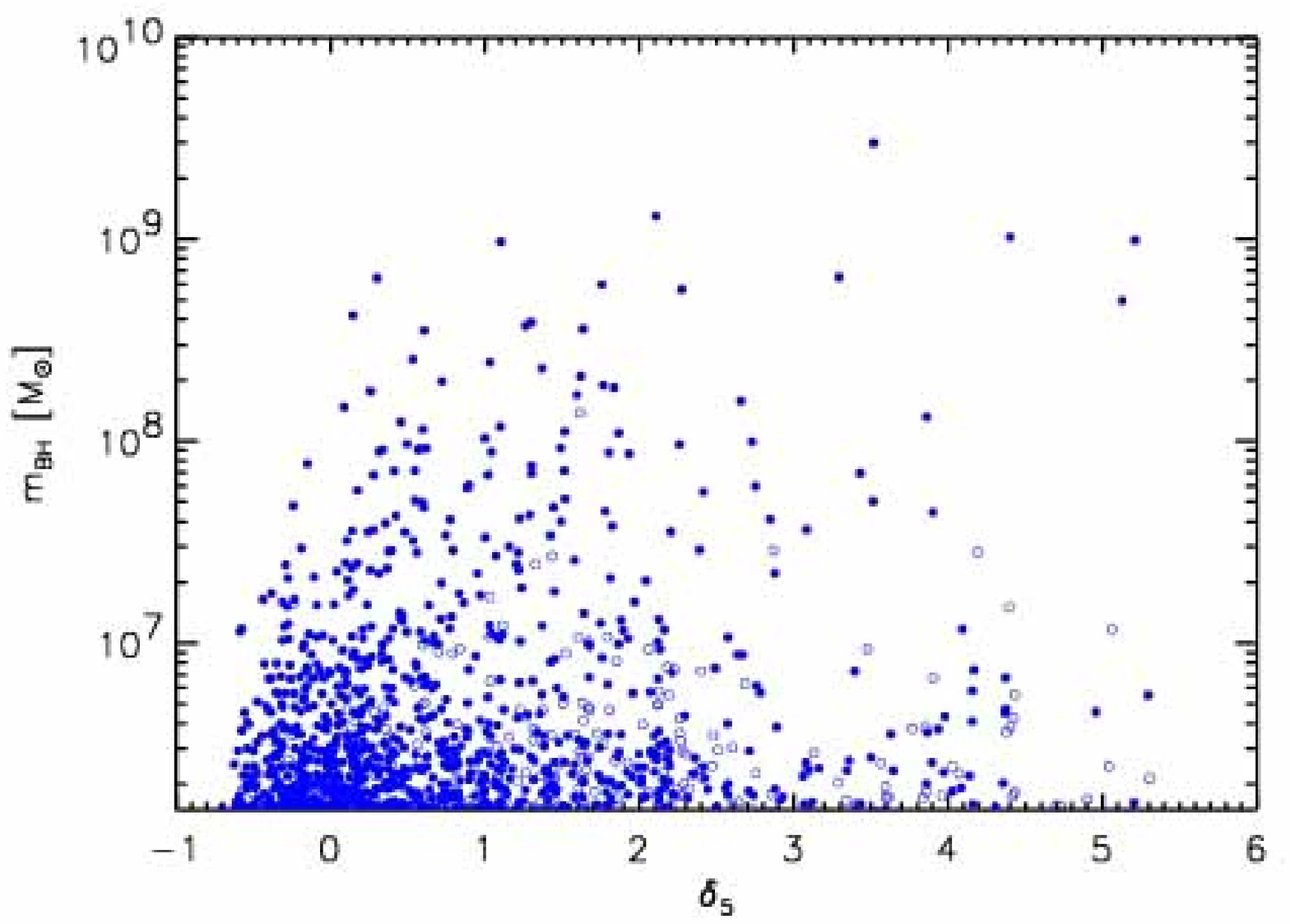}
        \end{center}
      \end{minipage}
      \hspace{-0.9cm}
      \begin{minipage}{95mm}
        \begin{center}
          \includegraphics[width=95mm]{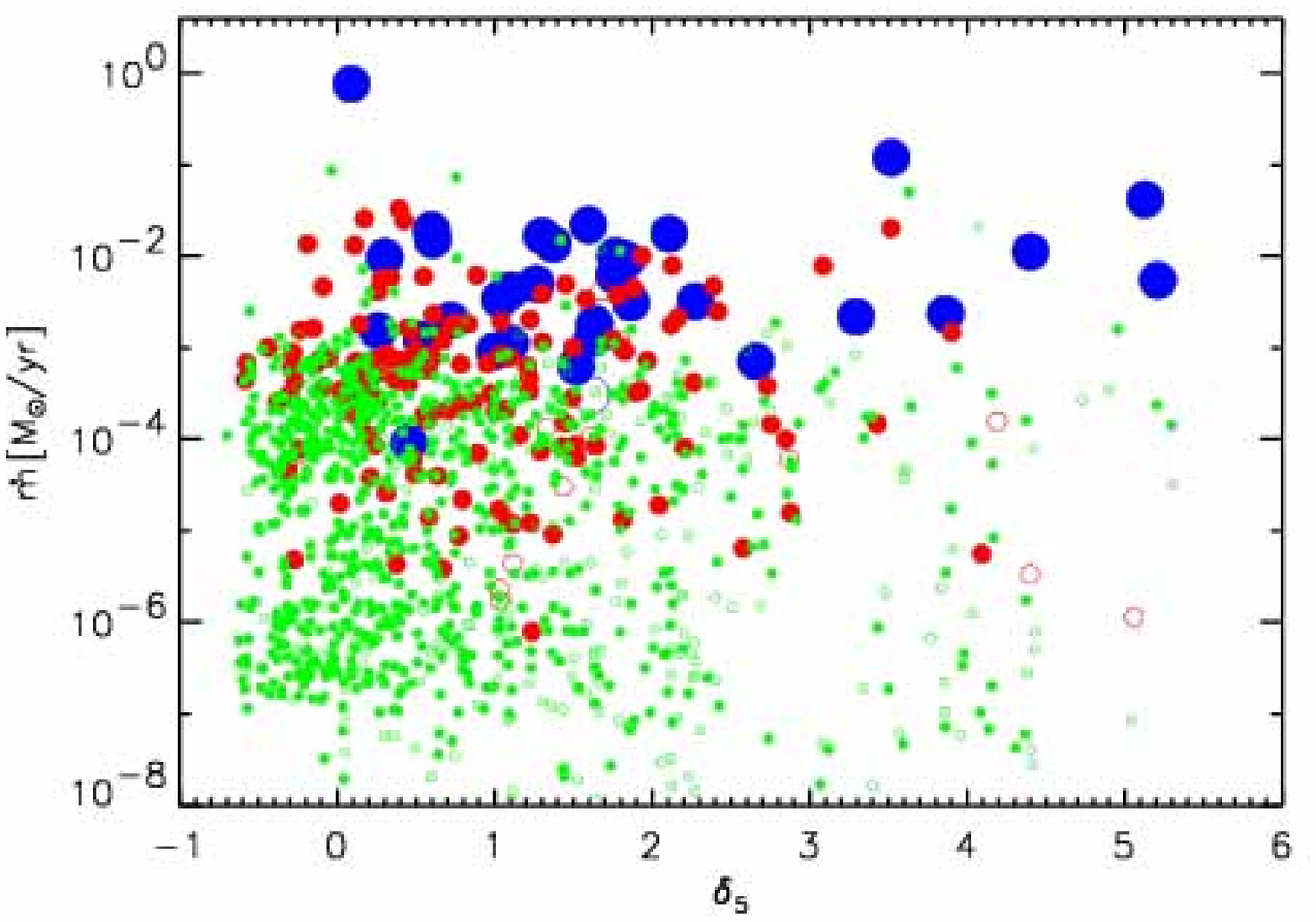}
        \end{center}
      \end{minipage}
    \end{tabular}
    \end{center}
    \vspace{-4mm}
  \caption{Left panel: $z=1$ BH overdensity $\delta_{5}$ versus mean BH 
           mass $m_{\rm BH}$ (using the same style as in
           Figures~\ref{fig:density_halomass} and \ref{fig:density_accretion})
           from the {\it BHCosmo} simulation. Right panel: The same for the mean
           accretion rate.}
\label{fig:d5}
\end{figure*}

\subsection{Environmental density estimates} \label{sec:density}

As seen in Figures~\ref{fig:density_halomass}, expressing the environment
through $r_{13}$ makes the most massive BHs lie predominantly in the densest
regions, while the lowest density regions are occupied mostly by small BHs.
As already indicated above, while observational studies have been increasingly
using an adaptive measure of density, in the past, studies often determined
the density around a position in space using spheres of {\it fixed} radius,
typically a few Megaparsecs (with 8\,$h^{-1}$\,Mpc a particularly common
choice). To what extent do our results depend on our choice of an adaptive
measurement of density? To investigate this question, we here repeat some of
our analysis by adopting a different measure of density. We define the
quantity $\delta_5$, the overdensity in spheres of radius 5\,$h^{-1}$\,Mpc,
around the position of a BH\footnote{Given the somewhat limited size of our
simulation volume, we cannot use significantly larger spheres than this.},
{\it which is equivalent to smoothing the mass distribution with a Top Hat
kernel of size 5\,$h^{-1}$\,Mpc}. The two panels in Figure~\ref{fig:d5} show
plots equivalent to the upper left panels in
Figures~\ref{fig:density_halomass} and \ref{fig:density_accretion}, but with
$\delta_5$ replacing $r_{13}$. The relatively tight relationship between the
densest region and the most massive BHs from Figure~\ref{fig:density_halomass}
disappears. This is not surprising and it is due to the fact that
5\,$h^{-1}$\,Mpc is much larger than the virial radius of one of the large
haloes in the simulation, so $\delta_5$ is a probe of the mass {\it inside and
around} these haloes. In other words, while $\delta_5$ contains information
about a scale of 5\,$h^{-1}$\,Mpc, which is a typical scale for black holes
sitting in a small group or filament like enviroment, it is way too large for
black holes in larger groups/clusters. The right panel of Figure~\ref{fig:d5}
shows most the density dependence from Figure~\ref{fig:density_accretion}
washed out, with virtually no dependence of the accretion rate on
environment. As already mentioned, using $\delta_5$ amounts to smoothing the
mass distribution with a filter of that size, thus erasing all information on
smaller scales. We are thus led to conclude that there exist environmental
trends for the black holes in our simulation volume, with the scales on which
these trends can be found being below 5\,$h^{-1}$\,Mpc. In other words, the
environmental differences found in our simulation volume exist on scales which
are {\it non--linear}, a situation comparable to what is found observationally
(see, for example, Kauffmann et al. 2004, Blanton \& Berlind 2007, Park et
al. 2007 -- all for $z=0$). This is no surprise also because both star
formation and black hole accretion depend strongly on local density enhancements.


%
%
\begin{figure*}
  \begin{center}
    \begin{tabular}{cc}
      \hspace{0.85cm}
      \begin{minipage}{51mm}
        \begin{center}
          \includegraphics[width=51mm]{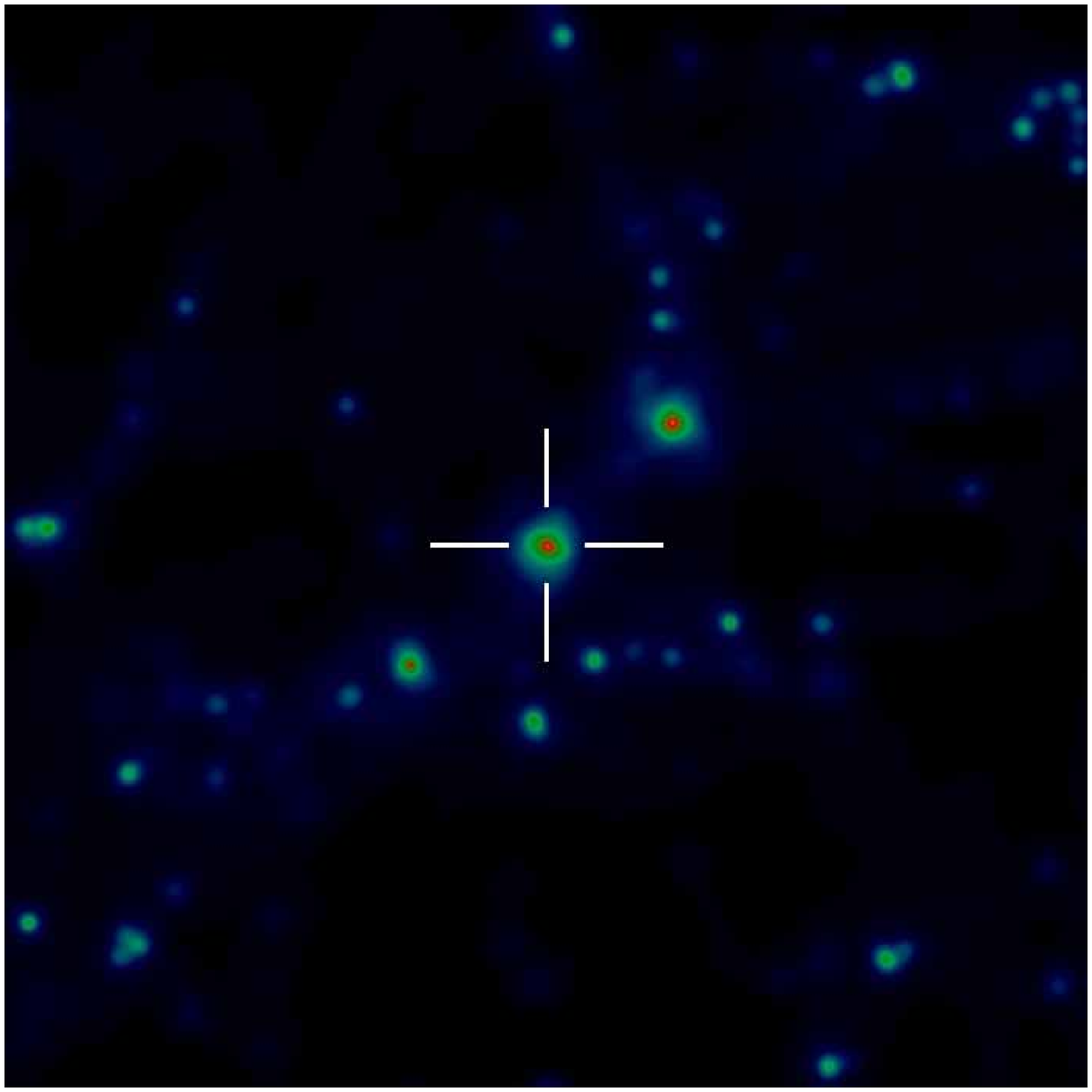}
        \end{center}
      \end{minipage}
      \hspace{0.3cm}
      \begin{minipage}{51mm}
        \begin{center}
          \includegraphics[width=51mm]{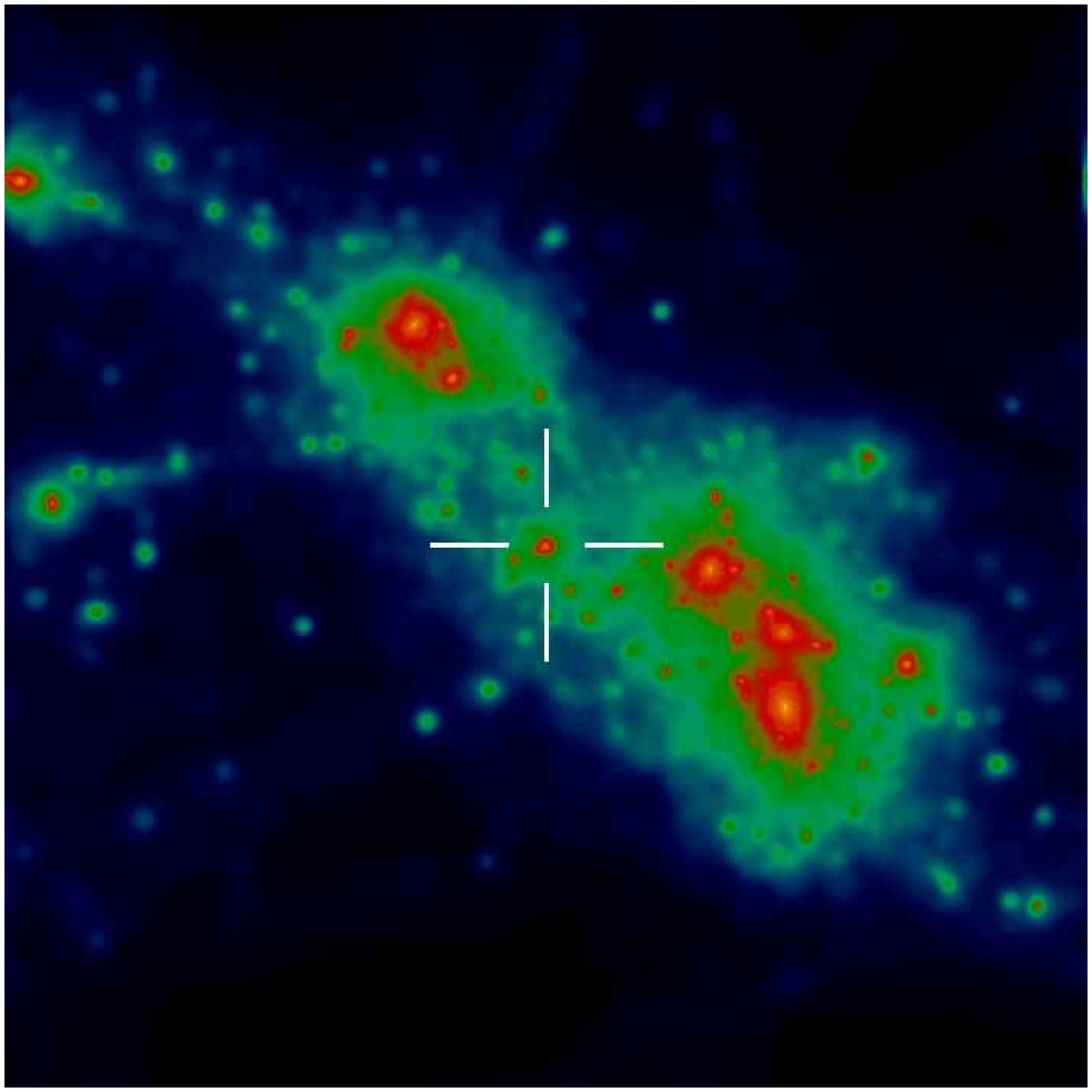}
        \end{center}
      \end{minipage}
      \hspace{0.3cm}
      \begin{minipage}{51mm}
        \begin{center}
          \includegraphics[width=51mm]{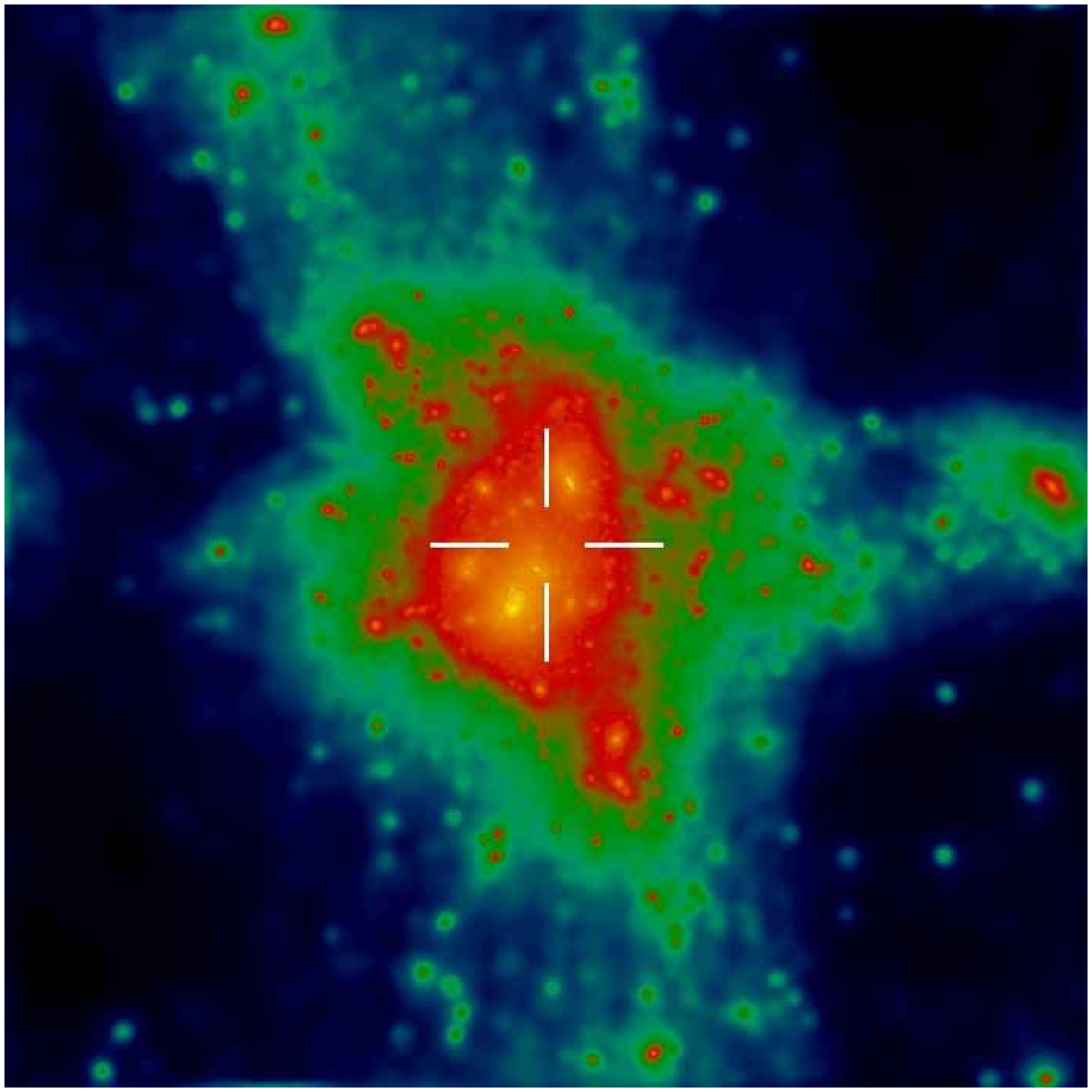}
        \end{center}
      \end{minipage}
    \end{tabular}
    \vspace{1mm}
    \begin{tabular}{cc}
      \begin{minipage}{65mm}
        \begin{center}
          \includegraphics[width=65mm]{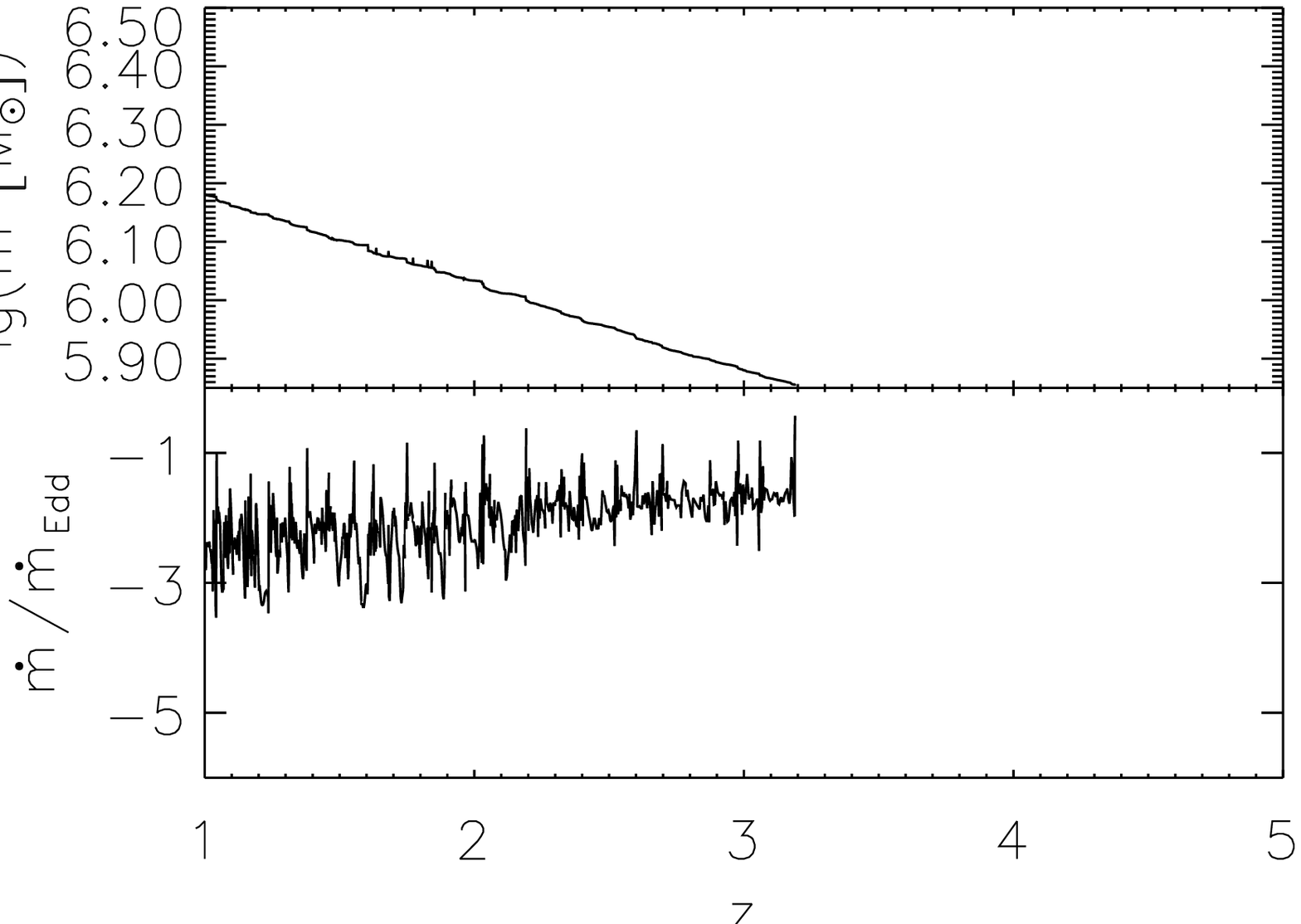}
        \end{center}
      \end{minipage}
      \hspace{-1.1cm}
      \begin{minipage}{65mm}
        \begin{center}
          \includegraphics[width=65mm]{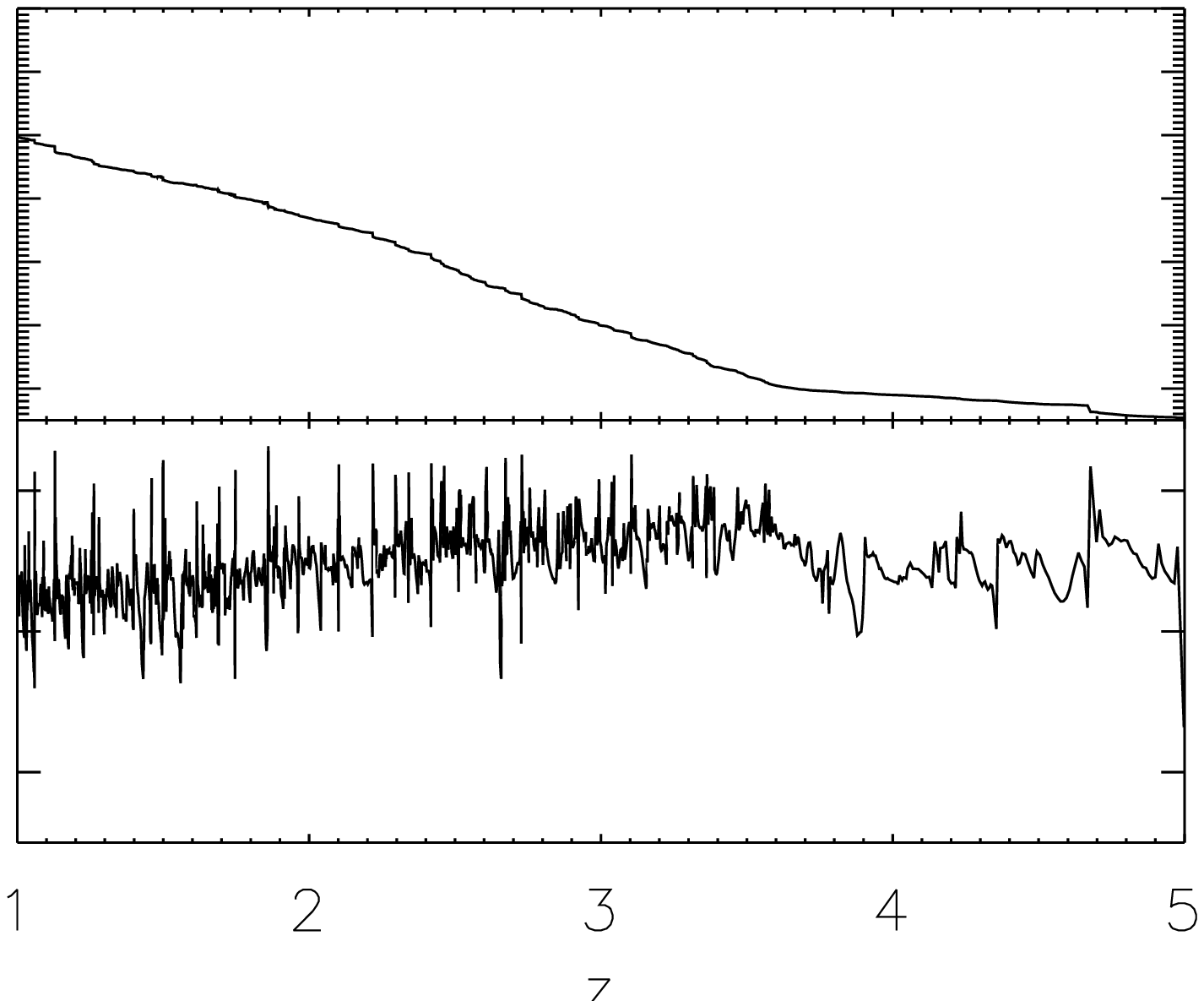}
        \end{center}
      \end{minipage}
      \hspace{-1.1cm}
      \begin{minipage}{65mm}
        \begin{center}
          \includegraphics[width=65mm]{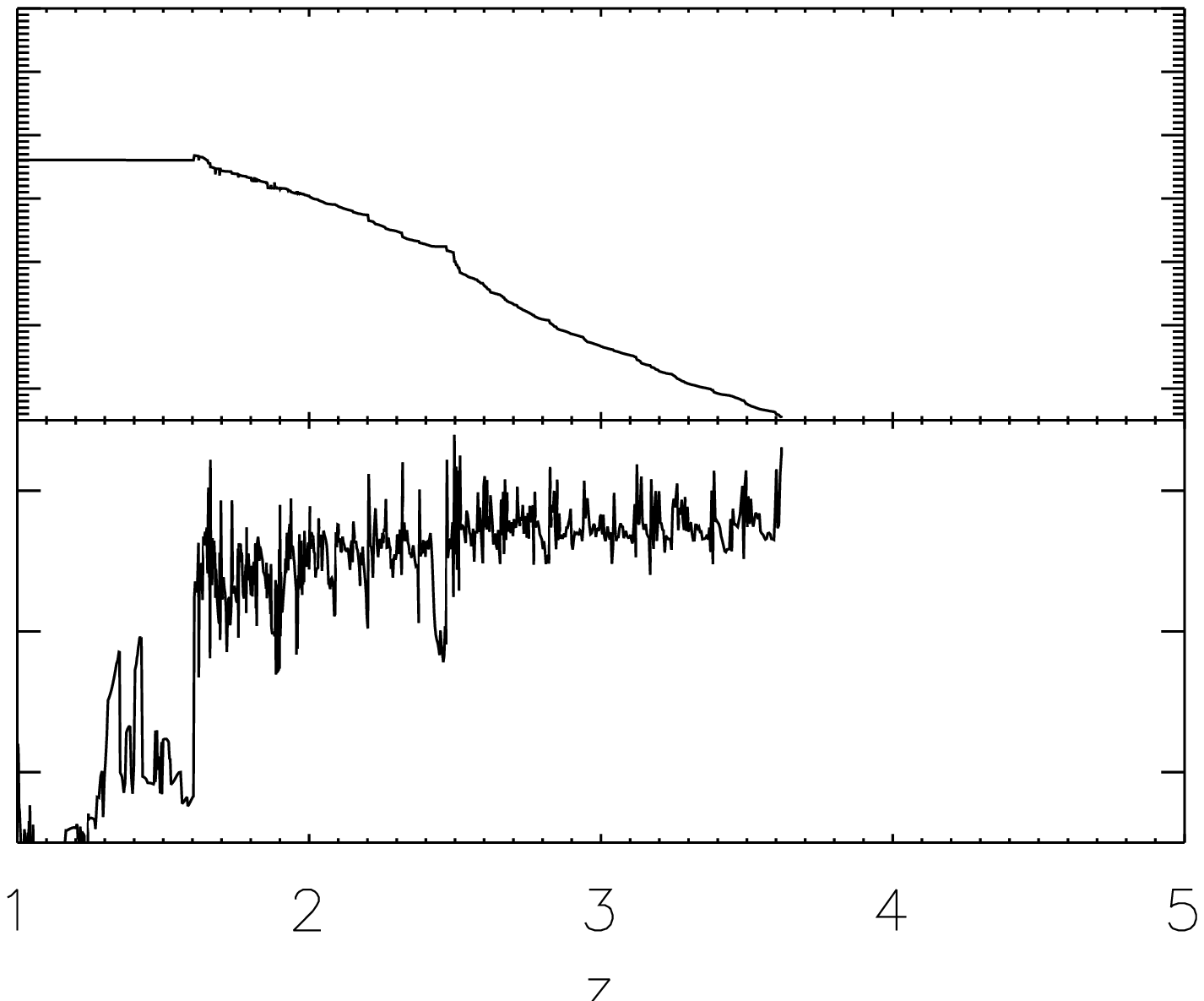}
        \end{center}
      \end{minipage}
    \end{tabular}
    \vspace{-3mm}
    \begin{tabular}{cc}
      \hspace{-1.0cm}
      \begin{minipage}{76mm}
        \begin{center}
          \includegraphics[width=76mm]{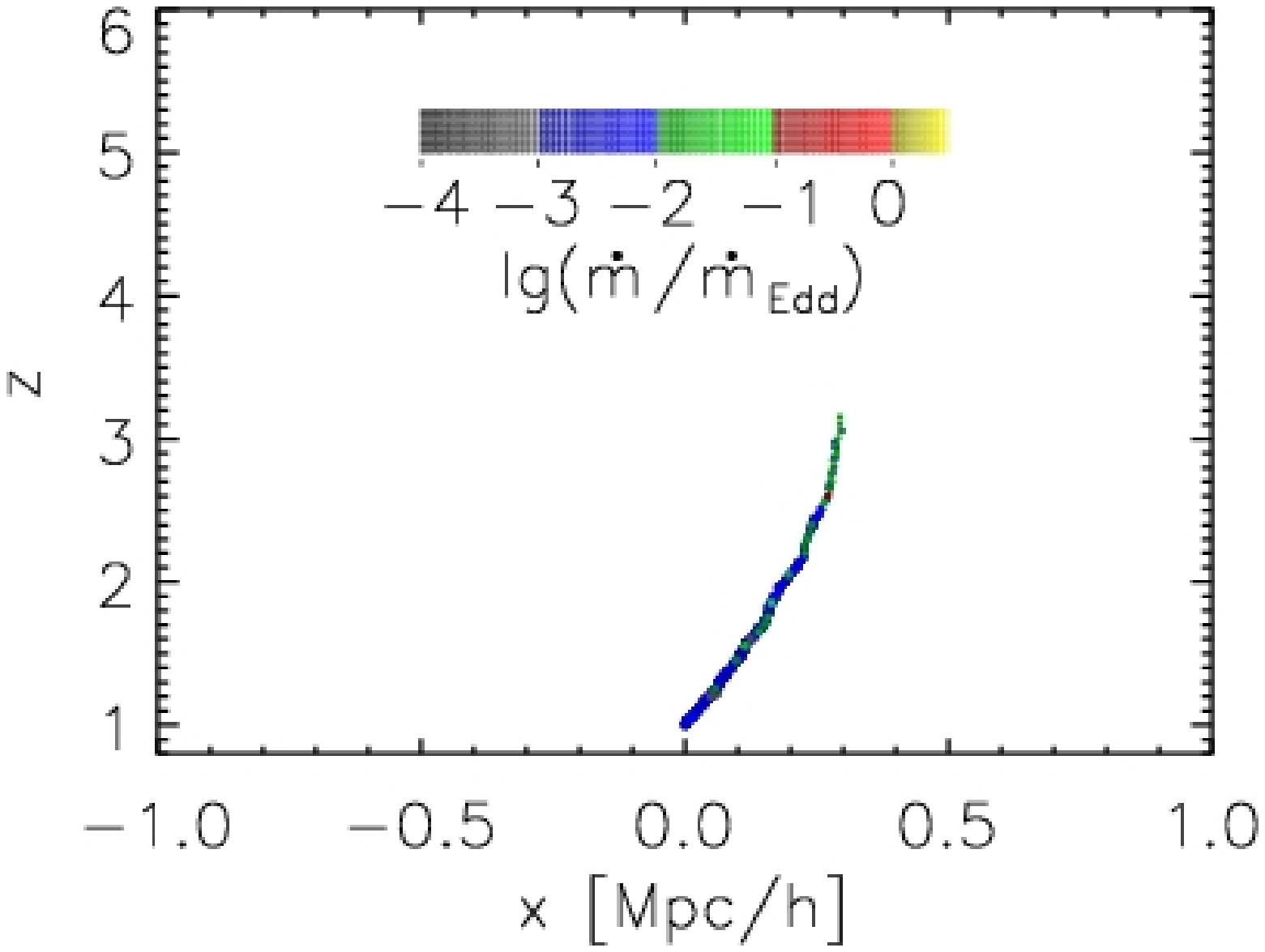}
        \end{center}
      \end{minipage}
      \hspace{-2.2cm}
      \begin{minipage}{76mm}
        \begin{center}
          \includegraphics[width=76mm]{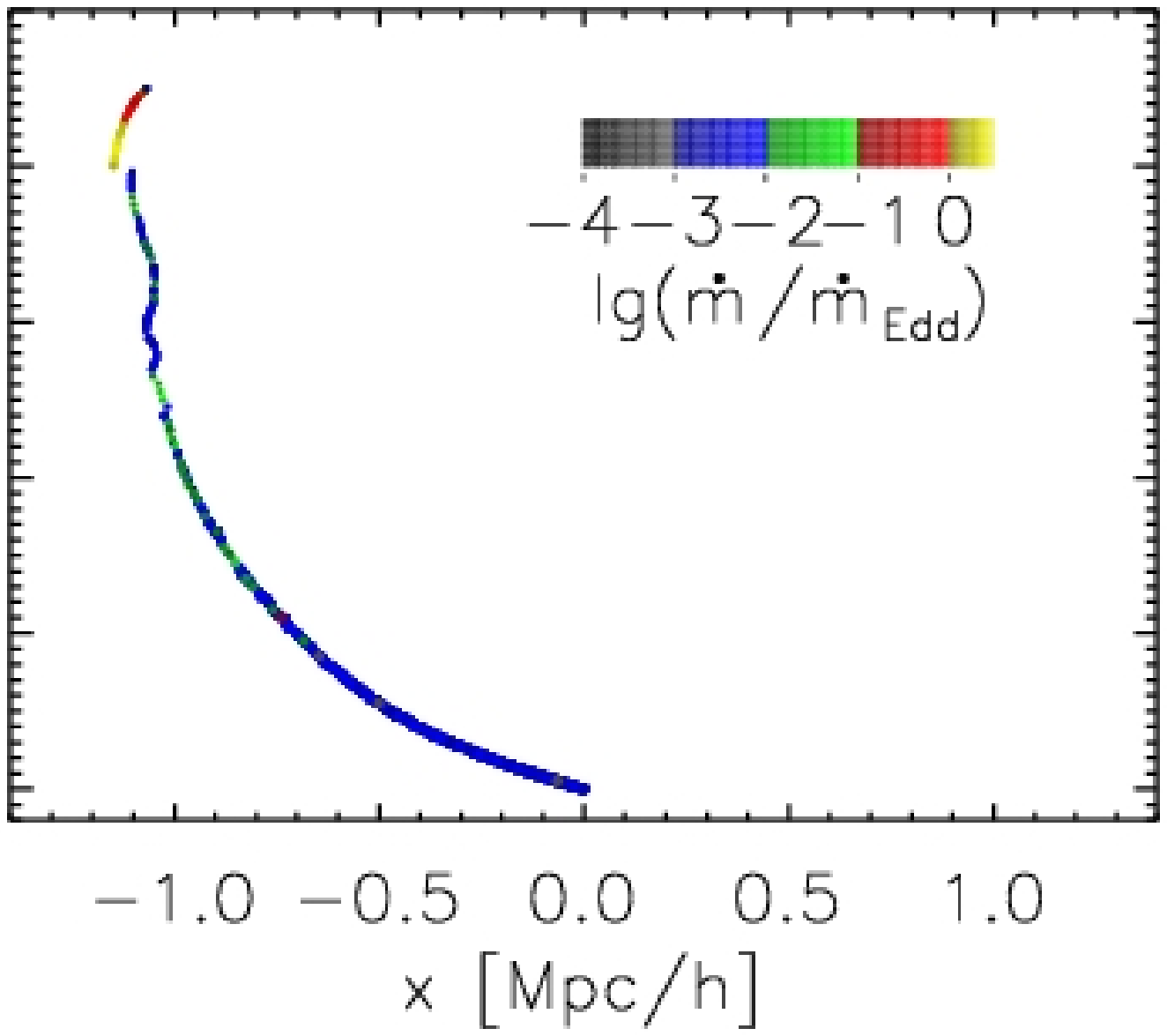}
        \end{center}
      \end{minipage}
      \hspace{-2.2cm}
      \begin{minipage}{76mm}
        \begin{center}
          \includegraphics[width=76mm]{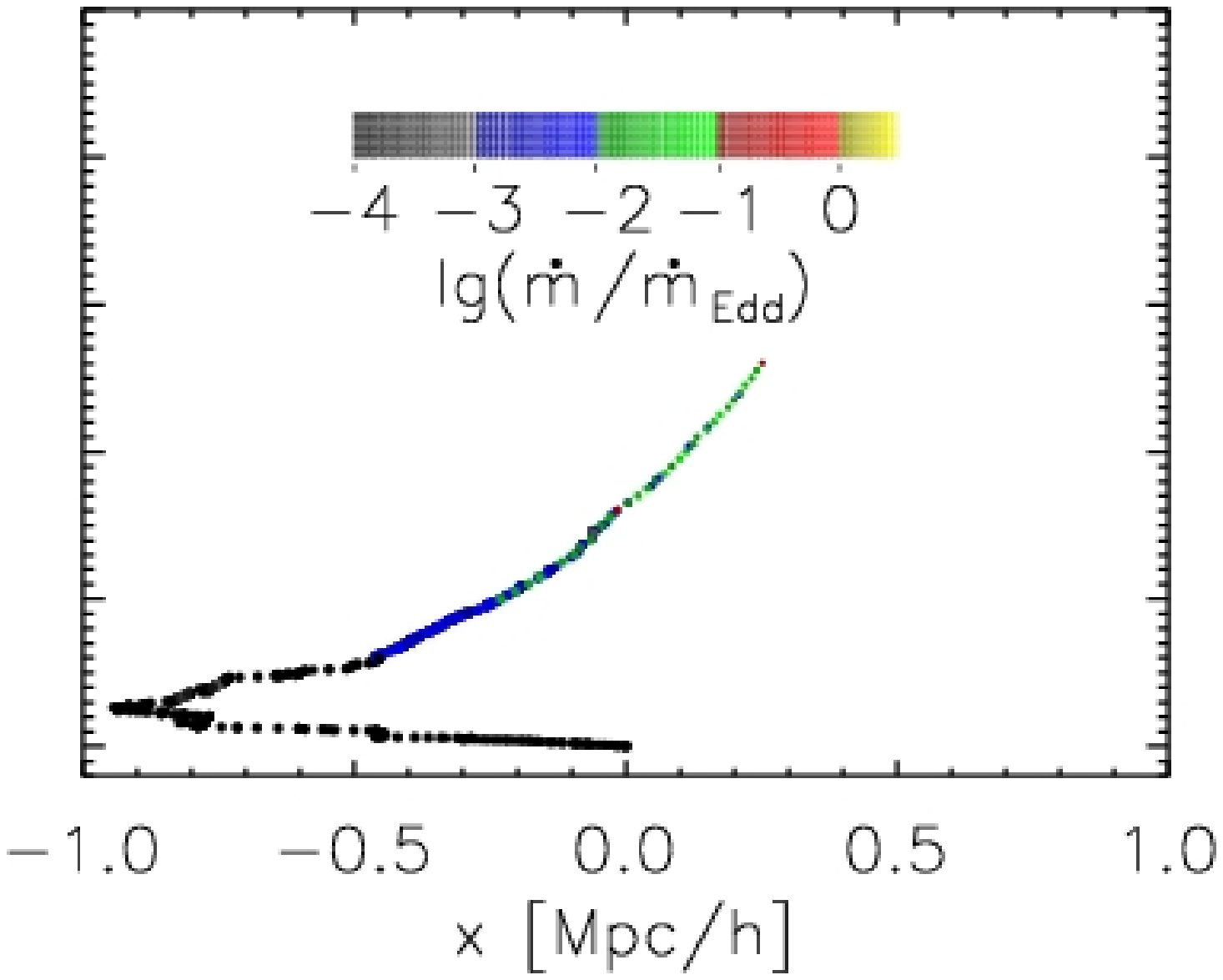}
        \end{center}
      \end{minipage}
    \end{tabular}
   \end{center}
   \vspace{-0.3cm}
  \caption{Three low--mass BH in regions of different density. From left to
           right, depicted are a low--density region, a high--density region,
           and a cluster. The halo masses and the local densities are
           $1.63\cdot 10^{11}$\,M$_\odot$ and $r_{13}=4.0\,h^{-1}$\,Mpc (left
           column), $2.8\cdot 10^{11}$\,M$_\odot$ and $r_{13}=1.2\,h^{-1}$\,Mpc
           (central column), and $6.25\cdot 10^{13}$\,M$_\odot$ and
           $r_{13}=0.27\,h^{-1}$\,Mpc (right column). The images show the 
           adaptively smoothed dark--matter density in volumes of size 
           ($3.0\,h^{-1}$\,Mpc)$^3$. Each image is centered on the location
           of the black hole, which is also marked with cross hairs.}
\label{fig:regions}
\end{figure*}

\subsection{Low--Mass BHs and their environments} \label{sec:starvation}

In Figure~\ref{fig:zf_vs_halo_mass}, there exists an interesting population of
black holes with formation redshifts higher than 2.5. This population contains
only a small number of the intermediate--mass black holes and about half of
the high--mass ones, but large numbers of black holes with masses lower than
$10^{7}$\,M$_\odot$. These latter BHs are composed of two populations, main
BHs present in small (less massive than $\approx 10^{12}$\,M$_\odot$) haloes
and satellite BHs embedded in larger haloes.

The non--satellite low--mass BHs are an interesting population of objects, 
especially since they cover the full range of local densities but a fairly
narrow range of host halo masses. For these black holes to have such low
masses and high formation redshifts, they cannot have been accreting 
much gas (or experience many mergers). Such conditions are expected to
exist in low--density regions such as voids, where the growth of haloes is 
truncated towards low--mass haloes, with very slow growth with redshift 
(Gottl\"ober et al. 2003). However, Croton et al. (2007) have shown 
that in the neighbourhoods of massive systems, and hence in high--density 
environments, most of the material is accreted onto those large haloes, 
leaving smaller haloes nearby to starve. We now investigate such stifled 
growth in the population of low--mass black holes.

To illustrate directly the variety of environments of the population of 
low--mass BHs in the simulation, in Figure~\ref{fig:regions} we show 
three black holes located in small haloes in a low--density region 
(left column), a high--density region (middle column), and inside a 
much larger halo (right column). The local densities and the halo 
masses for these black holes are $r_{13} = 4.0\,h^{-1}$\,Mpc, 
$r_{13} = 1.2\,h^{-1}$\,Mpc, and $r_{13} = 0.27\,h^{-1}$\,Mpc and 
$m_{\rm h} = 1.63\cdot 10^{11}$\,M$_\odot$, $m_{\rm h} = 2.8\cdot 
10^{11}$\,M$_\odot$, and $m_{\rm h} = 6.25\cdot 10^{13}$\,M$_\odot$ for 
the left, middle, and right columns, respectively. We have focused on 
the extreme ends of $r_{13}$, picking objects with roughly the same mass. 
In each column, the top panel shows the location of the black holes in 
the adaptively smoothed dark--matter distribution in a volume of size 
($3.0\,h^{-1}$\,Mpc)$^3$, with the panel centered on the black hole (the 
locations of the black holes are also marked with cross hairs). The central 
panel shows the mass growth (top parts) and accretion histories (bottom parts) 
of these black holes. Just like in Figure~\ref{fig:massive}, the bottom panel 
contains the merger trees.

Interestingly, there is virtually no merger activity for any of the three
black holes shown here. Their merger trees are reduced to a single branch
each.  In addition, accretion rates typically remain below 10\%
$\dot{m}_{Edd}$ most of the time. The BH mass growth histories of the objects
are very similar, too. It needs to be added, though, that some black holes in
this mass range do experience mergers. About 20\% of the black holes in the
lowest mass range have at least one merger. This has to be compared with rates
of 90\% and 100\% in the intermediate and high mass ranges, respectively (also
see Colberg et al. 2007). The merger ``trees'' show one minor 
difference between these black holes. In the low--density regions 
neither BHs or their halos undergo any mergers, so they move relatively small distances. 
The black hole in the high--density region and especially that accreted 
onto a larger halo have crossed larger distances (as the bottom left panel 
in Figure~\ref{fig:regions} shows, the trajectory of an accreted black 
hole might undergo quite severe changes). This is simply a result of dynamical
friction that leads the BH to sink into the central of the potential.

The right column of Figure~\ref{fig:regions} shows one of the most extreme
cases of low--mass BHs. This black hole is located inside a massive halo, so
its host halo merged with another halo at some point in the past.  Because of
the small number of simulation snapshots, we cannot pinpoint the merging time
accurately, but the accretion history provides a clue.  Accretion drops by
several orders of magnitude at a redshift of around $z=1.6$, to remain below
$10^{-3}\,\dot{m}_{Edd}$ until the final simulation output (incidentally, the
black holes trajectory also changes noticeably at around that time). As
discussed above (compare Figure~\ref{fig:density_accretion}), this is because
all gas was stripped from the black hole's host halo when it fell into the
larger halo. The rather abrupt shutting off of accretion onto the satellite BH
is remarkable implying that satellite galaxies in clusters of galaxies should
contain dormant black holes.


The existence of this population of low--mass black holes across a wide range
of environments is quite remarkable, especially since it extends all the way
into the lowest--density regions. The discovery of actively growing BHs in the
centers of galaxies inside voids, recently announced by Constantin et
al. (2007), is thus accounted for in the model discussed here (also compare 
Gallo et al. 2007 for a recent $z=0$ survey that includes low--mass BHs in the
mass range discussed here).  While
Constantin et al. (2007) use SDSS data at $z <1$, no major changes can be
expected for the black holes in our lowest--density regions (again, see
Gottl\"ober et al. 2003 for the growth of haloes in voids). In fact our $z=1$
population of low--mass BHs in underdense regions could be taken as the
progenitors of the void AGNs observed by Constantin et al. (2007). Our
simulation indicates that Constantin et al. do not observe an unusual category
of objects but merely the low--density environment tail of BHs, as was
also suggested when looking at semi--analytical models (Croton \& Farrar 2008).

%
%
\begin{figure}
\begin{tabular}{cc}
  \hspace{-0.6cm}
  \begin{minipage}{90mm}
    \begin{center}
      \includegraphics[width=90mm]{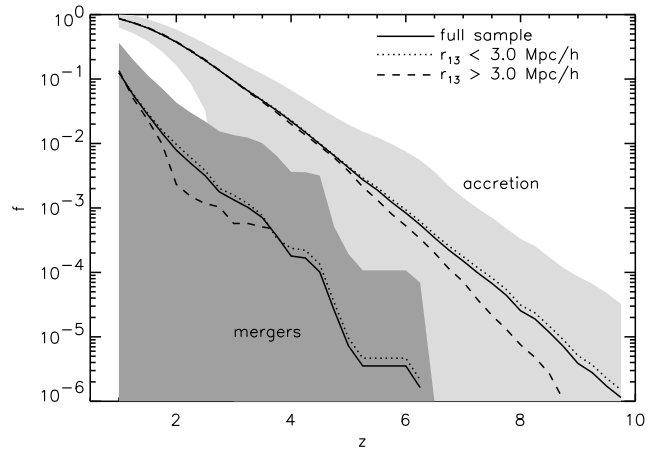}
    \end{center}
  \end{minipage}
\end{tabular}
\vspace{-0.4cm}
\caption{The cumulative growth of the black hole mass as a function of
         redshift. Shown is the growth of the main progenitor, averaged over
         each sample, and the growth is divided into growth through
         accretion and mergers. Black holes of all masses are combined in  
         this Figure. While the solid lines show the full samples, the 
         dotted and dashed lines give black holes in overdense ($r_{13} 
         < 3\,h^{-1}$\,Mpc) and underdense ($r_{13} \ge 3\,h^{-1}$\,Mpc) 
         regions, respectively. For the full sample, the superimposed 
         solid area shows the scatter around the mean.}
\label{fig:assembly}
\end{figure}

\subsection{BH growth by mergers and accretion in Different Environments} \label{sec:assembly}

Given the findings from the previous Section, it an important question to
see how universal they are. In particular, we now investigate the question 
of how the environment may affect the modes of black hole growth: 
accretion and mergers. In order to do this we study the growth of the most 
massive progenitor of each black hole as a function of redshift, and 
calculate the contributions to the BH mass due to accretion and mergers 
separately and how they depend on local environments.

Figure~\ref{fig:assembly} shows the {\it averaged} cumulative fraction $f$ of
the final BH mass built up through mergers and accretion as a function of
redshift (at $z=1$ the sum of the two fractions equals one). The solid lines
show the respective fractions for the whole BH sample while the dotted and
dashed lines show $f$ in overdense ($r_{13} < 3\,h^{-1}$\,Mpc and underdense
($r_{13} \ge 3\,h^{-1}$\,Mpc) regions respectively. The {\it rms} around the
mean for the full sample as shaded regions -- the scatter for subsamples is of
comparable size. Figure~\ref{fig:assembly} shows a mild dependence of the mode
of black hole growth with environment. In particular, underdense regions show
reduced accretion at $z > 3$ and somewhat less mergers at $z < 3$ with respect
to the high density regions (or the full mean). The reduced merger activity in
underdense regions is fully compatible with the evolution of the mass function
of haloes in voids (see, for example, Gottl\"ober et al. 2003 or Goldberg \&
Vogeley 2004). In voids, the mass function of haloes covers a reduced mass
range, and its growth is delayed compared with the growth of the mass function
in the $\Lambda$CDM cosmology. It is also expected that the fraction of BH
mass contributed by accretion will also be smaller in underdense regions. 
Overall, the environments affects the mode of BH mass assembly reducing
accretion at high--$z$ and mergers at intermediate redshift. However, the final 
black hole mass fraction at $z=1$ contributed by these two mode remains very 
similar across dense or underdense environments. Our finding is independent of
BH mass -- if split up into mass ranges as above the same trend is visible, the
only differences between the mass ranges being the times at which the different
mass BHs start to form, of course.

A comparison with the mass assembly histories studied in Maulbetsch et al. (2007)
is instructive -- keep in mind that they use a fixed--radius measure for
density, though. Maulbetsch et al. find that for $z=1$ (and higher), average
mass aggregation rates of haloes are higher in high--density regions -- a finding 
that agrees with what we see for the black holes in our simulation. 
Figure~\ref{fig:assembly} can also be compared with Figure~9 in Malbon et al. (2007). 
Since we do not split the BHs into mass bins Figure~\ref{fig:assembly} is dominated
by lower--mass BHs. Taking the different final redshifts into account, Malbon et
al.'s Figure~9 and Figure~\ref{fig:assembly} agree qualitatively: For low--mass
BHs, accretion is dominant at any redshift (and in any environment). However, it
is worthwhile to note how our results indicate that BH growth, both via
accretion and mergers, is relevant at redshifts higher than than those seen in
Malbon et al. (2007), probably due to the much better mass resolution in our
direct simulation.

%
%
\begin{figure}
\begin{tabular}{cc}
  \hspace{-0.6cm}
  \begin{minipage}{90mm}
    \begin{center}
      \includegraphics[width=90mm]{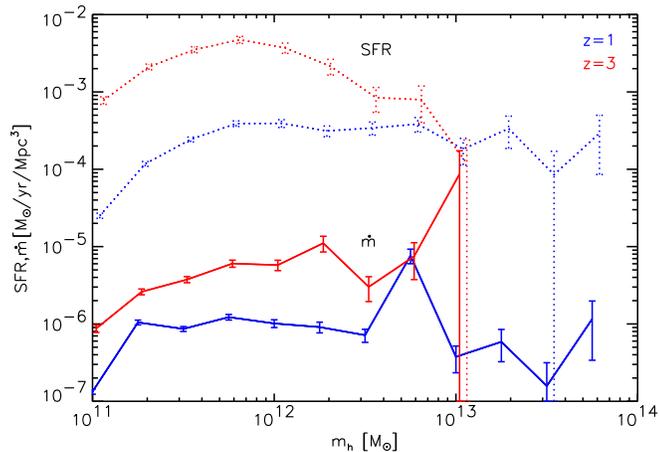}
    \end{center}
  \end{minipage}
\end{tabular}
\vspace{-0.4cm}
\caption{Star formation (dotted lines) and accretion density (solid lines) 
         as a function of halo mass for redshifts $z=1$ (blue) and $z=3$.
         See main text for full details.}
\label{fig:downsizing}
\end{figure}

\section{Summary and Discussion} \label{sec:summary}
We have used high--resolution cosmological hydrodynamic simulations, the
first of its kind to fully model massive black holes in cosmological volumes,
to investigate the effects of the environment on the growth and evolution of
BHs. We have focused our analysis on the {\it BHCosmo} simulation 
and complemented it with a second, larger volume (E6) to achieve higher 
statistics at high $z$ (see Di Matteo et al. 2007). In this work, we have 
made used of the full merger trees of each of the black holes in the 
simulations, comprising four and a half million progenitors in the more 
than three and a half thousand black holes in the simulation. Our results 
can be summarized as follows.

\begin{itemize}
\item While there is a well--defined correlation between the masses of dark matter
haloes and of supermassive black holes at low redshifts, there are some
important deviations from the simple hierarchical picture.  A large population
of the most massive black holes are formed {\it anti--hierarchically}, with 
formation redshifts $z_f$ from $z \sim 2$ to $z \sim 4$, while a residual 
population of the massive BHs is assembled at lower redshift completely 
consistent with hierarchical assembly.
\item Comparing the growth histories of the masses of the most massive 
black holes and their host haloes' we have shown that between redshifts of 
around $z \simeq 6$ and $z \simeq 2$ black holes grow at a much faster rate 
than their host haloes.
\item We have found quite a tight relation between black hole mass
$m_{\mbox{\scriptsize{BH}}}$ and host halo mass $m_{\mbox{\scriptsize{h}}}$ at
$z=1$, with $m_{\rm BH} \propto m_{\rm h}^{\sim 1.2}$ (for 
$m_{\mbox{\scriptsize{h}}} \ge 10^7$\,M$_\odot$), consistent with 
observational results by, e.g., Shankar \& Mathur (2007). The relation 
becomes slightly shallower ($m_{\rm BH} \propto m_{\rm h}^{\sim 1.0}$) 
if we include less massive black holes. The BH accretion rate $\dot{m}$ 
displays a very similar behaviour ($\dot{m} \propto m_{\rm h}^{\sim 1.0}$), 
albeit with much larger scatter. In particular, for the most active black 
holes ($\dot{m} \ge 10^{-2}$\,M$_\odot$/year), the correlation of 
accretion rate with halo mass disappears.
\item Using $r_{13}$, an adaptive measure of the local density, we find that
with the exception of satellites, more massive black holes live in more
massive haloes, and more massive haloes tend to live in denser environments
(roughly $M_{\rm BH} \propto (\rho/\rho_{mean})^{-1.5}$ in
Fig.~\ref{fig:density_halo_mass}). A similar overall density dependency is
found for black hole accretion rates, albeit with a far larger scatter.
Therefore, in general the largest and most active black holes are found in
group environments. Lower mass ($m_{\rm BH} < 10^8\,\Msun$) and
low accretion black holes are common in all environments, from cluster/group
to voids. Accretion rates typically increase with increasing redshift across
all environments.  The overall decrease in the mean accretion rates is
(Fig.~\ref{fig:density_accretion}) accompanied by a shallower density
dependency (Fig.~\ref{fig:density_accretion}) with decreasing redshift is in
line with observed downsizing in the AGN population (e.g. Steffen et al. 2003,
Ueda et al. 2003, Hasinger et al. 2005).
\item Figure~\ref{fig:accretion_SFR} links accretion rates of black holes and
SFRs of their hosts. For fixed accretion (SFR), there is a wide range in SFR
(accretion), across the full range of black hole mass ranges. This finding
indicates that the relationship between black hole activity and SFR is more
complicated than a simple one--to--one correlation (for recent observational
work see, e.g., Alonso--Herrero et al. 2007). Consistent with our
previous results (Di Matteo et al 2007), the difference in BH accretion and
SFR density dependeces further emphasizes a difference in the main triggering
mechanisms for quasars and star formation. Even though mergers lead both star
formation and black hole accretion, quasar activity is likely to require a
major merger. High gas density regions lead to star formation irrespective of
a major merger.  To further demonstrate this, in Figure~\ref{fig:downsizing},
we plot the mean accretion rate density (in M$_\odot$/yr/Mpc$^3$) as a
function of halo mass for redshifts $z=1$ (blue solid line) and $z=3$ (red
solid line), using Poisson error bars. This illustrates clearly that the peak
of the SFR occurs at lower halo masses than those where most quasar activity
takes place.  Whereas accretion rate tracks the major merger history
(effectively the growth of relatively massive halos), the SFR follows the
global build up of the relatively low mass halos hosting the most rapidly
star-forming galaxies (those with the highest concentration and highest
density at high $z$; see also Hopkins et al. 2007).  As both SFR and accretion
rate decline at lower redshift as a result of the declining gas fraction and
merger rates the two better trace each other's environments. Note that since
we do not track the merger histories of either dark matter haloes or of the
BH's host galaxies, we lack the data to distinguish between star formation 
triggered by a merger/interaction or by quiescent star formation.

\item There is a significant number of black holes with masses below 
$10^{7}$\,M$_\odot$ in haloes of mass $10^{12}$\,M$_\odot$ or less, 
spread out over the full range of environments, with formation redshifts of 
2.5 or higher. Their growth histories are very similar, with only minor 
differences between them. Mergers are rare (only 20\% of these black holes 
ever experience at least one merger). Thus, at $z=1$ there exists a fairly 
large population of quite inactive BH, all in small galaxy--sized haloes, 
across al cosmic environments, including, of course, the most underdense 
regions, the voids. These low--mass BHs and their slow growth account for 
the recently observed sample of such AGNs in voids (Constantin et al. 2007).
\end{itemize}

Measurements of the spatial clustering of quasars as a function of redshift
show a rapid increase of QSO bias with redshift (Croom et al. 2005, Myers et
al. 2006, Shen et al. 2007, Francke et al. 2008; also see White et al. 2007), 
from which then minimum halo masses can be
estimated. Croom et al. (2005) report a constant halo mass for their 2dF QSOs
of $m = (3.0 \pm 1.6)\cdot h^{-1}\,10^{12}$\,M$_\odot$, consistent with the
results by Shen et al. (2007). We have shown that the range of halo masses for
bright AGN in the simulation is broadly consistent with these results (see
Fig.~\ref{fig:BH_vs_halo_1}). Figure~\ref{fig:downsizing}, also summarizes
these results. Note that at both redshifts, the respective high--mass ends
are noisy because of the small number of objects. The distributions at both
these redshift are not really peaked, but are broadly consistent with the halo
masses derived from the quasars in the 2dF and Deep2 survey (Coil et
al. 2006). They also agree with the results obtained by Hopkins et al. (2007),
whose model shows the same mass range for quasar
environments at all redshifts. Please note, that while their model has to rely
on using constraints on halo/subhalo and halo occupation distributions from
simulations (plus assumptions on what constitutes a major merger), our
simulation contains all of these ingredients as additional results.

Regardless of what all of these black holes would do between $z=1$ and $z=0$,
their behaviour up to $z=1$ is very consistent with observed observational
patterns. For more detailed studies, there clearly is need for further
simulations, especially in a larger cosmological volume up to $z=0$. But the
basic agreement of the simulations discussed here (and in Di Matteo et
al. 2007 and Sijacki et al. 2007) with observations supports the simple model
with which the formation and evolution of BHs is treated in the simulation
code. Studies of the formation and evolution of BHs in cosmological volumes
thus appear feasible.

\section*{Acknowledgments}

The simulations presented here were performed at the Pittsburgh
Supertcomputer Center (PSC). This work has been supported in part
through NSF AST 06-07819 and NSF OCI 0749212.
We thank Debora Sijacki and Volker Springel for valuable comments, which helped
to improve this work. JMC also thanks Darren Croton, Neal Katz, Ravi Sheth, 
and Michael Vogeley for discussions and suggestion.

\label{lastpage}

\end{document}